\documentclass[aps,prx,reprint,superscriptaddress]{revtex4-2}
\usepackage[T1]{fontenc}
\usepackage{hyperref}
\hypersetup{
  colorlinks   = true,
  urlcolor     = blue,
  linkcolor    = blue,    
  citecolor    = blue
}

\usepackage{siunitx}
\usepackage{physics,amsfonts,amsmath,graphicx,dcolumn,bm,array,color,amssymb,setspace,empheq,braket,multirow,cases,enumitem,xcolor}
\usepackage[ruled,lined]{algorithm2e}

\begin{document}

\title{Mode Multiplexing for Scalable Cavity-Enhanced Operations in \\
Neutral-Atom Arrays}

\author{Ziv Aqua}
\thanks{These authors contributed equally to this work.}
\affiliation{Department of Physics, MIT-Harvard Center for Ultracold Atoms and Research Laboratory of Electronics, Massachusetts Institute of Technology, Cambridge, Massachusetts 02139, USA}

\author{Matthew L. Peters}
\thanks{These authors contributed equally to this work.}
\affiliation{Department of Physics, MIT-Harvard Center for Ultracold Atoms and Research Laboratory of Electronics, Massachusetts Institute of Technology, Cambridge, Massachusetts 02139, USA}

\author{David C. Spierings}
\affiliation{Department of Physics, MIT-Harvard Center for Ultracold Atoms and Research Laboratory of Electronics, Massachusetts Institute of Technology, Cambridge, Massachusetts 02139, USA}

\author{Guoqing Wang}
\affiliation{Department of Physics, MIT-Harvard Center for Ultracold Atoms and Research Laboratory of Electronics, Massachusetts Institute of Technology, Cambridge, Massachusetts 02139, USA}
\affiliation{International Center for Quantum Materials, School of Physics, Peking University, Beijing 100871, China}

\author{Edita Bytyqi}
\affiliation{Department of Physics, MIT-Harvard Center for Ultracold Atoms and Research Laboratory of Electronics, Massachusetts Institute of Technology, Cambridge, Massachusetts 02139, USA}

\author{Thomas Propson}
\affiliation{Department of Physics, MIT-Harvard Center for Ultracold Atoms and Research Laboratory of Electronics, Massachusetts Institute of Technology, Cambridge, Massachusetts 02139, USA}

\author{Vladan Vuleti\'c}
\thanks{Contact author: vuletic@mit.edu}
\affiliation{Department of Physics, MIT-Harvard Center for Ultracold Atoms and Research Laboratory of Electronics, Massachusetts Institute of Technology, Cambridge, Massachusetts 02139, USA}

\date{\today}
             
\begin{abstract}
Neutral atom arrays provide a versatile platform for quantum information processing. However, in large-scale arrays, efficient photon collection remains a bottleneck for key tasks such as fast, nondestructive qubit readout and remote entanglement distribution. We propose a cavity-based approach that enables fast, parallel operations over many atoms using multiple modes of a single optical cavity. By selectively shifting the relevant atomic transitions, each atom can be coupled to a distinct cavity mode, allowing independent simultaneous processing. We present practical system designs that support cavity-mode multiplexing with up to 50 modes, enabling rapid mid-circuit syndrome extraction and significantly enhancing entanglement distribution rates between remote atom arrays. This approach offers a scalable solution to core challenges in neutral atom arrays, advancing the development of practical quantum technologies.
\end{abstract}

\maketitle
             
\section{Introduction}
\label{sec:intro}

Neutral-atom arrays are a leading platform for fault-tolerant quantum computing, offering high-fidelity single- and two-qubit gates~\cite{levine2022dispersive,evered2023highfidelity,tsai2025benchmarking,radnaev2025universal,peper2025spectroscopy,muniz2025highfidelity} with arbitrary connectivity enabled by coherent transport of qubits~\cite{beugnon2007twodimensional,bluvstein2022quantum}. These capabilities have already supported demonstrations of fault-tolerant logical operations in small quantum error correction (QEC) codes~\cite{bluvstein2024logical,bluvstein2025faulttolerant}. At the same time, systems with increasingly large array sizes have been realized~\cite{manetsch2025tweezer}, including recent demonstrations of continuous operation with up to 3000 atoms~\cite{gyger2024continuous,norcia2024iterative,chiu2025continuous}. Despite these advances, utility-level atomic quantum processors are expected to require millions of qubits operating under high-rate QEC cycles~\cite{zhou2025resource,fowler2012surface,ogorman2017quantum,gidney2021how,gidney2025how}.

Progress toward this regime can be advanced by improving two essential hardware-level operations: (i) nondestructive qubit readout for mid-circuit measurements (MCMs), enabling entropy removal in QEC codes~\cite{preskill1998reliable,terhal2015quantum,lis2023midcircuit,norcia2023midcircuit,bluvstein2025faulttolerant}, and (ii) remote entanglement distribution, enabling scalability through a network of interconnected quantum processing modules~\cite{cirac1999distributed,jiang2007distributed,nickerson2013topological,nickerson2014freely,monroe2014largescale,covey2023quantum,ramette2024faulttolerant,main2025distributed}. Both operations depend on collecting photons scattered by individual atoms in the array. In current systems, this is typically implemented using high numerical-aperture microscope objectives, but their limited collection efficiency ($\lesssim\!10\%$) restricts performance. Consequently, nondestructive readout and remote entanglement generation are generally bounded to timescales of several milliseconds~\cite{lis2023midcircuit,norcia2023midcircuit,bluvstein2025faulttolerant,young2022architecture,sinclair2025faulttolerant}, introducing a bottleneck for QEC cycle rates and hindering the feasibility of modular scale-up.

Optical cavities provide a promising solution, enabling high photon collection efficiencies by enhancing atomic emission into a well-defined mode via the Purcell effect~\cite{purcell1946proceedings}, while also supporting fast operations through cavity-mediated light-matter interactions. A key figure of merit in this context is the cooperativity, denoted by $\eta$, a dimensionless parameter characterizing the strength of the atom–cavity interaction. Recent experiments have demonstrated fast, high-fidelity, nondestructive qubit readout in systems with $\eta\!>\!1$~\cite{bochmann2010lossless,gehr2010cavitybased,deist2022midcircuit,hu2025siteselective,grinkemeyer2025errordetected}.
Cavities have also been used to realize atom–photon and remote atom–atom entanglement in both neutral-atom~\cite{wilk2007singleatom,weber2009photonphoton,lettner2011remote,reiserer2014quantum,reiserer2015cavitybased,ritter2012elementary,nolleke2013efficient,langenfeld2021quantum,daiss2021quantumlogic} and trapped-ion~\cite{stute2012tunable,krutyanskiy2019lightmatter,schupp2021interface,kobel2021deterministic,krutyanskiy2023entanglement,krutyanskiy2023telecomwavelength,krutyanskiy2024multimode} systems. Theoretical proposals further suggest that significantly higher entanglement generation rates can be achieved with heralded schemes and improved system parameters~\cite{young2022architecture,gao2023optimization,li2024highrate,sinclair2025faulttolerant}.

However, integrating optical cavities with large-scale neutral-atom arrays presents a significant technical challenge.
To date, cavity operations have been restricted to serial execution, with only one atom–cavity interaction occurring at a time. This limitation creates a bottleneck for protocols involving many atoms and hinders scalability to large arrays. Moreover, a fundamental trade-off arises from the choice of cavity geometry. Since cooperativity scales inversely with the mode area, reducing the cavity mode size generally enhances photon collection efficiency and strengthens atom–photon coupling, enabling faster cavity-mediated operations. On the other hand, tighter mode confinement restricts the number of atoms that can be simultaneously coupled to the cavity mode. As a result, implementations based on small cavities often require transporting atoms into and out of the cavity mode, introducing additional time overhead for large-scale operations. One alternative scaling strategy is to employ multiple cavities, each coupled to an individual atom~\cite{menon2024integrated,shaw2025cavity}. While this approach circumvents the single-mode constraints, it typically introduces different challenges, such as restricted optical access and reduced cavity performance. More broadly, an ideal integration approach must preserve the full functionality of neutral-atom arrays. In particular, the performance of two-qubit Rydberg gates in compact cavities can be compromised by stray electric fields arising from charges accumulating on nearby surfaces ($<\!\SI{100}{\micro m}$)~\cite{kubler2010coherent}, and mitigation strategies are active areas of research~\cite{wang2025can}. Collectively, these considerations motivate the need for a scalable cavity integration approach that simultaneously preserves the advantages of high cooperativity, accommodates large atom numbers, keeps the mirror surfaces at sufficient distance from the atoms, and remains generally compatible with the geometrical and other constraints of neutral-atom arrays.

In this work, we introduce an approach that enables parallel cavity-enhanced operations on many atoms via multiple modes of a single optical cavity.
Interfacing atoms with near-degenerate multimode cavities has previously been explored for various purposes, including tracking the motion of single atoms~\cite{horak2002optical,puppe2004singleatom}, mediating tunable-range atom-atom interactions ~\cite{vaidya2018tunablerange,kroeze2025directly}, realizing Laughlin states of light~\cite{clark2020observation}, and enhancing cooperativity~\cite{kroeze2023high}.
Here, we propose to selectively couple atoms to nondegenerate cavity modes by leveraging local light shifts -- building on techniques demonstrated in Ref.~\cite{hu2025siteselective}.
Addressing distinct longitudinal and transverse modes enables both frequency and spatial multiplexing of operations, a concept we term \textit{cavity-mode multiplexing} (CMM).
This approach substantially increases operational throughput, alleviating the single-mode trade-off between cavity size and processing rate. It thus allows the use of larger cavities that are naturally compatible with the geometry of atom arrays and can accommodate many qubits, thereby reducing requirements for atom transport. While CMM enables a wide range of applications, in this work we focus on two use cases: (i) rapid syndrome extraction in QEC cycles through parallelized, adaptive qubit measurements, and (ii) fast, fault-tolerant links between logical qubits in distinct modules, achieved via high-rate remote entanglement in a design compatible with intracavity Rydberg gates. For both tasks, when applied to large-scale atom arrays, CMM offers a potential speedup of approximately two orders of magnitude compared to architectures relying on free-space photon collection. Together, these capabilities establish CMM as a promising approach to fast and scalable modular quantum computing with neutral-atom arrays.

This paper is organized as follows. In Sec.~\ref{sec:concept}, we describe a system overview of CMM with neutral atoms in more detail, including the role of light shifts and mode sorting. We then present two applications of CMM: Sec.~\ref{sec:syndrome} outlines a cavity design tailored for fast syndrome extraction across thousands of qubits, while Sec.~\ref{sec:modular} introduces a different design optimized for high-rate remote entanglement generation and support for teleported CNOT gates between qubits in distant modules. Finally, Sec.~\ref{sec:diss} provides concluding remarks and discusses future directions.

\section{System Overview}
\label{sec:concept}

\begin{figure}
    \centering
    \includegraphics[width=\linewidth]{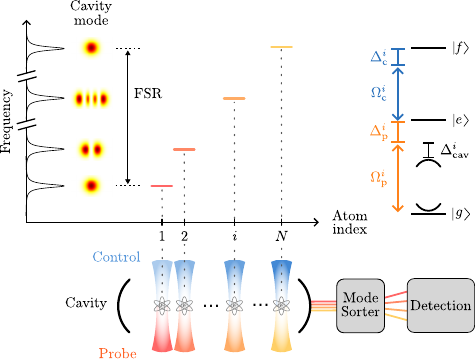}
    \caption{Conceptual depiction of cavity-mode multiplexing in an atomic register. The bare cavity spectrum is shown on the left vertical axis, where distinct longitudinal and transverse modes appear at different frequencies. Site-selective control beams (blue) are used to induce light shifts on the excited state $\ket{e}$ via the $\ket{e}\!\rightarrow\!\ket{f}$ transition, thereby tuning the $\ket{g}\!\rightarrow\!\ket{e}$ transition frequency of each atom to match a different cavity resonance. External probe beams at the corresponding frequencies (red-orange-yellow) excite the atoms, and the emitted photons are coupled into their respective cavity modes. At the cavity output, the different modes are separated by a mode sorter for detection.} 
    \label{fig:concept}
\end{figure}

We begin by outlining the physical configuration for implementing CMM with neutral atoms. Consider a register of single atoms trapped in optical tweezers inside a  Fabry–Pérot optical cavity, as illustrated in Fig.~\ref{fig:concept}.
Longitudinal modes of a fixed transverse profile (e.g., $\text{TEM}_{00}$) are evenly separated in frequency by the free spectral range (FSR) of the cavity. Within each FSR, higher-order transverse modes exhibit distinct resonance frequencies due to their differing spatial profiles and associated Guoy phase shift~\cite{siegman1986lasers}. Together, this structure provides a versatile set of spectrally resolved cavity modes that can be harnessed for multiplexing. 

The minimal atomic level configuration consists of three states: a ground state $\ket{g}$, an excited state $\ket{e}$, and a higher-lying excited state $\ket{f}$. The cavity mode couples to the $\ket{g}\!\rightarrow\!\ket{e}$ atomic transition. To locally control this atomic transition frequency, additional beams addressing each atom couple the excited state $\ket{e}$ to $\ket{f}$ with Rabi frequency $\Omega_c$ and detuning $\Delta_c$, inducing a light shift on $\ket{e}$. Since the control beams are far detuned from any ground-state transitions, $\ket{g}$ remains essentially unaffected. Local light shifts induced via excited-to-excited state transitions have been employed in several experimental settings~\cite{clark2020observation,burgers2022controlling,baum2023optical,norcia2023midcircuit,bornet2024enhancing,orsi2024cavity,bluvstein2025faulttolerant,chiu2025continuous}. In particular, this technique has recently been used to selectively decouple atoms from a single cavity mode~\cite{hu2025siteselective}. With additional polarization control to address specific magnetic sublevels, such shifts have also been proposed as a means to modify cavity-mediated photon–atom gate dynamics~\cite{aqua2025atommediated}.  

In our setting, the induced shift allows the $\ket{g}\!\rightarrow\!\ket{e}$ transition of each atom to be independently coupled near-resonantly to a separate cavity mode, effectively providing each atom with its own dedicated cavity mode. This configuration supports multiplexed operations across the entire atomic register. As an illustrative example, we consider the case where photons scattered by different atoms are simultaneously collected through separate cavity modes. As depicted in Fig.~\ref{fig:concept}, for each atom $i$ in the register, a probe field drives the shifted transition with Rabi frequency $\Omega^i_{p}$ and detuning $\Delta^i_{p}$. We denote the detuning of the corresponding cavity mode from the probe frequency by $\Delta^i_{\mathrm{cav}}$. Setting $\Delta^i_{\mathrm{cav}}\!=\!0$ ensures maximal photon collection efficiency. In this way, photons scattered by all atoms in the register are simultaneously collected through their assigned cavity modes with high efficiency.

Differentiating the light scattered by each atom requires resolving the corresponding transverse and longitudinal cavity modes. Transverse mode separation can be realized with multiplane light conversion (MPLC), which performs arbitrary unitary transformations of spatial modes through a sequence of phase masks separated by optical Fourier transform or free-space propagation~\cite{morizur2010programmable}. This technique has been employed to convert colocated high-order spatial modes into spatially distinct Gaussian beams, scaling to more than a thousand modes~\cite{fontaine2019laguerregaussian,fontaine2021hermitegaussian,zhang2023multiplane}. The MPLC can be implemented using a programmable spatial light modulator (SLM); however, for static mode sorting, where the required phase remains fixed, custom-fabricated phase masks offer higher spatial resolution at lower optical losses~\cite{zhang2023multiplane,choi2024quantum}. 
Longitudinal modes, separated in frequency by the cavity FSR, can be spatially separated using a high–resolving power dispersive element such as a virtually imaged phased array (VIPA)~\cite{shirasaki1996large,xiao2004dispersion}, which enables extremely high frequency resolution in excess of $\Delta f/f\sim10^6$ \cite{limbach2019fully,leonov2021application}. Alternatively, the re-imaging phased array (RIPA) spectrometer based on a self-imaging VIPA scheme can offer improved spatial profiles for better collection efficiency~\cite{wei202610}.
Depending on the intended application, the separated modes can then be imaged onto a camera or coupled into single-mode fibers and directed to single-photon detectors for high-efficiency counting.

Achieving multiplexing across many modes while maintaining low crosstalk requires the application of large light shifts to many atoms individually. To keep the required optical powers at practical levels, we consider on-resonance control light ($\Delta_c = 0$) that dresses the states $\ket{e}$ and $\ket{f}$, producing new eigenstates $\ket{e_\pm}\!=\!(\ket{e}\pm\ket{f})/\sqrt{2}$. This results in ground-to-dressed-state transitions with frequencies $\mp \Omega_c$ relative to the original $\ket{g}\!\rightarrow\!\ket{e}$ transition. We constrain the maximum frequency shift, $\Omega_c$, such that intensity fluctuations of 0.1\% produce frequency variations on the order of the excited-state linewidth $\Gamma$, which permits shifts of up to $2000\Gamma$. The mixing of $\ket{e}$ and $\ket{f}$ also reduces the atom–cavity coupling strengths by a factor of 2, and modifies the dressed-state linewidths compared to the bare state $\ket{e}$. Additionally, we note that beam pointing fluctuations and finite atomic temperature can lead to heating from dipole force fluctuations due to the large $\sim$GHz potential depth differences between ground states and excited dressed state $\ket{e_+}$, which we assume are well-controlled (see Appendix~\ref{app:control}). While the following discussion focuses on implementing CMM with $^{87}$Rb atoms, the concepts presented are readily applicable to other alkali and alkaline–earth species, as well as to ions.

\section{Syndrome extraction}
\label{sec:syndrome}

\begin{figure*}[t]
    \centering
    \includegraphics[width=\linewidth]{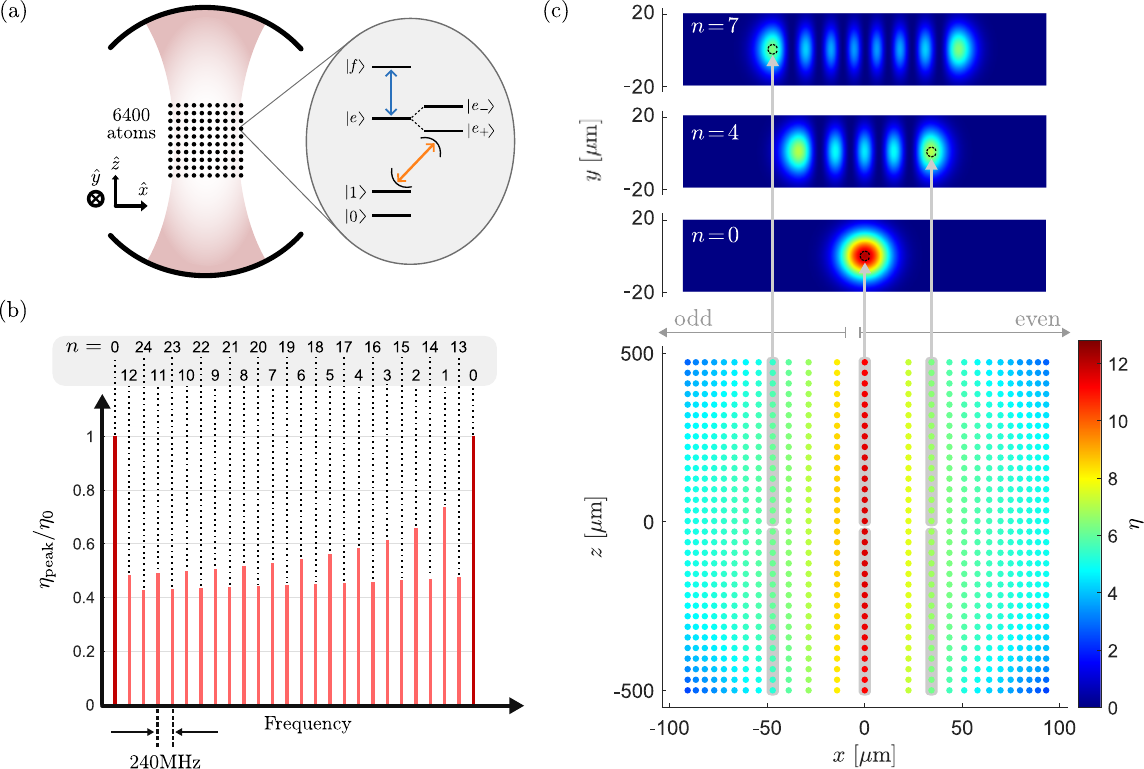}
    \caption{Cavity and atom-array configuration for syndrome readout using cavity-mode multiplexing with $\mathrm{HG}_{n,0}$ modes. (a)~Left: atom array with up to 6400 atoms, each coupled to a cavity mode. Right: atomic level scheme used for qubit readout. (b)~Cavity spectrum showing the peak cooperativity $\eta_{\text{peak}}$ of 25 equally spaced $\mathrm{HG}$ modes spanning the free spectral range. Cooperativity values are normalized to $\eta_0$, the peak cooperativity of the $\mathrm{HG}_{0,0}$ mode.
    (c)~Top: spatial profiles of $\mathrm{HG}_{n,0}$ modes with indices $n=0,4,7$. Bottom: cooperativity per atom across the array (every eighth atom shown). Columns of atoms are placed at the intensity maxima of the $\mathrm{HG}_{n,0}$ modes as indicated by the dashed circles, with odd (even) values of $n$ assigned to the positive (negative) $x$-axis. Each column is divided into two registers, which couple to two longitudinal cavity modes with the same transverse profile, separated in frequency by one free spectral range.} 
    \label{fig:readout_config}
\end{figure*}

Error-corrected quantum computing relies on MCM of syndrome qubits to identify and correct errors occurring in the data qubits encoding the logical quantum information. In neutral-atom arrays, syndrome extraction via free-space imaging typically requires several milliseconds, constituting a substantial fraction of the QEC cycle duration. Optical cavities, in contrast, enable nondestructive qubit readout on the microsecond timescale by coupling one of the qubit states, $\ket{1}$, to an excited state, $\ket{e}$, via the cavity mode, as illustrated in Fig.~\ref{fig:readout_config}(a).
Common techniques include fluorescence detection, where probe light addressing the $\ket{1} \!\rightarrow\! \ket{e}$ transition is applied via an external beam and the scattered photons are collected through the cavity~\cite{bochmann2010lossless,deist2022midcircuit,hu2025siteselective}; and transmission detection, where the presence of an atom in $\ket{1}$ suppresses resonant cavity transmission~\cite{boozer2006cooling,gehr2010cavitybased,deist2022midcircuit,grinkemeyer2025errordetected}. However, even with microsecond-scale qubit readout, sequentially measuring thousands of syndrome qubits extends the total duration to the millisecond range. In this regime, the added complexity of integrating a cavity offers limited benefit compared to parallel free-space imaging, which can simultaneously read out large qubit arrays.

For syndrome extraction, the ability to couple multiple qubits simultaneously to the same cavity mode provides a significant advantage, enabling more efficient readout strategies than sequential readout~\cite{hu2025siteselective}. In practical quantum computers, the physical qubit error rates are expected to be small. This implies that the vast majority of syndrome qubits will occupy the $\ket{0}$ state (indicating the absence of an error) while only a small fraction will be in the $\ket{1}$ state (signifying the presence of an error). This inherent bias can be exploited to perform global checks: rather than measuring each syndrome qubit individually, one can probe an entire batch of $n_\text{batch}$ atoms simultaneously to determine whether any are in $\ket{1}$. If no error is detected---an outcome that is likely given the low error probability---the procedure advances directly to the next batch. If an error is detected, an adaptive binary search can be performed by selectively decoupling subsets of atoms from the cavity mode through locally controlled light shifts, successively dividing the batch into two subsets and narrowing the set of candidate qubits until the faulty syndrome qubit is identified. In cases where multiple erroneous syndromes are present within the same batch, the global check and binary search sequence are repeated until all errors are located.
For an independent probability of a faulty syndrome $p_\text{synd}$, the expected number of queries $m$ required per batch is,
\begin{equation}
\label{eq:batch}
    m = 1 + n_{\mathrm{batch}} p_{\mathrm{synd}} \left[ 1 + \log_{2} \left( n_{\mathrm{batch}} \right) \right],
\end{equation}
where the first term corresponds to the single global check that is always performed on the full batch. The second term accounts for the additional queries required when faulty syndromes are present: each faulty syndrome triggers a binary search to localize it within the batch, followed by a global check to confirm that no further faulty syndromes remain within the batch. This process requires $1 + \log_{2}(n_{\mathrm{batch}})$ queries per faulty syndrome, and is multiplied by the expected number of faulty syndromes per batch, $n_{\mathrm{batch}} p_{\mathrm{synd}}$.

With a single cavity mode, the expected number of steps required to process $N$ syndrome qubits is  
\begin{equation}
    \# \; \text{steps} = \left\lceil \frac{N}{n_\text{batch}} \right\rceil \times m,
\end{equation}
where $\lceil N / n_\text{batch} \rceil$ is the number of batches that must be processed sequentially.  
For $p_{\mathrm{synd}}\!=\!5 \!\times\!10^{-3}$, a maximum batch size of $n_{\mathrm{batch}}\!=\!256$, and a query time of $\SI{10}{\micro s}$, the total readout duration enters the millisecond regime when $N\!\approx\!2000$ atoms, not accounting for additional time overhead from transporting atom batches in and out of the cavity.  
Therefore, even with adaptive searches, a single cavity mode does not provide a net advantage over parallel free-space imaging.
If, however, adaptive searches can be performed in parallel across multiple registers—each coupled to a distinct cavity mode—the scaling improves to,
\begin{equation}
\label{eq:scale}
     \# \; \text{steps} = \lceil \frac{N}{n_\text{batch}n_\text{modes}}\rceil \times m,
\end{equation}
where $n_{\mathrm{modes}}$ denotes the number of available modes. 

To realize this parallelism using CMM, we propose the following design, tailored to $^{87}$Rb atoms. We define the relevant atomic states as  
\begin{align}
    &\ket{0} \equiv \ket{5S_{1/2},F=1,m_F=1}, \\
    &\ket{1} \equiv \ket{5S_{1/2},F=2,m_F=2}, \nonumber \\
    &\ket{e} \equiv \ket{5P_{3/2},F'=3,m_F=3}, \nonumber \\
    &\ket{f} \equiv \ket{4D_{5/2},F''=4,m_F=4}. \nonumber
\end{align}
These qubit states are employed solely during readout; the qubit need not remain encoded in these states throughout computation. Since the measurement projects the qubit into either $\ket{0}$ or $\ket{1}$, coherence preservation is not required at this stage, which simplifies changing the basis from the computational encoding—typically the clock states with $m_F=0$—to the readout encoding. It is important to note that the use of the cycling transition for $\ket{e}\!\rightarrow \!\ket{f}$ prevents mixing with other states that could induce readout errors via depumping channels from $\ket{1}$ to $\ket{0}$ or out of the logical subspace.

For this atomic configuration, we consider a Fabry–Pérot optical cavity characterized by the parameters summarized in Table~\ref{tab:cavity_params_readout}. We tune the cavity such that the higher-order Hermite–Gauss modes along the transverse x-axis, $\mathrm{HG}_{n,0}$ with $n \in [0,24]$, are arranged to be equally spaced in frequency, with a $\SI{240}{MHz}$ separation between adjacent modes. Using these HG modes across two FSRs provides access to a total of 50 equally spaced cavity modes (see Fig.~\ref{fig:readout_config}(b)). The control beams, resonant with the $\ket{e}\!\rightarrow\!\ket{f}$ transition, are used to dress the excited states, allowing to tune the $\ket{1}\!\rightarrow\!\ket{e_+}$ transition of each atom to couple to the appropriate cavity mode. The total frequency range spans $12\,\mathrm{GHz}$, and with $0.1\%$ intensity stability of the control field, the resulting variation of the ground-to-dressed-state transition frequency is limited to the linewidth of $\ket{e}$, $\Gamma/(2\pi)=\SI{6}{MHz}$.

\begin{table}[h]
\centering
\caption{Parameters of the optical cavity used for syndrome extraction.}
\label{tab:cavity_params_readout}
\begin{tabular}{l @{\hspace{1cm}} c}
\hline\hline
Cavity length & $\SI{25}{mm}$ \\
Mode waist & $\SI{20}{\micro m}$ \\
Rayleigh range & $\SI{1.6}{mm}$\\
Finesse & $6 \times 10^{4}$ \\
FSR & $\SI{6}{GHz}$ \\
Linewidth (FWHM) & $2\pi\times\SI{100}{kHz}$ \\
\hline\hline
\end{tabular}
\end{table}

In addition to parallelized syndrome extraction, this architecture increases the number of atoms that can be accommodated within the cavity by exploiting the spatially separated intensity maxima of higher-order HG modes. Rather than being restricted to the spatial extent of the fundamental $\mathrm{HG}_{00}$ mode, columns of atoms are arranged along the $z$-axis at the intensity maxima of each higher-order HG mode, positioned at different points along the $x$-axis (Fig.~\ref{fig:readout_config}(c)). To reduce crosstalk and increase the spatial separation between adjacent columns of atoms, we assign modes with odd (even) $n$ to the positive (negative) $x$-axis, taking advantage of the symmetry of the intensity peaks about the $y$-axis. The resulting square array geometry is compatible with parallel control of atomic motion using two-dimensional acousto-optic deflectors (AODs). With a minimum spacing of about $\SI{4}{\micro m}$ between adjacent atoms, this arrangement allows up to $6400$ atoms to couple to some mode of the cavity at once, substantially exceeding what is achievable in a one-dimensional arrangement. As a result, simultaneous operations can be performed on 50 distinct batches of $128$ atoms each.

Figure~\ref{fig:readout_config}(d) shows the cooperativity $\eta$ per atom for the $\ket{1}\!\rightarrow\!\ket{e_{+}}$ transition, accounting for (i) the reduction in atom–cavity coupling and decay rate due to the dressed-state nature of $\ket{e_+}$, (ii) the Gaussian mode Rayleigh range, which reduces coupling for atoms farther from the cavity center along $z$, (iii) the electric field profile of higher-order modes, which results in lower atom-cavity coupling as $n$ increases, and (iv) the spatial extent of the atomic probability distribution in the tweezers for a temperature of $\SI{10}{\micro K}$ (see Appendix~\ref{app:temp}). The resulting average cooperativity is $\bar{\eta}\!=\!5.8$, with a standard deviation of $\sigma_\eta\!=\!1.6$ across the array. This configuration enables fast, high-fidelity fluorescence readout. Atoms are individually addressed by probe beams with Rabi frequency $\Omega_{\mathrm{p}}/(2\pi)=\SI{22}{MHz}$, detuned from the dressed atomic transition by $\Delta_{\mathrm{p}}/(2\pi)=\SI{120}{MHz}$, while the corresponding cavity mode is kept resonant with the probe field ($\Delta_{\mathrm{cav}}=0$). Under these conditions, atoms in state $\ket{1}$ scatter photons efficiently into the cavity mode, enabling state-selective readout on a timescale of about $\SI{10}{\micro s}$ (see Appendix~\ref{app:readout} for details).

The combination of individual probe addressing and frequency separation between cavity modes suppresses crosstalk to below $10^{-5}$. Additionally, assuming $p_{\mathrm{synd}}\!=\!5 \!\times\! 10^{-3}$, the dispersive frequency shift of a given cavity mode due to the presence of faulty syndrome qubits in other modes is smaller than $10\%$ of the cavity linewidth, leading to less than a $10\%$ decrease in the collection efficiency. Furthermore, the frequency stability of the dressed state ensures that readout times vary by less than $10\%$, as the frequency fluctuations are small compared to the probe detuning $\Delta_{\mathrm{p}}$. Alternatively, other readout techniques could be employed to achieve comparable or improved performance by coupling light into the different cavity modes via the mode sorter; however, a detailed description of such methods lies beyond the scope of this work. 

\begin{figure}
    \centering
    \includegraphics[width=\linewidth]{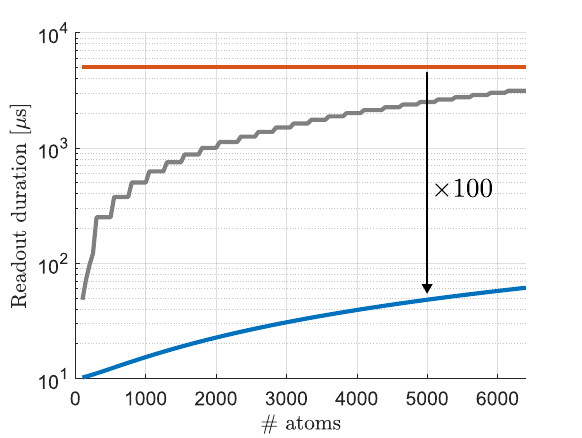}
    \caption{Scaling of syndrome readout duration using free-space imaging (orange), a single cavity mode (gray), and cavity-mode multiplexing over 50 cavity modes (blue), assuming $p_{\mathrm{synd}}\!=\!5 \!\times\!10^{-3}$ and $\SI{10}{\micro s}$ query time. Compared to free-space readout, cavity-mode multiplexing offers a speed-up of two orders of magnitude, allowing readout of $5000$ syndrome qubits within $\SI{50}{\micro s}$.} 
    \label{fig:readout_search}
\end{figure}

With 50 available modes and a query time of $\SI{10}{\micro s}$, this design yields a substantial improvement in syndrome extraction speed. Following Eqs.~\ref{eq:batch}–\ref{eq:scale}, Fig.~\ref{fig:readout_search} compares the scaling of syndrome readout duration with the number of atoms for three cases: a $\SI{5}{ms}$ free-space readout; adaptive searches utilizing a single cavity mode; and adaptive searches employing 50 cavity modes. As discussed above, a single-mode system reaches the millisecond timescale at approximately $2000$ atoms. However, CMM with 50 modes in the proposed design achieves an improvement of more than two orders of magnitude over free-space imaging, completing syndrome extraction in $\SI{50}{\micro s}$ for $5000$ atoms. Beyond this speedup, cavity readout also strongly suppresses decoherence of the data qubits from stray photon scattering: in free-space imaging a large fraction of photons are emitted into uncontrolled directions, including toward the data qubits, whereas in the cavity configuration over $80\%$ of the photons are collected into the cavity modes, thereby strongly reducing this effect. Moreover, the cavity-enhanced photon collection efficiency allows fewer photons to be scattered during readout, which minimizes heating and ensures that the measured atoms can be readily reused for subsequent operations.

Our scheme is compatible with a zoned architecture for neutral-atom arrays. Entanglement via Rydberg gates can be performed within the cavity, and the readout zone described in this design spans only $\SI{200}{\micro m}$ along the $x$-axis. This zone can be further restricted to a subset of the available sites while still maintaining an average cooperativity of $\bar{\eta} > 6$. Additionally, we note that to fully exploit the benefits of adaptive search, individual control over the light dressing beams in two dimensions is required, a capability actively pursued by several research efforts~\cite{knoernschild2010independent,graham2023multiscale,menssen2023scalable,zhang2024scaled,christen2025integrated,lin2025optical,zhao2025integrated,li2025fiber}. In the following application involving a one-dimensional array, this level of control is readily available using AODs.

\section{Modular connectivity}
\label{sec:modular}

As atom array sizes advance toward thousands of qubits~\cite{manetsch2025tweezer,chiu2025continuous}, a modular architecture, in which fixed-size atom-array nodes are interconnected via optical links to enable distributed quantum computation, would significantly simplify scaling.
Such an approach combines scalability with flexibility, allowing each module to be optimized for high-fidelity local operations while avoiding the physical constraints associated with very large monolithic arrays.

At the core of a modular architecture is the ability to generate remote entanglement between nodes. Shared Bell pairs, together with local operations, provide the essential resources for intermodule gates~\cite{maunz2009heralded,daiss2021quantumlogic,main2025distributed}. As a concrete example, teleported CNOT gates~\cite{gottesman1999demonstrating} can be used to fault-tolerantly connect logical qubits encoded in surface codes across separate modules~\cite{ramette2024faulttolerant}. In this protocol, a remote Bell state is first distributed between two \emph{communication qubits}, which are then locally entangled with \emph{code qubits} at their respective nodes. Subsequent measurement of the communication qubits and classical feedforward complete the teleported operation between the remote code qubits. Overall, this procedure combines remote Bell-state generation, local single- and two-qubit gates, and qubit readout, all of which must be performed on short timescales to support high-rate QEC cycles.

Several approaches exist for generating Bell states between remote atomic qubits~\cite{li2024highrate,vanleent2022entangling,ritter2012elementary,singh2024modular}. Here we focus on a heralded scheme in which atom–photon entanglement is first generated independently at each node by emitting a single photon from each communication atom. The photons, each entangled with its local qubit, are then routed to a probabilistic Bell-state measurement (BSM), which heralds the creation of an atom–atom Bell state~\cite{hofmann2012heralded,rosenfeld2017eventready,stephenson2020highrate,oreilly2024fast}. The main benefit of this scheme is its heralding property, retaining only successful event and ensuring that subsequent operations act on a known entangled state. However, since the BSM relies on detection of both photons, the success probability scales quadratically with the photon collection efficiency,
\begin{equation}
\label{eq:P_s}
P_\mathrm{s}=\frac{1}{2}\bigl(\alpha_\mathrm{interface}\times\alpha_\mathrm{setup}\bigr)^2,
\end{equation}
where the factor of $1/2$ reflects the intrinsic BSM success probability, $\alpha_\mathrm{interface}$ is the probability of obtaining a photon at the cavity output, and $\alpha_\mathrm{setup}$ accounts for the efficiency of the optical setup, including fiber coupling, optical elements in the path, and detection efficiency.

In free-space implementations, limited collection efficiency restricts Bell-pair generation rates to the few-hundred-hertz range~\cite{stephenson2020highrate,young2022architecture,oreilly2024fast}. An advantage of that approach is the ability to attempt entanglement generation in parallel across many communication qubits, leading to a total rate that scales linearly with their number. In principle, large arrays could therefore reach \SI{}{MHz}-scale entanglement rates by allocating thousands of atoms to communication. However, this comes at the expense of dedicating a substantial fraction of the array to communication rather than computation, and even with such parallelization, intermodule operations remain limited by millisecond-scale qubit readout times.

Optical cavities provide a more efficient atom–photon interface, enabling higher Bell-pair generation rates. Arrays of microcavities have been proposed as a path toward entanglement rates in the tens of \SI{}{MHz}, with estimates for individual microcavities reaching up to \SI{2.4}{MHz}~\cite{sinclair2025faulttolerant}. However, within a single cavity mode, entanglement attempts must be performed sequentially, and in small-mode-volume cavities the attempt rate is often limited by atom transport through the cavity.
Additionally, in such compact architectures, local operations can further constrain the rate of teleported CNOT gates.
For instance, executing two-qubit Rydberg gates would likely require atom transport to a suitable distance away from the cavity structure, introducing additional operational overhead and complexity. Reliance on atom transport not only reduces operation speed but also requires additional measures to preserve coherence, such as dynamical decoupling~\cite{bluvstein2022quantum}. We note that cavity-array microscope architectures offer a viable route that minimizes atom transport, maintains Rydberg compatibility, and supports parallel operations~\cite{shaw2025cavity}.

\begin{figure*}[t]
    \centering
    \includegraphics[width=\linewidth]{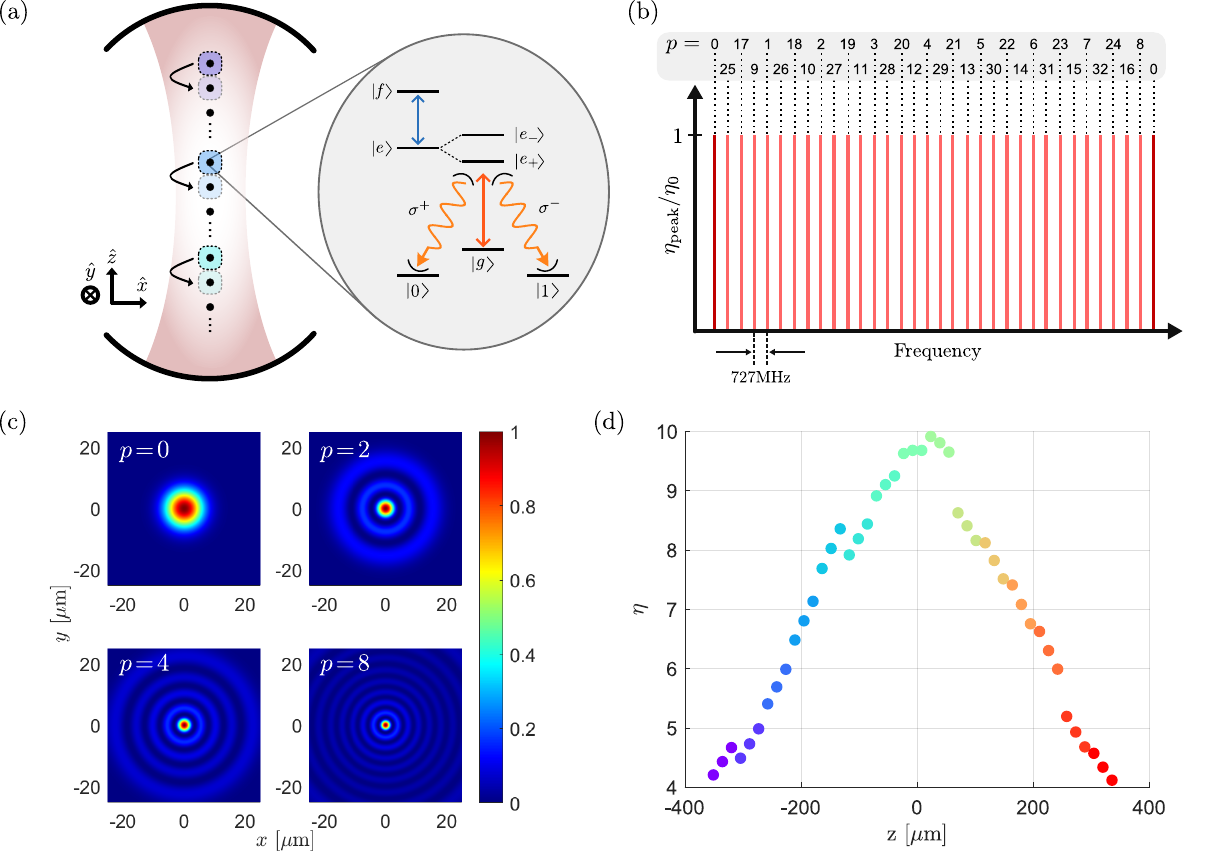}
    \caption{Cavity and atom-array configuration for remote atom–atom Bell-state generation using cavity-mode multiplexing with radial $\mathrm{LG}_{p,0}$ modes.(a)~Left: 1D array with up to 255 atoms, where each atom is coupled to one of the cavity modes. Groups of consecutive atoms form registers, each associated with a distinct cavity mode. A control beam scans across the array, sequentially coupling atoms in each register to their assigned mode. Right: atomic level scheme for atom–photon entanglement via vacuum-stimulated Raman adiabatic passage. (b)~Cavity spectrum showing the peak cooperativity of 33 equally spaced LG modes across the free spectral range, given relative to the peak cooperativity of the $\mathrm{LG}_{0,0}$ mode, denoted $\eta_0$. The peak cooperativity remains unchanged for the higher-order radial LG modes. (c)~Spatial profiles of $\mathrm{LG}_{p,0}$ modes with indices $p = 0, 2, 4, 8$. (d) Cooperativity per atom across the array (every fifth atom shown). Each color denotes a register of 15 consecutive atoms associated with a cavity mode at a distinct resonance frequency.}
    \label{fig:bell_config}
\end{figure*}

We propose an alternative approach based on CMM, which enables parallel Bell-state generation attempts within a single-cavity that hosts many qubits and remains compatible with intracavity Rydberg gates, thereby reducing the reliance on atomic motion. To illustrate this concept, we present a system design with $^{87}\mathrm{Rb}$ that implements atom–photon entanglement generation via vacuum-stimulated Raman adiabatic passage (vSTIRAP)~\cite{wilk2007singleatom,thomas2022efficient}. The relevant atomic states, shown in Fig.~\ref{fig:bell_config}(a), are defined as
\begin{align}
    &\ket{0} \equiv \ket{5S_{1/2},F=1,m_F=-1}, \\
    &\ket{1} \equiv \ket{5S_{1/2},F=1,m_F=1}, \nonumber \\
    &\ket{g} \equiv \ket{5S_{1/2},F=2,m_F=0}, \nonumber \\
    &\ket{e} \equiv \ket{5P_{3/2},F'=1,m_F=0}, \nonumber \\
    &\ket{f} \equiv \ket{4D_{5/2},F''=2,m_F=0}. \nonumber
\end{align}
The atom–photon entanglement protocol proceeds as follows. The atom is first initialized in $\ket{g}$. An external field then drives the $\ket{g}\!\rightarrow\!\ket{e_+}$ transition, inducing a vSTIRAP process that generates a single photon in the cavity whose polarization ($\sigma^+$ or $\sigma^-$) is entangled with the atomic qubit state:
\begin{equation}
    \ket{\psi_\text{final}} = \frac{1}{\sqrt{2}}\bigl(\ket{0,\sigma^+} + \ket{1,\sigma^-}\bigr).
\end{equation}
In this scheme, the probability of obtaining a photon at the cavity output is given by
\begin{equation}
\label{eq:alpha_interface}
    \alpha_\text{interface} = \frac{\eta}{\eta + 1} \times \frac{\kappa_e}{\kappa_e + \kappa_i},
\end{equation}
where $\kappa_e$ is the cavity decay rate into the designated output mode and $\kappa_i$ is the decay rate associated with intrinsic cavity losses.
An important feature of this scheme is that the external drive can be used to shape the temporal wavefunction of the emitted photon~\cite{morin2019deterministic}. Such control over the photon envelope is essential for ensuring the indistinguishability of photons generated across different modules, thereby enabling high-fidelity BSMs even in the presence of variations in system parameters between modules.

We consider a Fabry–Pérot cavity characterized by the parameters summarized in Table~\ref{tab:cavity_params_bell}. The total cavity linewidth consists of the coupling rate to the output mode, $\kappa_e/(2\pi)\!=\!\SI{2}{MHz}$, and the internal loss rate, $\kappa_i/(2\pi)\!=\! \SI{38}{kHz}$, assuming mirror losses of \SI{10}{ppm}. In contrast to syndrome extraction, simultaneously coupling multiple atoms to the same cavity mode offers no advantage for atom–photon entanglement. We therefore employ higher-order radial Laguerre–Gauss modes, $\mathrm{LG}_{p,0}$ with $p \in [0,32]$, which offer the advantage that the peak field intensity at $(x,y)=(0,0)$ remains constant with $p$ (Fig.~\ref{fig:bell_config}(c)). This property ensures a mode-independent atom–cavity coupling at that location.
The cavity is tuned so that these modes are equally spaced in frequency across the FSR, with $\SI{727}{MHz}$ separation between adjacent modes. Similar to the syndrome extraction case, control beams resonant with the $\ket{e}\!\rightarrow\!\ket{f}$ transition dress the excited states, enabling the $\ket{0}\!\rightarrow\!\ket{e_+}$ and $\ket{1}\!\rightarrow\!\ket{e_+}$ transitions to be tuned into resonance with the desired cavity mode. A maximum frequency shift of $\SI{12}{GHz}$ with $0.1\%$ intensity stability keeps fluctuations of the dressed-state transition within the natural linewidth, $\Gamma/(2\pi) = \SI{6}{MHz}$. While this tuning range does not span the full FSR, it provides access to 17 cavity modes.

\begin{table}[h]
\centering
\caption{Parameters of the optical cavity used for modular connectivity.}
\label{tab:cavity_params_bell}
\begin{tabular}{l @{\hspace{1cm}} c}
\hline\hline
Cavity length & $\SI{6.25}{mm}$ \\
Mode waist & $\SI{8.6}{\micro m}$ \\
Rayleigh range & $\SI{0.3}{mm}$\\
Finesse & $10^{4}$ \\
FSR & $\SI{24}{GHz}$ \\
Linewidth (FWHM) & $2\pi \times \SI{2.0}{MHz}$\\
\hline\hline
\end{tabular}
\end{table}

We arrange $225$ atoms in a one-dimensional array along the $z$-axis with a spacing of approximately $\SI{3}{\micro m}$, such that the outermost atoms lie roughly one Rayleigh length from the cavity center. Individual atoms are addressed by control beams focused to a $\SI{2}{\micro m}$ waist. At this tight spacing, however, residual light shifts on neighboring atoms are unavoidable, compromising the selective addressing of individual atoms. For the maximum applied frequency shift of $\SI{12}{GHz}$, adjacent atoms experience residual shifts of approximately $\SI{1}{GHz}$. To mitigate the crosstalk between control beams, the array is divided into registers of consecutive atoms, with each register associated with a single cavity mode. Within each register, the control beam is scanned sequentially to couple individual atoms to the appropriate mode for atom–photon entanglement. This procedure is performed in parallel across all registers, as illustrated in Fig.~\ref{fig:bell_config}(a). By increasing the spacing between simultaneously active control beams, this scheme substantially suppresses the crosstalk between the beams. Nevertheless, neighboring atoms can still experience residual light shifts that may couple them to cavity modes outside their assigned register. To prevent such unintended interactions, cavity modes with resonance frequencies within the range of the maximum residual shift are excluded from use. In practice, this constraint eliminates the two lowest-frequency modes, while the next available mode remains sufficiently detuned to suppress unwanted coupling. Consequently, a total of $15$ cavity modes are available, enabling the $225$-atom array to be partitioned into $15$ registers of 15 atoms each.

\begin{figure}
    \centering
    \includegraphics[width=\linewidth]{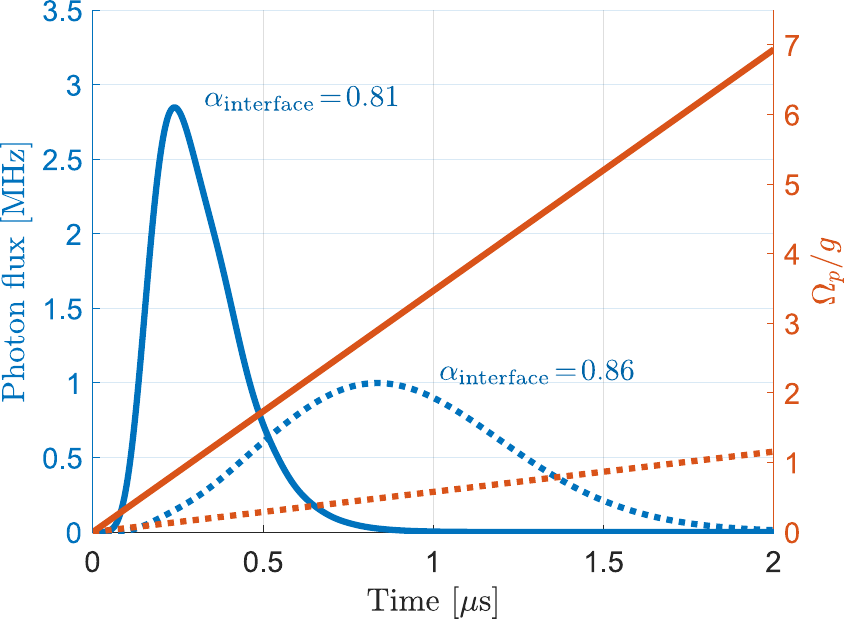}
    \caption{Photon generation using vacuum-stimulated Raman adiabatic passage. Simulated photon flux at the cavity output (blue) produced by a linearly increasing external probe field (orange) as a function of time. The simulation assumes $\eta\!=\!7$ and the cavity parameters described in the main text. The solid and dotted lines correspond to two different slopes of the external probe, resulting in distinct photon temporal profiles, with the solid line representing the probe drive chosen for our application. The dotted probe drive saturates the vacuum-stimulated Raman adiabatic passage adiabaticity condition, resulting in the photon generation probability given by Eq.~\ref{eq:alpha_interface}, but at the cost of a reduced generation rate.} 
    \label{fig:bell_vstirap}
\end{figure}

This configuration yields an average cooperativity of $\bar{\eta}\!=\!7$, with a standard deviation of $\sigma_\eta\!=\!1.8$ across the array (Fig. \ref{fig:bell_config}(d)). In the calculation, we account for the branching ratios of $\ket{e_+}$ to the atomic qubit states, the variation of atom–cavity coupling along the $z$-axis, and an assumed atomic temperature of $\SI{10}{\micro K}$. The latter plays an important role for higher-order LG modes, where the central lobe narrows with increasing mode number $p$. Consequently, the finite spatial extent of the atomic wavefunction reduces the effective atom–cavity coupling (see Appendix~\ref{app:temp}). 

With this cooperativity, it is possible to implement vSTIRAP using an external probe field resonant with the $\ket{g}\!\leftrightarrow\!\ket{e_+}$ transition and a cavity mode tuned to the $\ket{0,1}\!\rightarrow\!\ket{e_+}$ transitions. Using Eqs.~\ref{eq:P_s} and \ref{eq:alpha_interface}, and assuming an optical setup efficiency of $\alpha_{\text{setup}}\!=\!0.75$, the predicted atom–atom entanglement success probability is $P_s\!=\!0.21$. This value corresponds to the ideal case where the vSTIRAP adiabaticity condition is fully saturated and $\eta$ is uniform across the array. To assess the performance under realistic conditions, we simulate the process for the entire 225-atom array, including all relevant atomic levels, using a linearly increasing drive (see Appendix~\ref{app:vstirap} for details). As shown in Fig.~\ref{fig:bell_vstirap}, the external probe allows control over the temporal profile of the output photon. This reveals a trade-off between photon generation probability—maximized when the adiabaticity condition is saturated—and the achievable generation rate. Balancing these factors, our chosen parameters yield an average success probability of $\bar{P}_s = 0.175$ across the array. With CMM over the 15 modes, the corresponding atom–atom Bell-state generation rate is $\SI{3.75}{MHz}$. We note that the system parameters and drive profile can be further optimized to increase this rate; however, such optimization lies beyond the scope of this work.

The protocol remains robust against transition frequency fluctuations, where a $\SI{6}{MHz}$ atomic detuning leads to less than a $1\%$ change in the atom–atom Bell-state probability. This robustness allows operation in a magnetic field, which induces Zeeman shifts to the $\ket{0}$ and $\ket{1}$ qubit states, and is essential for implementing high-fidelity Rydberg gates. Additional imperfections arise from spontaneous decay out of $\ket{f}$ into other states in the $5P_{3/2}$ manifold. Most such decay channels only reduce the photon collection efficiency, but decay to the $\ket{F'=1,m_F=\pm 1}$ states can also produce a photon in the cavity while leaving the atom in $\ket{F=1,m_F=0}$. This occurs with a probability of less than $1\%$, and can be mitigated using a measurement that discriminates between the magnetic sublevels of the $F\!=\!1$ manifold. We further emphasize that, although the $\ket{e}\!\leftrightarrow\!\ket{f}$ transition is not a cycling transition, a $\pi$-polarized control beam prevents any unwanted mixing with other states. This is because the $\pi$-polarized transitions from $\ket{e}$ to $\ket{F''=1,3,4, m_F=0}$ are forbidden; the transition to $\ket{F''=1, m_F=0}$ is excluded by the $\Delta F = 0$, $\Delta m_F = 0$ selection rule, while transitions to $\ket{F''=3,4}$ are forbidden since they require $\Delta F > 1$. 
Under these conditions, the achievable atom–atom Bell-state fidelity is sufficient for modular architectures, since the threshold for stitching surface-code patches across modules can be as high as $10\%$~\cite{ramette2024faulttolerant}.

To evaluate the implications for fault-tolerant architectures, we connect our results to the cycle time of a distance-20 surface code, which requires on the order of 40 teleported CNOT gates per cycle between distant code patches. In our construction, the communication atoms remain fixed in place throughout the entire process, ensuring stable coupling and minimizing the use of atom transport. The sequence of operations proceeds as follows. First, the communication qubits are initialized in $\ket{g}$; we assume that state preparation, together with occasional recooling, requires $\SI{16}{\micro s}$ per cycle~\cite{young2022architecture}. Next, since the expected number of successful pairs equals the number of atoms times $P_s$, the 40 required Bell pairs shared between modules can be generated in a single multiplexed scan across the array. With a control-beam switching time of $\SI{100}{ns}$, the scan completes within $\SI{13}{\micro s}$. In the following step, Rydberg gates are performed between the successful communication qubits and the code qubits. This process involves qubit basis changes, transporting the code qubits, and the Rydberg interactions themselves. The dominant contribution here is the atomic motion, as Raman manipulations and Rydberg gates operate in the $\SI{}{MHz}$ regime~\cite{bluvstein2024logical}. Because Rydberg operations can be carried out directly within the cavity, the surface code patch can be positioned near the cavity mode, allowing short-range transport. We estimate this step at approximately $\SI{20}{\micro s}$. Finally, measurement of the 40 communication qubits is performed. With a $\SI{10}{\micro s}$ measurement time multiplexed over 15 modes, this requires roughly $\SI{25}{\micro s}$. Adding these contributions, the full sequence of 40 teleported CNOTs completes in about $\SI{70}{\micro s}$. This cycle time is roughly two orders of magnitude faster than free-space implementations, while requiring significantly fewer communication qubits.

\section{Discussion}
\label{sec:diss}

We have introduced mode-multiplexed optical cavities as a scalable interface for neutral atom arrays. In our approach, atoms are selectively coupled to multiple modes of a single cavity using light shifts induced via an excited-to-excited atomic transition. This enables parallel cavity-enhanced operations, allowing fast processing in cavities large enough to accommodate large-scale atom arrays. As a result, the need for atom transport in and out of the cavity is minimized and the configuration maintains full compatibility with existing capabilities of the neutral-atom platform.

We presented two applications of CMM in the context of quantum computing. First, for MCM of syndrome qubits, our approach reduces the readout duration by two orders of magnitude compared to free-space readout for systems with thousands of qubits. The scheme combines two key capabilities: the ability to couple many atoms to the same mode, which allows an adaptive search for faulty syndrome qubits, and the ability to perform this search in parallel across multiple cavity modes. Beyond the favorable logarithmic scaling of binary search, adaptive syndrome extraction has also been proposed as a strategy to enhance code performance and shorten QEC cycles by prioritizing specific syndrome sets in time~\cite{berthusen2025adaptive}, a capability that our scheme naturally supports. Second, we highlight the advantages of CMM in a modular atom array architecture. In this setting, the mode-multiplexed interface supports fast intermodule operations by enabling high-rate remote entanglement of atomic qubits, and fast local operations such as cavity-based readout and intracavity Rydberg gates. Together, these capabilities provide fast, fault-tolerant connectivity between logical qubits encoded in distant modules, supporting distributed quantum computing with neutral-atom arrays. While these two applications are optimized using different cavity designs, they are not fundamentally incompatible, and a crossed-cavity architecture could in principle integrate both functionalities within the same setup.

To fully realize the benefits of CMM, several challenging hardware requirements must be addressed. One key aspect is precise control over the cavity symmetry: either maintaining cylindrical symmetry to support LG modes or intentionally breaking it to produce HG modes in the desired orientation. Such control can be achieved by mounting one of the cavity mirrors on a 3D piezo translation stage, as demonstrated in Ref.~\cite{utama2021coupling}. Another crucial requirement is the ability to individually address atoms, enabling the application of local light shifts and the probing of atoms at distinct frequencies corresponding to their shifted transitions. This capability is being actively explored, and both experiments and theoretical proposals point toward feasible implementations~\cite{knoernschild2010independent,graham2023multiscale,menssen2023scalable,zhang2024scaled,christen2025integrated,lin2025optical,zhao2025integrated,li2025fiber}. We note that achieving the required stability across many control beams is technically demanding and represents an important experimental challenge; nevertheless, we are optimistic that continued advances in site-resolved atom-array addressing will make such capabilities accessible.

We note that, in addition to Fabry–Pérot cavities, this approach can be implemented in other cavity geometries. For instance, bow-tie cavities allow for small mode waists while supporting longer cavity lengths, enabling high-cooperativity designs with smaller FSRs. This configuration reduces reliance on high-order spatial modes, instead making greater use of longitudinal modes of lower-order transverse profiles. In the context of remote entanglement, this can help maintain a high average cooperativity at elevated atomic temperatures, thereby reducing the amount of required atom cooling.

\begin{acknowledgments}
We thank Ohad Lib and Nicolas Fontaine for useful discussions on spatial mode sorting. This work is supported by a collaboration between the US DOE and other Agencies. This material is based upon work supported by the U.S. Department of Energy, Office of Science, National Quantum Information Science Research Centers, Quantum Systems Accelerator. We acknowledge support from the MIT-Harvard Center for Ultracold Atoms (an NSF Frontier Center, Award No. PHY-2317134), the NSF Quantum Leap Challenge Institute (Award No. 2016244), and the NSF-funded QuSeC-TAQS program (Award No. 2326787).
\end{acknowledgments}

\appendix

\section{Effect of finite atom temperature on the cooperativity of higher-order cavity modes}
\label{app:temp}

\begin{figure*}[t]
    \centering
    \includegraphics[width=1\textwidth]{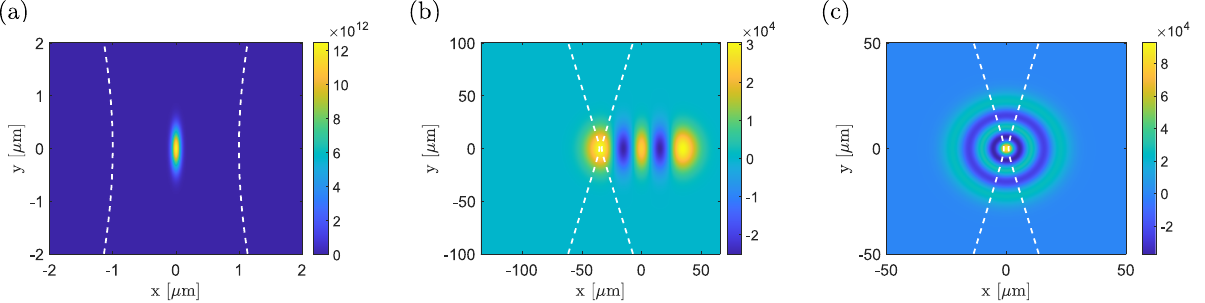}
    \caption{(a) Atomic spatial probability density for a temperature of $\SI{10}{\micro K}$ in the optical tweezer. Transverse intensity profiles of (b) the HG$_{4,0}$ mode with waist $w_{\rm HG}\!=\!\SI{20}{\micro m}$ and (c) the LG$_{4,0}$ mode with waist $w_{\rm LG}\!=\!\SI{8.6}{\micro m}$. The profile of the tweezer waist along $y$ is indicated by the white dashed lines.}
    \label{fig_app:spatial}
\end{figure*}

We calculate the cooperativity of higher-order HG and LG cavity modes for a single $^{87}$Rb atom confined in an optical tweezer. The tweezer operates at a wavelength $\lambda_{\rm trap}=\SI{852}{nm}$ with waist $w_{\rm trap}=\SI{1}{\micro m}$, and provides trap frequencies $\omega_x=\omega_z=2\pi\times\SI{100}{kHz}$ and $\omega_y=2\pi\times\SI{19.17}{kHz}$ in the coordinate system defined in Fig.~\ref{fig:readout_config}(a) and Fig.~\ref{fig:bell_config}(a). Over the temperature range considered here, the atomic motion in each direction is well described by a harmonic potential.

For a harmonically trapped atom at temperature $T$, the position distribution is Gaussian with variance
\begin{equation}
\sigma_i^2 = \frac{\hbar}{2m\omega_i}\coth\!\left(\frac{\hbar\omega_i}{2k_B T}\right),
\quad i \in \{x,y,z\}.
\end{equation}
An example of this spatial probability density is shown in Fig.~\ref{fig_app:spatial}(a) for $T\!=\!\SI{10}{\micro K}$.

The transverse cavity field at the cavity waist ($z=0$) is modeled using normalized HG$_{n,0}$ and LG$_{p,0}$ modes,
\begin{equation}
|u_n(x,y)| =
\sqrt{\frac{2/\pi}{w^2_{\rm HG} 2^n n!}} \,
H_n\!\left(\frac{\sqrt{2} x}{w_{\rm HG}}\right)
\exp\!\left(-\frac{x^2+y^2}{w_{\rm HG}^2}\right),
\end{equation}
and
\begin{equation}
|u_{p}(x,y)| =
\sqrt{\frac{2}{\pi w^2_{\rm LG}}} \,
L_p\!\left(\frac{2(x^2+y^2)}{w_{\rm LG}^2}\right)
\exp\!\left(-\frac{x^2+y^2}{w_{\rm LG}^2}\right),
\end{equation}
with waists $w_{\rm HG}=\SI{20}{\micro m}$ and $w_{\rm LG}=\SI{8.6}{\micro m}$, as considered in Secs.~\ref{sec:syndrome} and~\ref{sec:modular}, respectively.
Representative transverse intensity profiles of these modes are shown in Fig.~\ref{fig_app:spatial}(b) and~\ref{fig_app:spatial}(c).

The position-dependent atom-cavity coupling is proportional to the field amplitude
\begin{equation}
g(\mathbf{r}) \propto u_{\nu}(x,y)\cos(k z),
\end{equation}
where $k\!=\!2\pi/\lambda_{\rm cav}$ is the cavity wavevector, and we use $\lambda_{\rm cav}=\SI{780}{nm}$.

Thermal motion reduces the cooperativity through spatial averaging of the coupling strength. We separate transverse (x,y) and axial (z) contributions by defining reduction factors relative to the peak cooperativity of the fundamental mode, $\eta_0$,
\begin{align}
\frac{\eta_\perp}{\eta_0} &= \langle u_\nu(x,y)\rangle^2 /  [u_0(0,0)]^2  \\
\frac{\eta_z}{\eta_0} &= \langle \cos(k z)\rangle^2=\exp\big(-k^2 \sigma_z^2\big),\nonumber
\end{align}
with $\nu\in{n,p}$. The transverse factor captures the mode-dependent reduction arising in higher-order transverse cavity modes, while the axial factor accounts for motion along the cavity axis, which is common to all standing-wave cavities. For HG modes, the transverse coordinate system $(x,y)$ is shifted to the location of the mode maximum prior to thermal averaging, ensuring that the atomic distribution is centered at the relevant position, as illustrated in Fig.~\ref{fig_app:spatial}(b).

\begin{figure*}[t]
    \centering
    \includegraphics[width=1\textwidth]{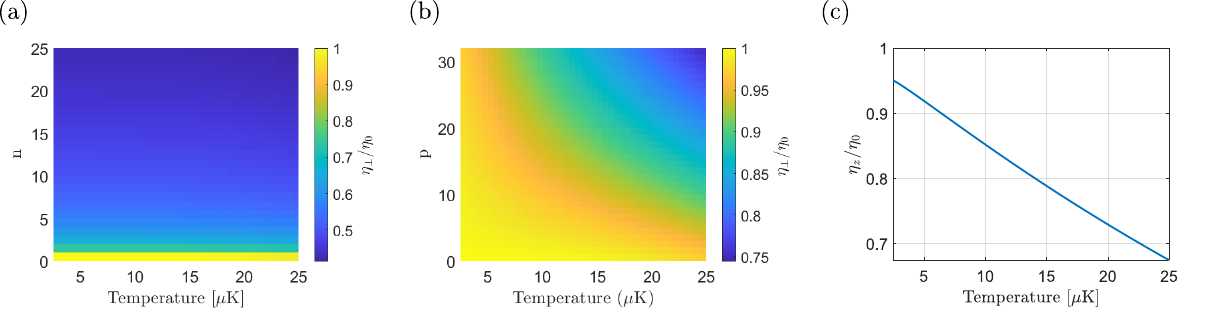}
    \caption{(a,b) Transverse cooperativity reduction vs mode index and temperature for (a) $\mathrm{HG}_{n,0}$ modes with waist $w_{\rm HG} = \SI{20}{\micro m}$ and (b) $\mathrm{LG}_{p,0}$ modes with waist $w_{\rm LG} = \SI{8.6}{\micro m}$. (c) Axial cooperativity reduction due to thermal motion along the cavity standing wave.}
    \label{fig_app:eta_ratio}
\end{figure*}

The transverse cooperativity reduction for HG$_{n,0}$ and LG$_{p,0}$ modes is shown as a function of temperature and mode index in Fig.~\ref{fig_app:eta_ratio}(a) and~\ref{fig_app:eta_ratio}(b), respectively.
Over the temperature range considered, the HG modes exhibit negligible temperature dependence ($<\!1\%$), whereas higher-order LG modes display a more pronounced reduction. This behavior reflects two key differences between the HG and LG modes. First, the larger mode waist for HG modes ($w_{\rm HG}=\SI{20}{\micro m}$) compared to that of the LG modes ($w_{\rm LG}=\SI{8.6}{\micro m}$) reduces the effect of thermal averaging. Second, the transverse structure scales differently with mode order: for HG$_{n,0}$ modes, the peak lobe size decreases along only one transverse direction, whereas for LG$_{p,0}$ modes the central feature contracts radially, reducing the relevant length scale along both transverse directions. Together, these effects make higher-order LG modes more sensitive to finite atomic temperature.
The axial cooperativity reduction factor, shown in Fig.~\ref{fig_app:eta_ratio}(c), is identical for both HG and LG modes, as it depends only on thermal motion along the cavity standing wave. Both transverse and axial reduction factors are included when calculating the cooperativities and rates in Secs.~\ref{sec:syndrome} and~\ref{sec:modular}.

\section{Cavity-enhanced fluorescence readout}
\label{app:readout}

We model cavity-enhanced fluorescence readout of a single atom by driving the effective cycling transition $\ket{1}\!\rightarrow\!\ket{e_+}$ discussed in Sec.~\ref{sec:syndrome} using an external probe beam. The system is described by the Hamiltonian
\begin{equation}
H = \Delta_c a^\dagger a + \Delta_a\sigma^\dagger\sigma
+ g\left(a^\dagger\sigma+a\sigma^\dagger\right)
+ \Omega_p\left(\sigma + \sigma^\dagger \right),
\end{equation}
where $a$ is the cavity annihilation operator, $\sigma=\ket{1}\!\bra{e_+}$ is the lowering operator of the atom, $\Delta_c$ and $\Delta_a$ are the cavity and atomic detunings relative to the probe, $g$ is the atom-cavity coupling strength, and $\Omega_p$ is the Rabi frequency of the external probe.

Dissipation is included via the collapse operators
\begin{equation}
C_1=\sqrt{\Gamma}\,\sigma,\qquad
C_2=\sqrt{\kappa_e}\,a,\qquad
C_3=\sqrt{\kappa_i}\,a,
\end{equation}
which represent free-space spontaneous emission, cavity output coupling, and intrinsic cavity loss, respectively.

In the regime where cavity-induced driving of the atomic transition is negligible,
\begin{equation}
\sqrt{n_a}\, g \ll \Omega_p,
\end{equation}
where $n_a=\langle a^\dagger a\rangle$ is the intracavity photon number, the excited-state population is dominated by the classical probe and is well approximated by the free-space steady-state value
\begin{equation}
\label{eq:fsrhoee}
\rho_{e_+e_+}^{(\mathrm{th})}
=\frac{R_{\mathrm{opt}}}{2R_{\mathrm{opt}}+\Gamma},\qquad
R_{\mathrm{opt}}=\frac{\Omega_p^2}{(\Gamma/2)^2+\Delta_a^2}.
\end{equation}
The photon flux emitted into the cavity output mode is then
\begin{equation}
R_{\mathrm{cav}} = \kappa_e\langle a^\dagger a\rangle
\approx \frac{\kappa_e}{\kappa} \,\eta\,\Gamma\,\rho_{e_+e_+},
\end{equation}
where $\eta=4g^2/(\kappa\Gamma)$ is the single-atom cooperativity and $\kappa=\kappa_e+\kappa_i$.

However, this steady-state expression does not account for finite pulse duration and cavity bandwidth effects. To capture the full dynamics, we therefore solve the Lindblad master equation numerically. The readout drive is applied as a smoothed square pulse, and the total number of photons emitted into the cavity output is obtained from the integrated flux
\begin{equation}
N_{\mathrm{cav}}=\int dt\,\kappa_e\langle a^\dagger a\rangle(t).
\end{equation}

\begin{figure}[b]
    \centering
    \includegraphics[width=\linewidth]{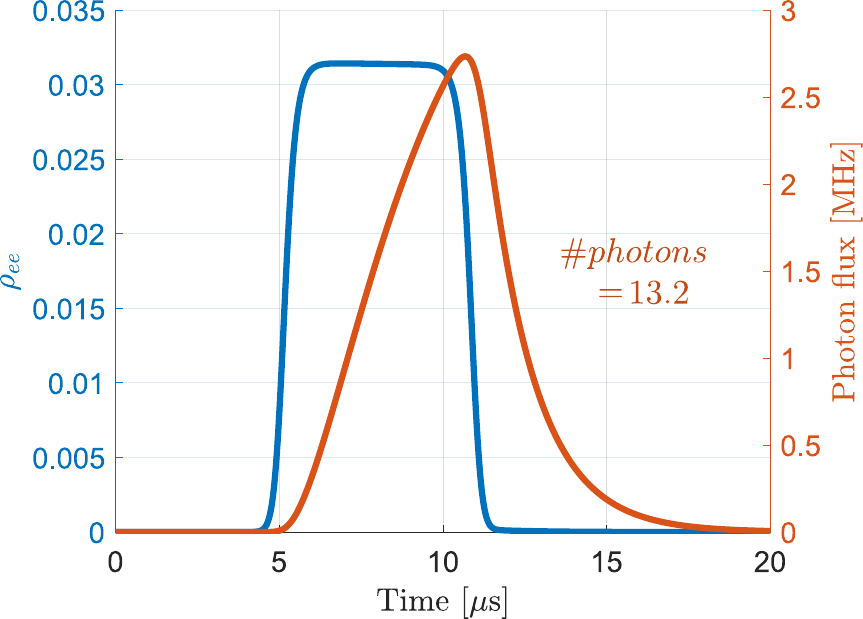}
    \caption{Numerical simulation of fluorescence scattering into the cavity mode. Excited-state population (blue) and photon flux at the cavity output (orange) as a function of time, showing the readout pulse dynamics used to calculate the detected photon number.} 
    \label{fig_app:readout_sim}
\end{figure}

The parameters used in the simulations are
\begin{align}
&(g,\kappa_e,\kappa_i,\Gamma,\Omega_p,\Delta_c,\Delta_a)/(2\pi) \\
&=(0.79,0.1,0.01,3.9,22,0,120)\ \mathrm{MHz},\nonumber
\end{align}
where $\Gamma$ accounts for equal mixing of the states $\ket{e}$ and $\ket{f}$ (associated with the $5P_{3/2}$ and $4D_{5/2}$ manifolds) in the dressed state $\ket{e_+}$, such that $\Gamma=(\Gamma_e+\Gamma_f)/2$. The atom-cavity coupling $g$ is determined from the average cooperativity in the readout configuration, $\bar{\eta}\!=\!5.8$, and $\kappa_i$ corresponds to mirror losses of approximately $\SI{10}{ppm}$.

The simulated excited-state population agrees with the free-space prediction in Eq.~\eqref{eq:fsrhoee}, yielding $\rho_{e_+e_+}\,=\,3.1\%$, confirming that the scattering dynamics are dominated by the probe rather than by cavity-induced excitation. The intracavity photon number remains below $n_a\!<\!4.4$, satisfying the weak-cavity-driving condition. The dynamical simulation yields $13.2$ photons emitted into the cavity output during the $\sim\!\SI{10}{\micro s}$ readout window, as shown in Fig.~\ref{fig_app:readout_sim}.
This enables state discrimination with an infidelity at the $10^{-4}$ level using a simple photon-count threshold and assuming a low dark-count rate. Adaptive readout protocols, in which the measurement is terminated once a detection threshold is reached~\cite{hu2025siteselective}, can further reduce the required readout time and minimize heating due to excess photon scattering.

\section{Photon generation using vSTIRAP}
\label{app:vstirap}

We simulate vSTIRAP photon generation in an atom-cavity system using a Lindblad master equation. The atomic basis consists of the four states
$\{\ket{g_1},\ket{g_2},\ket{e},\ket{f}\}$ that describe the ideal process. The atom is initialized in $\ket{g_1}$ and undergoes vSTIRAP to $\ket{g_2}$ via a dressed excited state
\begin{equation}
\ket{e_+} = \frac{1}{\sqrt{2}}\left(\ket{e}+\ket{f}\right),
\end{equation}
formed by resonant dressing of $\ket{e}$ and $\ket{f}$ with a strong classical drive of Rabi frequency $\Omega_c$. The cavity couples the $\ket{g_2}\leftrightarrow\ket{e}$ transition, while a time-dependent classical control field drives $\ket{g_1}\leftrightarrow\ket{e}$.
For simplicity, we do not explicitly distinguish decay into two ground states via orthogonally polarized cavity modes, as shown in Fig.~\ref{fig:bell_config}, since here we focus solely on the photon-generation probability and error channels that are independent of polarization mismatch between cavity modes.

To capture loss and error processes outside the ideal manifold, we include three auxiliary states
$\{\ket{g_{\mathrm{aux}}},\ket{e_{\mathrm{aux}}},\ket{e'_{\mathrm{aux}}}\}$. The state $\ket{g_{\mathrm{aux}}}$ represents decay from $\ket{e}$ to ground states that do not result in cavity photon emission, while $\ket{e_{\mathrm{aux}}}$ represents decay from $\ket{f}$ to excited states that cannot emit into the cavity. The state $\ket{e'_{\mathrm{aux}}}$ corresponds to decay from $\ket{f}$ to excited states $\ket{F'\!=\!1,m_F\!=\!\pm1}$ that can emit a cavity photon but project the atom into an incorrect final ground state $\ket{F\!=\!1,m_F\!=\!0}$.

Because the protocol is heralded, population loss into $\ket{g_{\mathrm{aux}}}$ and $\ket{e_{\mathrm{aux}}}$ reduces the success probability only, whereas population transferred through $\ket{e'_{\mathrm{aux}}}$ can reduce the fidelity of the heralded Bell state. This latter error can be mitigated by a measurement that discriminates between magnetic sublevels within the $F=1$ manifold.

In a rotating frame, the Hamiltonian is
\begin{equation}
H(t)=H_0+ \Omega_p(t)
\left(\ket{e}\!\bra{g_1}+\ket{g_1}\!\bra{e}\right),
\end{equation}
where the vSTIRAP control field is taken to be a linear ramp $\Omega_p(t)=A t$. The static Hamiltonian is
\begin{align}
H_0 &=
-\Delta_1 \ket{e}\!\bra{e}
-(\Delta_1+\Delta_2)\ket{f}\!\bra{f}
+\Delta_{g_2}\ket{g_2}\!\bra{g_2}
\nonumber\\ &\quad+\delta_c\, a^\dagger a
+\Omega_c\left(\ket{e}\!\bra{f}+\ket{f}\!\bra{e}\right)
\nonumber\\ &\quad+g\left(a\,\ket{e}\!\bra{g_2}+a^\dagger\,\ket{g_2}\!\bra{e}\right),
\end{align}
where $a$ is the cavity annihilation operator, $\Delta_{1,2}$ are the detunings of the $\ket{g_1}\!\leftrightarrow\!\ket{e}$ and $\ket{e}\!\leftrightarrow\!\ket{f}$ drives, $\Delta_{g_2}$ is the $\ket{g_2}$-$\ket{g_1}$ splitting in this frame, $\delta_c$ is the cavity detuning, and $g$ is the atom-cavity coupling.

Cavity loss is included via the collapse operators
\begin{equation}
C_{\kappa_e}=\sqrt{\kappa_e}\,a,\qquad
C_{\kappa_i}=\sqrt{\kappa_i}\,a,
\end{equation}
which represent output coupling and internal cavity losses, respectively. Spontaneous emission from the excited states $\ket{e}$ and $\ket{f}$ is modeled by the collapse operators
\begin{align}
C_{e\to g_1} &= \sqrt{\Gamma_e b_{g_1}}\,\ket{g_1}\!\bra{e}, \\
C_{e\to g_2} &= \sqrt{\Gamma_e b_{g_2}}\,\ket{g_2}\!\bra{e}, \nonumber\\
C_{e\to g_{\mathrm{aux}}} &= \sqrt{\Gamma_e b_{g_{\mathrm{aux}}}}\,\ket{g_{\mathrm{aux}}}\!\bra{e}, \nonumber\\
C_{f\to e} &= \sqrt{\Gamma_f b_{e}}\,\ket{e}\!\bra{f}, \nonumber\\
C_{f\to e_{\mathrm{aux}}} &= \sqrt{\Gamma_f b_{e_{\mathrm{aux}}}}\,\ket{e_{\mathrm{aux}}}\!\bra{f}, \nonumber\\
C_{f\to e'_{\mathrm{aex}}} &= \sqrt{\Gamma_f b_{e'_{\mathrm{aex}}}}\,\ket{e'_{\mathrm{aux}}}\!\bra{f},\nonumber
\end{align}
where $\Gamma_{e}$ and $\Gamma_{f}$ are the total spontaneous emission rates from $\ket{e}$ and $\ket{f}$, respectively, and the coefficients $b_j$ encode the corresponding branching ratios.

The probability to generate a photon at the cavity output during a single attempt of duration $T$ is
\begin{equation}
P=\int_0^T \kappa_e\,\langle a^\dagger a\rangle(t)\,dt.
\end{equation}
To account for inhomogeneous atom-cavity coupling in a multiplexed array, as considered in Sec.~\ref{sec:modular}, we repeat the calculation for atoms indexed by $i$ with couplings
\begin{equation}
g_i = g_0\,[1+(z_i/z_R)^2]^{-1/2} f_{\mathrm{temp}}^{(i)},
\end{equation}
where $z_i$ is the axial position of atom $i$ relative to the cavity waist, $z_R$ is the cavity Rayleigh range, and $f_{\mathrm{temp}}^{(i)}$ accounts for thermal averaging at that site (see Appendix~\ref{app:temp}).

\section{Control beam considerations}
\label{app:control}

\subsection{Heating from dipole force fluctuations}
To implement cavity-mode multiplexing we locally shift the optical transition frequency of each atom using a control beam that couples the excited state $\ket{e}$ to an auxiliary state $\ket{f}$ with Rabi frequency $\Omega_c(\mathbf{r})$ and detuning $\Delta_c$. We focus on the case of on-resonant control light ($\Delta_c=0$), which produces dressed eigenstates
\begin{equation}
\ket{e_\pm}=\frac{\ket{e}\pm \ket{f}}{\sqrt{2}},
\qquad
\delta_\pm(\mathbf{r})=\mp \Omega_c(\mathbf{r}),
\end{equation}
so that the ground-to-dressed-state optical resonance is shifted by $- \Omega_c(\mathbf{r})$ relative to the bare transition. During readout, the atom experiences stochastic switching between dressed $\ket{1}$ and $\ket{e_+}$ potentials, which leads to momentum-space diffusion and therefore heating. We derive this heating rate for the case of an atom in qubit state $\ket{1}$ coupled to the dressed excited state $\ket{e_+}$. 

We work in the dressed-atom picture spanned by $\{\ket{e_+,n},\ket{1,n+1}\}$ for fixed photon number and follow a similar derivation as in Ref.~\cite{Dalibard}. Diagonalizing the atom--probe Hamiltonian in this manifold yields two probe-dressed eigenstates, which we label $\ket{a}$ and $\ket{b}$:
\begin{align}
    \ket{a;n,\mathbf{r}} &= \cos\theta(\mathbf{r})\,\ket{e_+,n} \nonumber \\
    &\quad + \sin\theta(\mathbf{r})\,\ket{1,n+1}, \nonumber \\[2ex]
    \ket{b;n,\mathbf{r}} &= - \sin\theta(\mathbf{r})\,\ket{e_+,n} \nonumber \\
    &\quad + \cos\theta(\mathbf{r})\,\ket{1,n+1},
    \label{eq:dressed_states_ab}
\end{align}
where the mixing angle $\theta(\mathbf{r})$ is defined by
\begin{equation}
\sin 2\theta(\mathbf{r})=\frac{2\Omega(\mathbf{r})}{Q(\mathbf{r})},
\qquad
\cos 2\theta(\mathbf{r})=-\frac{\Delta(\mathbf{r})}{Q(\mathbf{r})}.
\label{eq:mixing_angle_def}
\end{equation}
with $Q(\mathbf{r}) \equiv \sqrt{4\Omega(\mathbf{r})^2 + \Delta(\mathbf{r})^2}$ defined as the generalized Rabi frequency, $\Delta(\mathbf{r})=\Delta_p-\Omega_c(\mathbf{r})$ as the position-dependent detuning, and $\Omega(\mathbf{r})$ as the position-dependent probe Rabi frequency. The dressed energies are
\begin{equation}
E_a(\mathbf{r})=+\frac{\hbar}{2} Q(\mathbf{r}),
\qquad
E_b(\mathbf{r})=-\frac{\hbar}{2} Q(\mathbf{r}),
\end{equation}
so the corresponding dipole forces are
\begin{equation}
\mathbf{F}_{a,b}(\mathbf{r})\equiv -\nabla E_{a,b}(\mathbf{r}).
\label{eq:Fab_def}
\end{equation}
With the approximation that the detuning gradient is set by the control beam,
\begin{equation}
\nabla\Delta(\mathbf{r})\simeq -\nabla\Omega_c(\mathbf{r}),
\label{eq:gradDelta_control}
\end{equation}
and the generalized Rabi frequency satisfies
\begin{equation}
\nabla Q(\mathbf{r})=\frac{4\Omega(\mathbf{r})\,\nabla\Omega(\mathbf{r})+\Delta(\mathbf{r})\,\nabla\Delta(\mathbf{r})}{Q(\mathbf{r})}\approx \frac{\Delta(\mathbf{r})\,\nabla\Delta(\mathbf{r})}{Q(\mathbf{r})}
\label{eq:gradQ_general}
\end{equation}
for $|\Omega(\mathbf{r})\,\nabla\Omega(\mathbf{r})|\ll  |\Omega_c(\mathbf{r})\nabla\Omega_c(\mathbf{r})|$.

Spontaneous emission induces stochastic jumps between the probe-dressed states. We denote the transition rates by $r_{ab}(\mathbf{r})$ (for $\ket{a}\to\ket{b}$) and $r_{ba}(\mathbf{r})$ (for $\ket{b}\to\ket{a}$). The rates for these transitions are
\begin{equation}
\label{eq:rates_ab}
\begin{split}
    r_{ab}(\mathbf{r}) &= \Gamma_{e_+}\cos^4\theta(\mathbf{r}), \\
    r_{ba}(\mathbf{r}) &= \Gamma_{e_+}\sin^4\theta(\mathbf{r}), \\
    r_{\mathrm{pop}}(\mathbf{r}) &\equiv r_{ab}(\mathbf{r})+r_{ba}(\mathbf{r}),
\end{split}
\end{equation}
where $\Gamma_{e_+}$ is the effective transition linewidth.

At long times the dressed-state dynamics reach a steady state with populations
\begin{equation}
\Pi_a^{\mathrm{st}}(\mathbf{r})=\frac{r_{ba}(\mathbf{r})}{r_{\mathrm{pop}}(\mathbf{r})},
\qquad
\Pi_b^{\mathrm{st}}(\mathbf{r})=\frac{r_{ab}(\mathbf{r})}{r_{\mathrm{pop}}(\mathbf{r})}.
\label{eq:Pi_st}
\end{equation}
The instantaneous force experienced by the atom is 
$\mathbf{F}(t)\in\{\mathbf{F}_a(\mathbf{r}),\mathbf{F}_b(\mathbf{r})\}$.
Since $E_b(\mathbf{r})=-E_a(\mathbf{r})$, Eq.~\eqref{eq:Fab_def} implies
\begin{equation}
\mathbf{F}_b(\mathbf{r})=-\mathbf{F}_a(\mathbf{r})
\equiv
\mathbf{F}(\mathbf{r})
=
\frac{\hbar}{2}\nabla Q(\mathbf{r}).
\label{eq:F_def}
\end{equation}
The mean dipole force is therefore
\begin{equation}
\mathbf{f}_{\mathrm{dip}}(\mathbf{r})
\equiv
\langle \mathbf{F}(t)\rangle
=
\big[\Pi_b^{\mathrm{st}}(\mathbf{r})-\Pi_a^{\mathrm{st}}(\mathbf{r})\big]\,
\mathbf{F}(\mathbf{r}).
\label{eq:fdip_def}
\end{equation}

To find the heating rate, we require the momentum diffusion coefficient given by 
\begin{equation}
D
\equiv
\int_{0}^{\infty}\! d\tau\,
\big\langle
\delta  \mathbf{F}(t)\cdot \delta  \mathbf{F}(t+\tau)
\big\rangle,
\qquad
\delta \mathbf{F}(t)\equiv \mathbf{F}(t)-\mathbf{f}_{\mathrm{dip}}.
\label{eq:D_def}
\end{equation}

The force autocorrelation can be written as
\begin{equation}
\langle \mathbf{F}(t)\!\cdot\!\mathbf{F}(t+\tau)\rangle
=
\sum_{i,j\in\{a,b\}}
\Pi_i^{\mathrm{st}}\,
\Pi(j,\tau|i,0)\,
\mathbf{F}_i\!\cdot\!\mathbf{F}_j,
\label{eq:FF_general}
\end{equation}
where the conditional transition probabilities
$\Pi(j,\tau|i,0)$ $i,j\in\{a,b\}$ for the two-state Markov process are
\begin{equation}
\label{eq:cond_probs}
\begin{split}
\Pi(a,\tau|a,0) &= \Pi_a^{\mathrm{st}} + \Pi_b^{\mathrm{st}}\,e^{-r_{\mathrm{pop}}\tau},\\
\Pi(b,\tau|a,0) &= \Pi_b^{\mathrm{st}} - \Pi_b^{\mathrm{st}}\,e^{-r_{\mathrm{pop}}\tau},\\
\Pi(b,\tau|b,0) &= \Pi_b^{\mathrm{st}} + \Pi_a^{\mathrm{st}}\,e^{-r_{\mathrm{pop}}\tau},\\
\Pi(a,\tau|b,0) &= \Pi_a^{\mathrm{st}} - \Pi_a^{\mathrm{st}}\,e^{-r_{\mathrm{pop}}\tau}.
\end{split}
\end{equation}
Eq. (\ref{eq:FF_general}) simplifies using $\mathbf{F}_b=-\mathbf{F}_a=\mathbf{F}$ to
\begin{equation}
\langle \mathbf{F}(t)\!\cdot\!\mathbf{F}(t+\tau)\rangle-\mathbf{f}_{\mathrm{dip}}^{\,2}
=
4\,\Pi_a^{\mathrm{st}}\Pi_b^{\mathrm{st}}\,
\mathbf{F}(\mathbf{r})^{2}\,
e^{-r_{\mathrm{pop}}(\mathbf{r})\,\tau}.
\label{eq:FF_connected}
\end{equation}

Using Eq.~\eqref{eq:D_def} gives
\begin{equation}
D
=
\frac{4\,\Pi_a^{\mathrm{st}}\Pi_b^{\mathrm{st}}}{r_{\mathrm{pop}}}\,
\mathbf{F}^{2}.
\label{eq:Dp_markov}
\end{equation}
Substituting Eqs.~\eqref{eq:rates_ab} and \eqref{eq:F_def} into Eq.~\eqref{eq:Dp_markov} yields
\begin{equation}
D
=
\frac{\hbar^{2}}{\Gamma_{e_+}}\,
\frac{\sin^{4}\!\theta\,\cos^{4}\!\theta}
{\big[\sin^{4}\!\theta+\cos^{4}\!\theta\big]^{3}}\,
\big|\nabla Q\big|^{2},
\label{eq:Dp_theta}
\end{equation}
where $\Gamma_{e_+}$ is the linewidth of the dressed excited state $\ket{e_+}$.
In the detuning-gradient limit of Eq.~\eqref{eq:gradQ_general} this becomes
\begin{equation}
D
\simeq
\frac{\hbar^{2}}{\Gamma_{e_+}}\,
\frac{\sin^{4}\!\theta\,\cos^{4}\!\theta}
{\big(\sin^{4}\!\theta+\cos^{4}\!\theta\big)^{3}}\,
\left(\frac{\Delta(\mathbf{r})}{Q(\mathbf{r})}\right)^{2}
\big|\nabla \Delta(\mathbf{r})\big|^{2},
\label{eq:Dp_DeltaGrad}
\end{equation}
and using Eq.~\eqref{eq:mixing_angle_def} one may write the result explicitly as
\begin{equation}
D
\simeq
\frac{\hbar^{2}}{\,\Gamma_{e_+}}\,
\frac{\Omega(\mathbf{r})^{4}\,\Delta(\mathbf{r})^{2}}
{\big[2\Omega(\mathbf{r})^{2}+\Delta(\mathbf{r})^{2}\big]^{3}}\,
\big|\nabla \Delta(\mathbf{r})\big|^{2}.
\label{eq:Dp_general}
\end{equation}
When the detuning gradient is dominated by the control beam, Eq.~\eqref{eq:gradDelta_control} gives
\begin{equation}
D
\simeq
\frac{\hbar^{2}}{\,\Gamma_{e_+}}\,
\frac{\Omega(\mathbf{r})^{4}\,\Delta(\mathbf{r})^{2}}
{\big[2\Omega(\mathbf{r})^{2}+\Delta(\mathbf{r})^{2}\big]^{3}}\,
\big|\nabla \Omega_c(\mathbf{r})\big|^{2}.
\label{eq:Dp_control}
\end{equation}

This can be expressed in terms of the population of the bare dressed excited state $\ket{e_+}$ in the limit $\Delta(\mathbf{r})\gg\Omega(\mathbf{r})$ as $\rho_{e_+e_+}\approx\frac{\Omega(\mathbf{r})^2}{\Delta(\mathbf{r})^2}$ so that
\begin{equation}
\label{eq:rho_epeps}
\begin{split}
D
\simeq
\frac{\hbar^{2}}{\,\Gamma_{e_+}}\,
\rho_{e_+e_+}^2(\mathbf{r})\,
\big|\nabla \Omega_c(\mathbf{r})\big|^{2}.
\end{split}
\end{equation}

Finally, to connect momentum diffusion to a heating rate, we average over the atomic motion in a harmonic trap.
For a Gaussian control beam centered on the trap,
\begin{equation}
\Omega_c(\mathbf{r}_\perp)=\Omega_{c,0}\,e^{-\frac{|\mathbf{r_\perp}-\mathbf{r_0}|^{2}}{w_c^{2}}},
\qquad
r_\perp^{2}=x^{2}+y^{2},
\label{eq:Omegac_gauss}
\end{equation}
where $r_0$ is a pointing offset between the trap and the control beam. The gradient satisfies
\begin{equation}
\nabla \Omega_c(\mathbf{r}_\perp)
=
-\frac{2\,(\mathbf{r_\perp}-\mathbf{r_0})}{w_c^{2}}\,
\Omega_c(\mathbf{r}_\perp)
\approx
-\frac{2\,(\mathbf{r_\perp}-\mathbf{r_0})}{w_c^{2}}\Omega_{c,0},
\label{eq:gradOmegac_sq}
\end{equation}
where the final approximation holds for $r_\perp\ll w_c$.
For an atom with total transverse energy $E$ in an isotropic harmonic trap of frequency $\omega_x$,
equipartition gives $\langle|\mathbf{r_\perp}-\mathbf{r_0}|^{2}\rangle= \langle r_\perp^{2}\rangle +r_0^2= E/(m\omega_x^{2})+r_0^2$ and therefore
\begin{equation}
\Big\langle \big|\nabla \Omega_c(\mathbf{r}_\perp)\big|^{2}\Big\rangle
\simeq
\frac{4\,\Omega_{c,0}^{2}}{w_c^{4}}\,
(\frac{\langle E\rangle}{m\omega_x^{2}}+r_0^2).
\label{eq:avg_gradDelta_sq}
\end{equation}
Inserting Eq.~\eqref{eq:avg_gradDelta_sq} into Eq.~\eqref{eq:rho_epeps} gives
\begin{equation}
\label{eq:result_D}
\begin{split}
    D &= \frac{4\hbar^{2}}{m\omega_x^2\,\Gamma_{e_+}}\,
    \rho_{e_+e_+}^2\,
    \frac{\,\Omega_{c,0}^{2}}{w_c^{4}} \\
    &\quad \times (\langle E\rangle+m\omega_x^{2}r_0^2).
\end{split}
\end{equation}

Finally we have

\begin{equation}
\label{eq:heating_rate_HE}
\begin{split}
    \Big\langle\frac{dE}{dt}\Big\rangle &= \frac{D}{m} =
    \frac{1}{\tau}\,(\langle E\rangle+m\omega_x^2r_0^2), \\
    \tau &\equiv
    \left( \frac{4\hbar^{2}}{m^2\omega_x^2\,\Gamma_{e_+}}\,
    \rho_{e_+e_+}^2\,
    \frac{\Omega_{c,0}^{2}}{w_c^{4}} \right)^{-1}.
\end{split}
\end{equation}

For the maximally shifted atom (and therefore most extreme heating), parameters of $\omega_x/(2\pi)=\SI{100}{kHz}$, $w_c=\SI{2}{\micro m}$, $\Gamma_{e_+}\approx \frac{\Gamma + 2\pi\times \SI{1.9}{MHz}}{2}$,$\rho_{e_+e_+}=0.031$, $r_0=\SI{0}{nm}$, and $\Omega_c=2000\Gamma$, lead to an energy e-folding time of $\SI{13.2}{\micro s}$, slower than our readout pulse time of $\SI{5}{\micro s}$. Importantly, since the scatter rate $\propto \rho_{e_+e_+}$ and the heating rate is $\propto \rho_{e_+e_+}^2$, it is possible to decrease the heating rate for a fixed collected photon number by extending smaller values of $\rho_{e_+e_+}$ at the cost of slower readout speeds. 

A pointing error between the trapping and control beams of \SI{50}{nm} would lead to a $\sim0.1\%$ intensity shift error and a heating rate of \SI{\sim 1}{\micro K/ \micro s} for a zero-temperature atom, both of which are tolerable in our proposed scheme. Such precision is readily achievable in trapped ion system, where position measurements of $\lesssim$ \SI{10}{nm} have been achieved~\cite{poschinger2016ion6nm,felix2026ion10nm,vladan2013ion10nm}. Additionally, effects from finite ion temperature are likely to be negligible due to large trapping frequencies, so that dipole force fluctuations are purely dominated by misalignment. Similar pointing alignment precisions should be extensible to neutral-atom arrays as well using techniques such as position-dependent phase shifts \cite{Shaw2024MultiEnsemble} or directly utilizing the control beam shift as a means to obtain precise alignment. By measuring the light shift on the atom due to the control beam, alignment for a well-localized atom should be feasible to order $\sim \Gamma_{e_+}$ or $\sim \SI{30}{nm}$ for the maximally shifted atom. Supergaussian beams provide an additional route that would reduce the sensitivity to atomic temperature and pointing misalignment \cite{gillen2016supergaussian}. Alternatively, fiber-array architectures that codeliver trapping and addressing light through a common waveguide have recently been shown to provide robust common-mode suppression of pointing noise~\cite{li2025fiber}.

\subsection{Ground state light shifts}

Although the \SI{1529}{nm} control beam is near-resonant with the $\ket{e}\leftrightarrow\ket{f}$ transition, it is far detuned from the qubit ground manifold $5S_{1/2}$. We calculate the light shift that arises through off-resonant coupling dominated by the D lines. In the below calculations, we do not include small corrections from other transitions such as $5S_{1/2}\rightarrow6P_{3/2}$. In the far-detuned limit, the scalar light shift can be written as
\begin{equation}
U_{gs}(\mathbf{r})
\simeq
\frac{\pi c^{2}\Gamma}{2\omega_{0}^{3}}
\left(
\frac{2}{\Delta_{D2}}+\frac{1}{\Delta_{D1}}
\right)
I(\mathbf{r}),
\label{eq:Ugs_Dlines}
\end{equation}
where $\Delta_{D1,D2}=\omega_c-\omega_{D1,D2}$ are the detunings of the control light and $I(\mathbf{r})$ is the control-beam intensity.
The associated off-resonant scattering rate from the ground state is
\begin{equation}
\Gamma_{\mathrm{sc}}(\mathbf{r})
\simeq
\frac{\pi c^{2}\Gamma^{2}}{2\hbar\omega_{0}^{3}}
\left(
\frac{2}{\Delta_{D2}^{2}}+\frac{1}{\Delta_{D1}^{2}}
\right)
I(\mathbf{r}).
\label{eq:Gamma_sc_Dlines}
\end{equation}

For the maximally shifted atom we take $\Omega_c=2000\Gamma$ and $w_0=\SI{2}{\micro m}$. Using $\langle 5P_{3/2}||\mathbf{d}||4D_{5/2}\rangle \approx 10.90\,ea_0$ \cite{Shaw2024MultiEnsemble} and the stretched-state $\sigma^{+}$ coupling yields a required optical power of $P\simeq \SI{1.5}{mW}$. At the beam center this corresponds to a scalar ground-state shift magnitude $|U_{gs}|/h\simeq \SI{0.4}{MHz}$ and a scattering rate $\Gamma_{\mathrm{sc}}\simeq \SI{0.1}{s^{-1}}$, which implies a negligible scattering probability of $\sim 10^{-6}$ during a readout pulse.

Finally, the differential light shift between the $F=1$ and $F=2$ hyperfine manifolds is strongly suppressed because the ground hyperfine splitting is small compared to the detuning. Evaluating the difference in $U_{gs}$ for detunings offset by the hyperfine splitting gives $|\Delta U_{\mathrm{hf}}|/h \simeq \SI{10}{Hz}$ for the above parameters, which would not significantly impact coherence in the qubit manifold.

\color{black}
\bibliography{cmm_refs.bib}

\begin{thebibliography}{110}%
\makeatletter
\providecommand \@ifxundefined [1]{%
 \@ifx{#1\undefined}
}%
\providecommand \@ifnum [1]{%
 \ifnum #1\expandafter \@firstoftwo
 \else \expandafter \@secondoftwo
 \fi
}%
\providecommand \@ifx [1]{%
 \ifx #1\expandafter \@firstoftwo
 \else \expandafter \@secondoftwo
 \fi
}%
\providecommand \natexlab [1]{#1}%
\providecommand \enquote  [1]{``#1''}%
\providecommand \bibnamefont  [1]{#1}%
\providecommand \bibfnamefont [1]{#1}%
\providecommand \citenamefont [1]{#1}%
\providecommand \href@noop [0]{\@secondoftwo}%
\providecommand \href [0]{\begingroup \@sanitize@url \@href}%
\providecommand \@href[1]{\@@startlink{#1}\@@href}%
\providecommand \@@href[1]{\endgroup#1\@@endlink}%
\providecommand \@sanitize@url [0]{\catcode `\\12\catcode `\$12\catcode `\&12\catcode `\#12\catcode `\^12\catcode `\_12\catcode `\%12\relax}%
\providecommand \@@startlink[1]{}%
\providecommand \@@endlink[0]{}%
\providecommand \url  [0]{\begingroup\@sanitize@url \@url }%
\providecommand \@url [1]{\endgroup\@href {#1}{\urlprefix }}%
\providecommand \urlprefix  [0]{URL }%
\providecommand \Eprint [0]{\href }%
\providecommand \doibase [0]{https://doi.org/}%
\providecommand \selectlanguage [0]{\@gobble}%
\providecommand \bibinfo  [0]{\@secondoftwo}%
\providecommand \bibfield  [0]{\@secondoftwo}%
\providecommand \translation [1]{[#1]}%
\providecommand \BibitemOpen [0]{}%
\providecommand \bibitemStop [0]{}%
\providecommand \bibitemNoStop [0]{.\EOS\space}%
\providecommand \EOS [0]{\spacefactor3000\relax}%
\providecommand \BibitemShut  [1]{\csname bibitem#1\endcsname}%
\let\auto@bib@innerbib\@empty
\bibitem [{\citenamefont {Levine}\ \emph {et~al.}(2022)\citenamefont {Levine}, \citenamefont {Bluvstein}, \citenamefont {Keesling}, \citenamefont {Wang}, \citenamefont {Ebadi}, \citenamefont {Semeghini}, \citenamefont {Omran}, \citenamefont {Greiner}, \citenamefont {Vuleti{\'c}},\ and\ \citenamefont {Lukin}}]{levine2022dispersive}%
  \BibitemOpen
  \bibfield  {author} {\bibinfo {author} {\bibfnamefont {H.}~\bibnamefont {Levine}}, \bibinfo {author} {\bibfnamefont {D.}~\bibnamefont {Bluvstein}}, \bibinfo {author} {\bibfnamefont {A.}~\bibnamefont {Keesling}}, \bibinfo {author} {\bibfnamefont {T.~T.}\ \bibnamefont {Wang}}, \bibinfo {author} {\bibfnamefont {S.}~\bibnamefont {Ebadi}}, \bibinfo {author} {\bibfnamefont {G.}~\bibnamefont {Semeghini}}, \bibinfo {author} {\bibfnamefont {A.}~\bibnamefont {Omran}}, \bibinfo {author} {\bibfnamefont {M.}~\bibnamefont {Greiner}}, \bibinfo {author} {\bibfnamefont {V.}~\bibnamefont {Vuleti{\'c}}},\ and\ \bibinfo {author} {\bibfnamefont {M.~D.}\ \bibnamefont {Lukin}},\ }\bibfield  {title} {\bibinfo {title} {Dispersive optical systems for scalable {{Raman}} driving of hyperfine qubits},\ }\href {https://doi.org/10.1103/PhysRevA.105.032618} {\bibfield  {journal} {\bibinfo  {journal} {Phys. Rev. A}\ }\textbf {\bibinfo {volume} {105}},\ \bibinfo {pages} {032618} (\bibinfo {year} {2022})}\BibitemShut {NoStop}%
\bibitem [{\citenamefont {Evered}\ \emph {et~al.}(2023)\citenamefont {Evered}, \citenamefont {Bluvstein}, \citenamefont {Kalinowski}, \citenamefont {Ebadi}, \citenamefont {Manovitz}, \citenamefont {Zhou}, \citenamefont {Li}, \citenamefont {Geim}, \citenamefont {Wang}, \citenamefont {Maskara}, \citenamefont {Levine}, \citenamefont {Semeghini}, \citenamefont {Greiner}, \citenamefont {Vuleti{\'c}},\ and\ \citenamefont {Lukin}}]{evered2023highfidelity}%
  \BibitemOpen
  \bibfield  {author} {\bibinfo {author} {\bibfnamefont {S.~J.}\ \bibnamefont {Evered}}, \bibinfo {author} {\bibfnamefont {D.}~\bibnamefont {Bluvstein}}, \bibinfo {author} {\bibfnamefont {M.}~\bibnamefont {Kalinowski}}, \bibinfo {author} {\bibfnamefont {S.}~\bibnamefont {Ebadi}}, \bibinfo {author} {\bibfnamefont {T.}~\bibnamefont {Manovitz}}, \bibinfo {author} {\bibfnamefont {H.}~\bibnamefont {Zhou}}, \bibinfo {author} {\bibfnamefont {S.~H.}\ \bibnamefont {Li}}, \bibinfo {author} {\bibfnamefont {A.~A.}\ \bibnamefont {Geim}}, \bibinfo {author} {\bibfnamefont {T.~T.}\ \bibnamefont {Wang}}, \bibinfo {author} {\bibfnamefont {N.}~\bibnamefont {Maskara}}, \bibinfo {author} {\bibfnamefont {H.}~\bibnamefont {Levine}}, \bibinfo {author} {\bibfnamefont {G.}~\bibnamefont {Semeghini}}, \bibinfo {author} {\bibfnamefont {M.}~\bibnamefont {Greiner}}, \bibinfo {author} {\bibfnamefont {V.}~\bibnamefont {Vuleti{\'c}}},\ and\ \bibinfo {author} {\bibfnamefont {M.~D.}\ \bibnamefont {Lukin}},\ }\bibfield  {title} {\bibinfo {title}
  {High-fidelity parallel entangling gates on a neutral-atom quantum computer},\ }\href {https://doi.org/10.1038/s41586-023-06481-y} {\bibfield  {journal} {\bibinfo  {journal} {Nature}\ }\textbf {\bibinfo {volume} {622}},\ \bibinfo {pages} {268} (\bibinfo {year} {2023})}\BibitemShut {NoStop}%
\bibitem [{\citenamefont {Tsai}\ \emph {et~al.}(2025)\citenamefont {Tsai}, \citenamefont {Sun}, \citenamefont {Shaw}, \citenamefont {Finkelstein},\ and\ \citenamefont {Endres}}]{tsai2025benchmarking}%
  \BibitemOpen
  \bibfield  {author} {\bibinfo {author} {\bibfnamefont {R.~B.-S.}\ \bibnamefont {Tsai}}, \bibinfo {author} {\bibfnamefont {X.}~\bibnamefont {Sun}}, \bibinfo {author} {\bibfnamefont {A.~L.}\ \bibnamefont {Shaw}}, \bibinfo {author} {\bibfnamefont {R.}~\bibnamefont {Finkelstein}},\ and\ \bibinfo {author} {\bibfnamefont {M.}~\bibnamefont {Endres}},\ }\bibfield  {title} {\bibinfo {title} {Benchmarking and {{Fidelity Response Theory}} of {{High-Fidelity Rydberg Entangling Gates}}},\ }\href {https://doi.org/10.1103/PRXQuantum.6.010331} {\bibfield  {journal} {\bibinfo  {journal} {PRX Quantum}\ }\textbf {\bibinfo {volume} {6}},\ \bibinfo {pages} {010331} (\bibinfo {year} {2025})}\BibitemShut {NoStop}%
\bibitem [{\citenamefont {Radnaev}\ \emph {et~al.}(2025)\citenamefont {Radnaev}, \citenamefont {Chung}, \citenamefont {Cole}, \citenamefont {Mason}, \citenamefont {Ballance}, \citenamefont {Bedalov}, \citenamefont {Belknap}, \citenamefont {Berman}, \citenamefont {Blakely}, \citenamefont {Bloomfield}, \citenamefont {Buttler}, \citenamefont {Campbell}, \citenamefont {Chopinaud}, \citenamefont {Copenhaver}, \citenamefont {Dawes}, \citenamefont {Eubanks}, \citenamefont {Friss}, \citenamefont {Garcia}, \citenamefont {Gilbert}, \citenamefont {Gillette}, \citenamefont {Goiporia}, \citenamefont {Gokhale}, \citenamefont {Goldwin}, \citenamefont {Goodwin}, \citenamefont {Graham}, \citenamefont {Guttormsson}, \citenamefont {Hickman}, \citenamefont {Hurtley}, \citenamefont {Iliev}, \citenamefont {Jones}, \citenamefont {Jones}, \citenamefont {Kuper}, \citenamefont {Lewis}, \citenamefont {Lichtman}, \citenamefont {Majdeteimouri}, \citenamefont {Mason}, \citenamefont {McMaster}, \citenamefont {Miles}, \citenamefont
  {Mitchell}, \citenamefont {Murphree}, \citenamefont {{Neff-Mallon}}, \citenamefont {Oh}, \citenamefont {Omole}, \citenamefont {Parlo~Simon}, \citenamefont {Pederson}, \citenamefont {Perlin}, \citenamefont {Reiter}, \citenamefont {Rines}, \citenamefont {Romlow}, \citenamefont {Scott}, \citenamefont {Stiefvater}, \citenamefont {Tanner}, \citenamefont {Tucker}, \citenamefont {Vinogradov}, \citenamefont {Warter}, \citenamefont {Yeo}, \citenamefont {Saffman},\ and\ \citenamefont {Noel}}]{radnaev2025universal}%
  \BibitemOpen
  \bibfield  {author} {\bibinfo {author} {\bibfnamefont {A.}~\bibnamefont {Radnaev}}, \bibinfo {author} {\bibfnamefont {W.}~\bibnamefont {Chung}}, \bibinfo {author} {\bibfnamefont {D.}~\bibnamefont {Cole}}, \bibinfo {author} {\bibfnamefont {D.}~\bibnamefont {Mason}}, \bibinfo {author} {\bibfnamefont {T.}~\bibnamefont {Ballance}}, \bibinfo {author} {\bibfnamefont {M.}~\bibnamefont {Bedalov}}, \bibinfo {author} {\bibfnamefont {D.}~\bibnamefont {Belknap}}, \bibinfo {author} {\bibfnamefont {M.}~\bibnamefont {Berman}}, \bibinfo {author} {\bibfnamefont {M.}~\bibnamefont {Blakely}}, \bibinfo {author} {\bibfnamefont {I.}~\bibnamefont {Bloomfield}}, \bibinfo {author} {\bibfnamefont {P.}~\bibnamefont {Buttler}}, \bibinfo {author} {\bibfnamefont {C.}~\bibnamefont {Campbell}}, \bibinfo {author} {\bibfnamefont {A.}~\bibnamefont {Chopinaud}}, \bibinfo {author} {\bibfnamefont {E.}~\bibnamefont {Copenhaver}}, \bibinfo {author} {\bibfnamefont {M.}~\bibnamefont {Dawes}}, \bibinfo {author} {\bibfnamefont {S.}~\bibnamefont
  {Eubanks}}, \bibinfo {author} {\bibfnamefont {A.}~\bibnamefont {Friss}}, \bibinfo {author} {\bibfnamefont {D.}~\bibnamefont {Garcia}}, \bibinfo {author} {\bibfnamefont {J.}~\bibnamefont {Gilbert}}, \bibinfo {author} {\bibfnamefont {M.}~\bibnamefont {Gillette}}, \bibinfo {author} {\bibfnamefont {P.}~\bibnamefont {Goiporia}}, \bibinfo {author} {\bibfnamefont {P.}~\bibnamefont {Gokhale}}, \bibinfo {author} {\bibfnamefont {J.}~\bibnamefont {Goldwin}}, \bibinfo {author} {\bibfnamefont {D.}~\bibnamefont {Goodwin}}, \bibinfo {author} {\bibfnamefont {T.}~\bibnamefont {Graham}}, \bibinfo {author} {\bibfnamefont {C.}~\bibnamefont {Guttormsson}}, \bibinfo {author} {\bibfnamefont {G.}~\bibnamefont {Hickman}}, \bibinfo {author} {\bibfnamefont {L.}~\bibnamefont {Hurtley}}, \bibinfo {author} {\bibfnamefont {M.}~\bibnamefont {Iliev}}, \bibinfo {author} {\bibfnamefont {E.}~\bibnamefont {Jones}}, \bibinfo {author} {\bibfnamefont {R.}~\bibnamefont {Jones}}, \bibinfo {author} {\bibfnamefont {K.}~\bibnamefont {Kuper}}, \bibinfo
  {author} {\bibfnamefont {T.}~\bibnamefont {Lewis}}, \bibinfo {author} {\bibfnamefont {M.}~\bibnamefont {Lichtman}}, \bibinfo {author} {\bibfnamefont {F.}~\bibnamefont {Majdeteimouri}}, \bibinfo {author} {\bibfnamefont {J.}~\bibnamefont {Mason}}, \bibinfo {author} {\bibfnamefont {J.}~\bibnamefont {McMaster}}, \bibinfo {author} {\bibfnamefont {J.}~\bibnamefont {Miles}}, \bibinfo {author} {\bibfnamefont {P.}~\bibnamefont {Mitchell}}, \bibinfo {author} {\bibfnamefont {J.}~\bibnamefont {Murphree}}, \bibinfo {author} {\bibfnamefont {N.}~\bibnamefont {{Neff-Mallon}}}, \bibinfo {author} {\bibfnamefont {T.}~\bibnamefont {Oh}}, \bibinfo {author} {\bibfnamefont {V.}~\bibnamefont {Omole}}, \bibinfo {author} {\bibfnamefont {C.}~\bibnamefont {Parlo~Simon}}, \bibinfo {author} {\bibfnamefont {N.}~\bibnamefont {Pederson}}, \bibinfo {author} {\bibfnamefont {M.}~\bibnamefont {Perlin}}, \bibinfo {author} {\bibfnamefont {A.}~\bibnamefont {Reiter}}, \bibinfo {author} {\bibfnamefont {R.}~\bibnamefont {Rines}}, \bibinfo {author}
  {\bibfnamefont {P.}~\bibnamefont {Romlow}}, \bibinfo {author} {\bibfnamefont {A.}~\bibnamefont {Scott}}, \bibinfo {author} {\bibfnamefont {D.}~\bibnamefont {Stiefvater}}, \bibinfo {author} {\bibfnamefont {J.}~\bibnamefont {Tanner}}, \bibinfo {author} {\bibfnamefont {A.}~\bibnamefont {Tucker}}, \bibinfo {author} {\bibfnamefont {I.}~\bibnamefont {Vinogradov}}, \bibinfo {author} {\bibfnamefont {M.}~\bibnamefont {Warter}}, \bibinfo {author} {\bibfnamefont {M.}~\bibnamefont {Yeo}}, \bibinfo {author} {\bibfnamefont {M.}~\bibnamefont {Saffman}},\ and\ \bibinfo {author} {\bibfnamefont {T.}~\bibnamefont {Noel}},\ }\bibfield  {title} {\bibinfo {title} {Universal {{Neutral-Atom Quantum Computer}} with {{Individual Optical Addressing}} and {{Nondestructive Readout}}},\ }\href {https://doi.org/10.1103/66s8-jj18} {\bibfield  {journal} {\bibinfo  {journal} {PRX Quantum}\ }\textbf {\bibinfo {volume} {6}},\ \bibinfo {pages} {030334} (\bibinfo {year} {2025})}\BibitemShut {NoStop}%
\bibitem [{\citenamefont {Peper}\ \emph {et~al.}(2025)\citenamefont {Peper}, \citenamefont {Li}, \citenamefont {Knapp}, \citenamefont {Bileska}, \citenamefont {Ma}, \citenamefont {Liu}, \citenamefont {Peng}, \citenamefont {Zhang}, \citenamefont {Horvath}, \citenamefont {Burgers},\ and\ \citenamefont {Thompson}}]{peper2025spectroscopy}%
  \BibitemOpen
  \bibfield  {author} {\bibinfo {author} {\bibfnamefont {M.}~\bibnamefont {Peper}}, \bibinfo {author} {\bibfnamefont {Y.}~\bibnamefont {Li}}, \bibinfo {author} {\bibfnamefont {D.~Y.}\ \bibnamefont {Knapp}}, \bibinfo {author} {\bibfnamefont {M.}~\bibnamefont {Bileska}}, \bibinfo {author} {\bibfnamefont {S.}~\bibnamefont {Ma}}, \bibinfo {author} {\bibfnamefont {G.}~\bibnamefont {Liu}}, \bibinfo {author} {\bibfnamefont {P.}~\bibnamefont {Peng}}, \bibinfo {author} {\bibfnamefont {B.}~\bibnamefont {Zhang}}, \bibinfo {author} {\bibfnamefont {S.~P.}\ \bibnamefont {Horvath}}, \bibinfo {author} {\bibfnamefont {A.~P.}\ \bibnamefont {Burgers}},\ and\ \bibinfo {author} {\bibfnamefont {J.~D.}\ \bibnamefont {Thompson}},\ }\bibfield  {title} {\bibinfo {title} {Spectroscopy and {{Modeling}} of {{Yb}} 171 {{Rydberg States}} for {{High-Fidelity Two-Qubit Gates}}},\ }\href {https://doi.org/10.1103/PhysRevX.15.011009} {\bibfield  {journal} {\bibinfo  {journal} {Phys. Rev. X}\ }\textbf {\bibinfo {volume} {15}},\ \bibinfo {pages}
  {011009} (\bibinfo {year} {2025})}\BibitemShut {NoStop}%
\bibitem [{\citenamefont {Muniz}\ \emph {et~al.}(2025)\citenamefont {Muniz}, \citenamefont {Stone}, \citenamefont {Stack}, \citenamefont {Jaffe}, \citenamefont {Kindem}, \citenamefont {Wadleigh}, \citenamefont {{Zalys-Geller}}, \citenamefont {Zhang}, \citenamefont {Chen}, \citenamefont {Norcia}, \citenamefont {Epstein}, \citenamefont {Halperin}, \citenamefont {Hummel}, \citenamefont {Wilkason}, \citenamefont {Li}, \citenamefont {Barnes}, \citenamefont {Battaglino}, \citenamefont {Bohdanowicz}, \citenamefont {Booth}, \citenamefont {Brown}, \citenamefont {Brown}, \citenamefont {Cairncross}, \citenamefont {Cassella}, \citenamefont {Coxe}, \citenamefont {Crow}, \citenamefont {Feldkamp}, \citenamefont {Griger}, \citenamefont {Heinz}, \citenamefont {Jones}, \citenamefont {Kim}, \citenamefont {King}, \citenamefont {Kotru}, \citenamefont {Lauigan}, \citenamefont {Marjanovic}, \citenamefont {Megidish}, \citenamefont {Meredith}, \citenamefont {McDonald}, \citenamefont {Morshead}, \citenamefont {Narayanaswami},
  \citenamefont {Nishiguchi}, \citenamefont {Paule}, \citenamefont {Pawlak}, \citenamefont {Pudenz}, \citenamefont {P{\'e}rez}, \citenamefont {Ryou}, \citenamefont {Simon}, \citenamefont {Smull}, \citenamefont {Urbanek}, \citenamefont {Van De~Veerdonk}, \citenamefont {Vendeiro}, \citenamefont {Wu}, \citenamefont {Xie},\ and\ \citenamefont {Bloom}}]{muniz2025highfidelity}%
  \BibitemOpen
  \bibfield  {author} {\bibinfo {author} {\bibfnamefont {J.~A.}\ \bibnamefont {Muniz}}, \bibinfo {author} {\bibfnamefont {M.}~\bibnamefont {Stone}}, \bibinfo {author} {\bibfnamefont {D.~T.}\ \bibnamefont {Stack}}, \bibinfo {author} {\bibfnamefont {M.}~\bibnamefont {Jaffe}}, \bibinfo {author} {\bibfnamefont {J.~M.}\ \bibnamefont {Kindem}}, \bibinfo {author} {\bibfnamefont {L.}~\bibnamefont {Wadleigh}}, \bibinfo {author} {\bibfnamefont {E.}~\bibnamefont {{Zalys-Geller}}}, \bibinfo {author} {\bibfnamefont {X.}~\bibnamefont {Zhang}}, \bibinfo {author} {\bibfnamefont {C.-A.}\ \bibnamefont {Chen}}, \bibinfo {author} {\bibfnamefont {M.~A.}\ \bibnamefont {Norcia}}, \bibinfo {author} {\bibfnamefont {J.}~\bibnamefont {Epstein}}, \bibinfo {author} {\bibfnamefont {E.}~\bibnamefont {Halperin}}, \bibinfo {author} {\bibfnamefont {F.}~\bibnamefont {Hummel}}, \bibinfo {author} {\bibfnamefont {T.}~\bibnamefont {Wilkason}}, \bibinfo {author} {\bibfnamefont {M.}~\bibnamefont {Li}}, \bibinfo {author} {\bibfnamefont
  {K.}~\bibnamefont {Barnes}}, \bibinfo {author} {\bibfnamefont {P.}~\bibnamefont {Battaglino}}, \bibinfo {author} {\bibfnamefont {T.~C.}\ \bibnamefont {Bohdanowicz}}, \bibinfo {author} {\bibfnamefont {G.}~\bibnamefont {Booth}}, \bibinfo {author} {\bibfnamefont {A.}~\bibnamefont {Brown}}, \bibinfo {author} {\bibfnamefont {M.~O.}\ \bibnamefont {Brown}}, \bibinfo {author} {\bibfnamefont {W.~B.}\ \bibnamefont {Cairncross}}, \bibinfo {author} {\bibfnamefont {K.}~\bibnamefont {Cassella}}, \bibinfo {author} {\bibfnamefont {R.}~\bibnamefont {Coxe}}, \bibinfo {author} {\bibfnamefont {D.}~\bibnamefont {Crow}}, \bibinfo {author} {\bibfnamefont {M.}~\bibnamefont {Feldkamp}}, \bibinfo {author} {\bibfnamefont {C.}~\bibnamefont {Griger}}, \bibinfo {author} {\bibfnamefont {A.}~\bibnamefont {Heinz}}, \bibinfo {author} {\bibfnamefont {A.~M.~W.}\ \bibnamefont {Jones}}, \bibinfo {author} {\bibfnamefont {H.}~\bibnamefont {Kim}}, \bibinfo {author} {\bibfnamefont {J.}~\bibnamefont {King}}, \bibinfo {author} {\bibfnamefont
  {K.}~\bibnamefont {Kotru}}, \bibinfo {author} {\bibfnamefont {J.}~\bibnamefont {Lauigan}}, \bibinfo {author} {\bibfnamefont {J.}~\bibnamefont {Marjanovic}}, \bibinfo {author} {\bibfnamefont {E.}~\bibnamefont {Megidish}}, \bibinfo {author} {\bibfnamefont {M.}~\bibnamefont {Meredith}}, \bibinfo {author} {\bibfnamefont {M.}~\bibnamefont {McDonald}}, \bibinfo {author} {\bibfnamefont {R.}~\bibnamefont {Morshead}}, \bibinfo {author} {\bibfnamefont {S.}~\bibnamefont {Narayanaswami}}, \bibinfo {author} {\bibfnamefont {C.}~\bibnamefont {Nishiguchi}}, \bibinfo {author} {\bibfnamefont {T.}~\bibnamefont {Paule}}, \bibinfo {author} {\bibfnamefont {K.~A.}\ \bibnamefont {Pawlak}}, \bibinfo {author} {\bibfnamefont {K.~L.}\ \bibnamefont {Pudenz}}, \bibinfo {author} {\bibfnamefont {D.~R.}\ \bibnamefont {P{\'e}rez}}, \bibinfo {author} {\bibfnamefont {A.}~\bibnamefont {Ryou}}, \bibinfo {author} {\bibfnamefont {J.}~\bibnamefont {Simon}}, \bibinfo {author} {\bibfnamefont {A.}~\bibnamefont {Smull}}, \bibinfo {author}
  {\bibfnamefont {M.}~\bibnamefont {Urbanek}}, \bibinfo {author} {\bibfnamefont {R.~J.~M.}\ \bibnamefont {Van De~Veerdonk}}, \bibinfo {author} {\bibfnamefont {Z.}~\bibnamefont {Vendeiro}}, \bibinfo {author} {\bibfnamefont {T.-Y.}\ \bibnamefont {Wu}}, \bibinfo {author} {\bibfnamefont {X.}~\bibnamefont {Xie}},\ and\ \bibinfo {author} {\bibfnamefont {B.~J.}\ \bibnamefont {Bloom}},\ }\bibfield  {title} {\bibinfo {title} {High-{{Fidelity Universal Gates}} in the 171 {{Yb Ground-State Nuclear-Spin Qubit}}},\ }\href {https://doi.org/10.1103/PRXQuantum.6.020334} {\bibfield  {journal} {\bibinfo  {journal} {PRX Quantum}\ }\textbf {\bibinfo {volume} {6}},\ \bibinfo {pages} {020334} (\bibinfo {year} {2025})}\BibitemShut {NoStop}%
\bibitem [{\citenamefont {Beugnon}\ \emph {et~al.}(2007)\citenamefont {Beugnon}, \citenamefont {Tuchendler}, \citenamefont {Marion}, \citenamefont {Ga{\"e}tan}, \citenamefont {Miroshnychenko}, \citenamefont {Sortais}, \citenamefont {Lance}, \citenamefont {Jones}, \citenamefont {Messin}, \citenamefont {Browaeys},\ and\ \citenamefont {Grangier}}]{beugnon2007twodimensional}%
  \BibitemOpen
  \bibfield  {author} {\bibinfo {author} {\bibfnamefont {J.}~\bibnamefont {Beugnon}}, \bibinfo {author} {\bibfnamefont {C.}~\bibnamefont {Tuchendler}}, \bibinfo {author} {\bibfnamefont {H.}~\bibnamefont {Marion}}, \bibinfo {author} {\bibfnamefont {A.}~\bibnamefont {Ga{\"e}tan}}, \bibinfo {author} {\bibfnamefont {Y.}~\bibnamefont {Miroshnychenko}}, \bibinfo {author} {\bibfnamefont {Y.~R.~P.}\ \bibnamefont {Sortais}}, \bibinfo {author} {\bibfnamefont {A.~M.}\ \bibnamefont {Lance}}, \bibinfo {author} {\bibfnamefont {M.~P.~A.}\ \bibnamefont {Jones}}, \bibinfo {author} {\bibfnamefont {G.}~\bibnamefont {Messin}}, \bibinfo {author} {\bibfnamefont {A.}~\bibnamefont {Browaeys}},\ and\ \bibinfo {author} {\bibfnamefont {P.}~\bibnamefont {Grangier}},\ }\bibfield  {title} {\bibinfo {title} {Two-dimensional transport and transfer of a single atomic qubit in optical tweezers},\ }\href {https://doi.org/10.1038/nphys698} {\bibfield  {journal} {\bibinfo  {journal} {Nature Phys}\ }\textbf {\bibinfo {volume} {3}},\ \bibinfo
  {pages} {696} (\bibinfo {year} {2007})}\BibitemShut {NoStop}%
\bibitem [{\citenamefont {Bluvstein}\ \emph {et~al.}(2022)\citenamefont {Bluvstein}, \citenamefont {Levine}, \citenamefont {Semeghini}, \citenamefont {Wang}, \citenamefont {Ebadi}, \citenamefont {Kalinowski}, \citenamefont {Keesling}, \citenamefont {Maskara}, \citenamefont {Pichler}, \citenamefont {Greiner}, \citenamefont {Vuleti{\'c}},\ and\ \citenamefont {Lukin}}]{bluvstein2022quantum}%
  \BibitemOpen
  \bibfield  {author} {\bibinfo {author} {\bibfnamefont {D.}~\bibnamefont {Bluvstein}}, \bibinfo {author} {\bibfnamefont {H.}~\bibnamefont {Levine}}, \bibinfo {author} {\bibfnamefont {G.}~\bibnamefont {Semeghini}}, \bibinfo {author} {\bibfnamefont {T.~T.}\ \bibnamefont {Wang}}, \bibinfo {author} {\bibfnamefont {S.}~\bibnamefont {Ebadi}}, \bibinfo {author} {\bibfnamefont {M.}~\bibnamefont {Kalinowski}}, \bibinfo {author} {\bibfnamefont {A.}~\bibnamefont {Keesling}}, \bibinfo {author} {\bibfnamefont {N.}~\bibnamefont {Maskara}}, \bibinfo {author} {\bibfnamefont {H.}~\bibnamefont {Pichler}}, \bibinfo {author} {\bibfnamefont {M.}~\bibnamefont {Greiner}}, \bibinfo {author} {\bibfnamefont {V.}~\bibnamefont {Vuleti{\'c}}},\ and\ \bibinfo {author} {\bibfnamefont {M.~D.}\ \bibnamefont {Lukin}},\ }\bibfield  {title} {\bibinfo {title} {A quantum processor based on coherent transport of entangled atom arrays},\ }\href {https://doi.org/10.1038/s41586-022-04592-6} {\bibfield  {journal} {\bibinfo  {journal} {Nature}\
  }\textbf {\bibinfo {volume} {604}},\ \bibinfo {pages} {451} (\bibinfo {year} {2022})}\BibitemShut {NoStop}%
\bibitem [{\citenamefont {Bluvstein}\ \emph {et~al.}(2024)\citenamefont {Bluvstein}, \citenamefont {Evered}, \citenamefont {Geim}, \citenamefont {Li}, \citenamefont {Zhou}, \citenamefont {Manovitz}, \citenamefont {Ebadi}, \citenamefont {Cain}, \citenamefont {Kalinowski}, \citenamefont {Hangleiter}, \citenamefont {Bonilla~Ataides}, \citenamefont {Maskara}, \citenamefont {Cong}, \citenamefont {Gao}, \citenamefont {Sales~Rodriguez}, \citenamefont {Karolyshyn}, \citenamefont {Semeghini}, \citenamefont {Gullans}, \citenamefont {Greiner}, \citenamefont {Vuleti{\'c}},\ and\ \citenamefont {Lukin}}]{bluvstein2024logical}%
  \BibitemOpen
  \bibfield  {author} {\bibinfo {author} {\bibfnamefont {D.}~\bibnamefont {Bluvstein}}, \bibinfo {author} {\bibfnamefont {S.~J.}\ \bibnamefont {Evered}}, \bibinfo {author} {\bibfnamefont {A.~A.}\ \bibnamefont {Geim}}, \bibinfo {author} {\bibfnamefont {S.~H.}\ \bibnamefont {Li}}, \bibinfo {author} {\bibfnamefont {H.}~\bibnamefont {Zhou}}, \bibinfo {author} {\bibfnamefont {T.}~\bibnamefont {Manovitz}}, \bibinfo {author} {\bibfnamefont {S.}~\bibnamefont {Ebadi}}, \bibinfo {author} {\bibfnamefont {M.}~\bibnamefont {Cain}}, \bibinfo {author} {\bibfnamefont {M.}~\bibnamefont {Kalinowski}}, \bibinfo {author} {\bibfnamefont {D.}~\bibnamefont {Hangleiter}}, \bibinfo {author} {\bibfnamefont {J.~P.}\ \bibnamefont {Bonilla~Ataides}}, \bibinfo {author} {\bibfnamefont {N.}~\bibnamefont {Maskara}}, \bibinfo {author} {\bibfnamefont {I.}~\bibnamefont {Cong}}, \bibinfo {author} {\bibfnamefont {X.}~\bibnamefont {Gao}}, \bibinfo {author} {\bibfnamefont {P.}~\bibnamefont {Sales~Rodriguez}}, \bibinfo {author} {\bibfnamefont
  {T.}~\bibnamefont {Karolyshyn}}, \bibinfo {author} {\bibfnamefont {G.}~\bibnamefont {Semeghini}}, \bibinfo {author} {\bibfnamefont {M.~J.}\ \bibnamefont {Gullans}}, \bibinfo {author} {\bibfnamefont {M.}~\bibnamefont {Greiner}}, \bibinfo {author} {\bibfnamefont {V.}~\bibnamefont {Vuleti{\'c}}},\ and\ \bibinfo {author} {\bibfnamefont {M.~D.}\ \bibnamefont {Lukin}},\ }\bibfield  {title} {\bibinfo {title} {Logical quantum processor based on reconfigurable atom arrays},\ }\href {https://doi.org/10.1038/s41586-023-06927-3} {\bibfield  {journal} {\bibinfo  {journal} {Nature}\ }\textbf {\bibinfo {volume} {626}},\ \bibinfo {pages} {58} (\bibinfo {year} {2024})}\BibitemShut {NoStop}%
\bibitem [{\citenamefont {Bluvstein}\ \emph {et~al.}(2025)\citenamefont {Bluvstein}, \citenamefont {Geim}, \citenamefont {Li}, \citenamefont {Evered}, \citenamefont {Bonilla~Ataides}, \citenamefont {Baranes}, \citenamefont {Gu}, \citenamefont {Manovitz}, \citenamefont {Xu}, \citenamefont {Kalinowski}, \citenamefont {Majidy}, \citenamefont {Kokail}, \citenamefont {Maskara}, \citenamefont {Trapp}, \citenamefont {Stewart}, \citenamefont {Hollerith}, \citenamefont {Zhou}, \citenamefont {Gullans}, \citenamefont {Yelin}, \citenamefont {Greiner}, \citenamefont {Vuleti{\'c}}, \citenamefont {Cain},\ and\ \citenamefont {Lukin}}]{bluvstein2025faulttolerant}%
  \BibitemOpen
  \bibfield  {author} {\bibinfo {author} {\bibfnamefont {D.}~\bibnamefont {Bluvstein}}, \bibinfo {author} {\bibfnamefont {A.~A.}\ \bibnamefont {Geim}}, \bibinfo {author} {\bibfnamefont {S.~H.}\ \bibnamefont {Li}}, \bibinfo {author} {\bibfnamefont {S.~J.}\ \bibnamefont {Evered}}, \bibinfo {author} {\bibfnamefont {J.~P.}\ \bibnamefont {Bonilla~Ataides}}, \bibinfo {author} {\bibfnamefont {G.}~\bibnamefont {Baranes}}, \bibinfo {author} {\bibfnamefont {A.}~\bibnamefont {Gu}}, \bibinfo {author} {\bibfnamefont {T.}~\bibnamefont {Manovitz}}, \bibinfo {author} {\bibfnamefont {M.}~\bibnamefont {Xu}}, \bibinfo {author} {\bibfnamefont {M.}~\bibnamefont {Kalinowski}}, \bibinfo {author} {\bibfnamefont {S.}~\bibnamefont {Majidy}}, \bibinfo {author} {\bibfnamefont {C.}~\bibnamefont {Kokail}}, \bibinfo {author} {\bibfnamefont {N.}~\bibnamefont {Maskara}}, \bibinfo {author} {\bibfnamefont {E.~C.}\ \bibnamefont {Trapp}}, \bibinfo {author} {\bibfnamefont {L.~M.}\ \bibnamefont {Stewart}}, \bibinfo {author} {\bibfnamefont
  {S.}~\bibnamefont {Hollerith}}, \bibinfo {author} {\bibfnamefont {H.}~\bibnamefont {Zhou}}, \bibinfo {author} {\bibfnamefont {M.~J.}\ \bibnamefont {Gullans}}, \bibinfo {author} {\bibfnamefont {S.~F.}\ \bibnamefont {Yelin}}, \bibinfo {author} {\bibfnamefont {M.}~\bibnamefont {Greiner}}, \bibinfo {author} {\bibfnamefont {V.}~\bibnamefont {Vuleti{\'c}}}, \bibinfo {author} {\bibfnamefont {M.}~\bibnamefont {Cain}},\ and\ \bibinfo {author} {\bibfnamefont {M.~D.}\ \bibnamefont {Lukin}},\ }\bibfield  {title} {\bibinfo {title} {A fault-tolerant neutral-atom architecture for universal quantum computation},\ }\href {https://doi.org/10.1038/s41586-025-09848-5} {\bibfield  {journal} {\bibinfo  {journal} {Nature}\ } (\bibinfo {year} {2025})}\BibitemShut {NoStop}%
\bibitem [{\citenamefont {Manetsch}\ \emph {et~al.}(2025)\citenamefont {Manetsch}, \citenamefont {Nomura}, \citenamefont {Bataille}, \citenamefont {Lv}, \citenamefont {Leung},\ and\ \citenamefont {Endres}}]{manetsch2025tweezer}%
  \BibitemOpen
  \bibfield  {author} {\bibinfo {author} {\bibfnamefont {H.~J.}\ \bibnamefont {Manetsch}}, \bibinfo {author} {\bibfnamefont {G.}~\bibnamefont {Nomura}}, \bibinfo {author} {\bibfnamefont {E.}~\bibnamefont {Bataille}}, \bibinfo {author} {\bibfnamefont {X.}~\bibnamefont {Lv}}, \bibinfo {author} {\bibfnamefont {K.~H.}\ \bibnamefont {Leung}},\ and\ \bibinfo {author} {\bibfnamefont {M.}~\bibnamefont {Endres}},\ }\bibfield  {title} {\bibinfo {title} {A tweezer array with 6,100 highly coherent atomic qubits},\ }\href {https://doi.org/10.1038/s41586-025-09641-4} {\bibfield  {journal} {\bibinfo  {journal} {Nature}\ }\textbf {\bibinfo {volume} {647}},\ \bibinfo {pages} {60} (\bibinfo {year} {2025})}\BibitemShut {NoStop}%
\bibitem [{\citenamefont {Gyger}\ \emph {et~al.}(2024)\citenamefont {Gyger}, \citenamefont {Ammenwerth}, \citenamefont {Tao}, \citenamefont {Timme}, \citenamefont {Snigirev}, \citenamefont {Bloch},\ and\ \citenamefont {Zeiher}}]{gyger2024continuous}%
  \BibitemOpen
  \bibfield  {author} {\bibinfo {author} {\bibfnamefont {F.}~\bibnamefont {Gyger}}, \bibinfo {author} {\bibfnamefont {M.}~\bibnamefont {Ammenwerth}}, \bibinfo {author} {\bibfnamefont {R.}~\bibnamefont {Tao}}, \bibinfo {author} {\bibfnamefont {H.}~\bibnamefont {Timme}}, \bibinfo {author} {\bibfnamefont {S.}~\bibnamefont {Snigirev}}, \bibinfo {author} {\bibfnamefont {I.}~\bibnamefont {Bloch}},\ and\ \bibinfo {author} {\bibfnamefont {J.}~\bibnamefont {Zeiher}},\ }\bibfield  {title} {\bibinfo {title} {Continuous operation of large-scale atom arrays in optical lattices},\ }\href {https://doi.org/10.1103/PhysRevResearch.6.033104} {\bibfield  {journal} {\bibinfo  {journal} {Phys. Rev. Res.}\ }\textbf {\bibinfo {volume} {6}},\ \bibinfo {pages} {033104} (\bibinfo {year} {2024})}\BibitemShut {NoStop}%
\bibitem [{\citenamefont {Norcia}\ \emph {et~al.}(2024)\citenamefont {Norcia}, \citenamefont {Kim}, \citenamefont {Cairncross}, \citenamefont {Stone}, \citenamefont {Ryou}, \citenamefont {Jaffe}, \citenamefont {Brown}, \citenamefont {Barnes}, \citenamefont {Battaglino}, \citenamefont {Bohdanowicz}, \citenamefont {Brown}, \citenamefont {Cassella}, \citenamefont {Chen}, \citenamefont {Coxe}, \citenamefont {Crow}, \citenamefont {Epstein}, \citenamefont {Griger}, \citenamefont {Halperin}, \citenamefont {Hummel}, \citenamefont {Jones}, \citenamefont {Kindem}, \citenamefont {King}, \citenamefont {Kotru}, \citenamefont {Lauigan}, \citenamefont {Li}, \citenamefont {Lu}, \citenamefont {Megidish}, \citenamefont {Marjanovic}, \citenamefont {McDonald}, \citenamefont {Mittiga}, \citenamefont {Muniz}, \citenamefont {Narayanaswami}, \citenamefont {Nishiguchi}, \citenamefont {Paule}, \citenamefont {Pawlak}, \citenamefont {Peng}, \citenamefont {Pudenz}, \citenamefont {Rodr{\'i}guez~P{\'e}rez}, \citenamefont {Smull},
  \citenamefont {Stack}, \citenamefont {Urbanek}, \citenamefont {Van De~Veerdonk}, \citenamefont {Vendeiro}, \citenamefont {Wadleigh}, \citenamefont {Wilkason}, \citenamefont {Wu}, \citenamefont {Xie}, \citenamefont {{Zalys-Geller}}, \citenamefont {Zhang},\ and\ \citenamefont {Bloom}}]{norcia2024iterative}%
  \BibitemOpen
  \bibfield  {author} {\bibinfo {author} {\bibfnamefont {M.~A.}\ \bibnamefont {Norcia}}, \bibinfo {author} {\bibfnamefont {H.}~\bibnamefont {Kim}}, \bibinfo {author} {\bibfnamefont {W.~B.}\ \bibnamefont {Cairncross}}, \bibinfo {author} {\bibfnamefont {M.}~\bibnamefont {Stone}}, \bibinfo {author} {\bibfnamefont {A.}~\bibnamefont {Ryou}}, \bibinfo {author} {\bibfnamefont {M.}~\bibnamefont {Jaffe}}, \bibinfo {author} {\bibfnamefont {M.~O.}\ \bibnamefont {Brown}}, \bibinfo {author} {\bibfnamefont {K.}~\bibnamefont {Barnes}}, \bibinfo {author} {\bibfnamefont {P.}~\bibnamefont {Battaglino}}, \bibinfo {author} {\bibfnamefont {T.~C.}\ \bibnamefont {Bohdanowicz}}, \bibinfo {author} {\bibfnamefont {A.}~\bibnamefont {Brown}}, \bibinfo {author} {\bibfnamefont {K.}~\bibnamefont {Cassella}}, \bibinfo {author} {\bibfnamefont {C.-A.}\ \bibnamefont {Chen}}, \bibinfo {author} {\bibfnamefont {R.}~\bibnamefont {Coxe}}, \bibinfo {author} {\bibfnamefont {D.}~\bibnamefont {Crow}}, \bibinfo {author} {\bibfnamefont {J.}~\bibnamefont
  {Epstein}}, \bibinfo {author} {\bibfnamefont {C.}~\bibnamefont {Griger}}, \bibinfo {author} {\bibfnamefont {E.}~\bibnamefont {Halperin}}, \bibinfo {author} {\bibfnamefont {F.}~\bibnamefont {Hummel}}, \bibinfo {author} {\bibfnamefont {A.~M.~W.}\ \bibnamefont {Jones}}, \bibinfo {author} {\bibfnamefont {J.~M.}\ \bibnamefont {Kindem}}, \bibinfo {author} {\bibfnamefont {J.}~\bibnamefont {King}}, \bibinfo {author} {\bibfnamefont {K.}~\bibnamefont {Kotru}}, \bibinfo {author} {\bibfnamefont {J.}~\bibnamefont {Lauigan}}, \bibinfo {author} {\bibfnamefont {M.}~\bibnamefont {Li}}, \bibinfo {author} {\bibfnamefont {M.}~\bibnamefont {Lu}}, \bibinfo {author} {\bibfnamefont {E.}~\bibnamefont {Megidish}}, \bibinfo {author} {\bibfnamefont {J.}~\bibnamefont {Marjanovic}}, \bibinfo {author} {\bibfnamefont {M.}~\bibnamefont {McDonald}}, \bibinfo {author} {\bibfnamefont {T.}~\bibnamefont {Mittiga}}, \bibinfo {author} {\bibfnamefont {J.~A.}\ \bibnamefont {Muniz}}, \bibinfo {author} {\bibfnamefont {S.}~\bibnamefont
  {Narayanaswami}}, \bibinfo {author} {\bibfnamefont {C.}~\bibnamefont {Nishiguchi}}, \bibinfo {author} {\bibfnamefont {T.}~\bibnamefont {Paule}}, \bibinfo {author} {\bibfnamefont {K.~A.}\ \bibnamefont {Pawlak}}, \bibinfo {author} {\bibfnamefont {L.~S.}\ \bibnamefont {Peng}}, \bibinfo {author} {\bibfnamefont {K.~L.}\ \bibnamefont {Pudenz}}, \bibinfo {author} {\bibfnamefont {D.}~\bibnamefont {Rodr{\'i}guez~P{\'e}rez}}, \bibinfo {author} {\bibfnamefont {A.}~\bibnamefont {Smull}}, \bibinfo {author} {\bibfnamefont {D.}~\bibnamefont {Stack}}, \bibinfo {author} {\bibfnamefont {M.}~\bibnamefont {Urbanek}}, \bibinfo {author} {\bibfnamefont {R.~J.~M.}\ \bibnamefont {Van De~Veerdonk}}, \bibinfo {author} {\bibfnamefont {Z.}~\bibnamefont {Vendeiro}}, \bibinfo {author} {\bibfnamefont {L.}~\bibnamefont {Wadleigh}}, \bibinfo {author} {\bibfnamefont {T.}~\bibnamefont {Wilkason}}, \bibinfo {author} {\bibfnamefont {T.-Y.}\ \bibnamefont {Wu}}, \bibinfo {author} {\bibfnamefont {X.}~\bibnamefont {Xie}}, \bibinfo {author}
  {\bibfnamefont {E.}~\bibnamefont {{Zalys-Geller}}}, \bibinfo {author} {\bibfnamefont {X.}~\bibnamefont {Zhang}},\ and\ \bibinfo {author} {\bibfnamefont {B.~J.}\ \bibnamefont {Bloom}},\ }\bibfield  {title} {\bibinfo {title} {Iterative {{Assembly}} of 171 {{Yb Atom Arrays}} with {{Cavity-Enhanced Optical Lattices}}},\ }\href {https://doi.org/10.1103/PRXQuantum.5.030316} {\bibfield  {journal} {\bibinfo  {journal} {PRX Quantum}\ }\textbf {\bibinfo {volume} {5}},\ \bibinfo {pages} {030316} (\bibinfo {year} {2024})}\BibitemShut {NoStop}%
\bibitem [{\citenamefont {Chiu}\ \emph {et~al.}(2025)\citenamefont {Chiu}, \citenamefont {Trapp}, \citenamefont {Guo}, \citenamefont {Abobeih}, \citenamefont {Stewart}, \citenamefont {Hollerith}, \citenamefont {Stroganov}, \citenamefont {Kalinowski}, \citenamefont {Geim}, \citenamefont {Evered}, \citenamefont {Li}, \citenamefont {Lyu}, \citenamefont {Peters}, \citenamefont {Bluvstein}, \citenamefont {Wang}, \citenamefont {Greiner}, \citenamefont {Vuleti{\'c}},\ and\ \citenamefont {Lukin}}]{chiu2025continuous}%
  \BibitemOpen
  \bibfield  {author} {\bibinfo {author} {\bibfnamefont {N.-C.}\ \bibnamefont {Chiu}}, \bibinfo {author} {\bibfnamefont {E.~C.}\ \bibnamefont {Trapp}}, \bibinfo {author} {\bibfnamefont {J.}~\bibnamefont {Guo}}, \bibinfo {author} {\bibfnamefont {M.~H.}\ \bibnamefont {Abobeih}}, \bibinfo {author} {\bibfnamefont {L.~M.}\ \bibnamefont {Stewart}}, \bibinfo {author} {\bibfnamefont {S.}~\bibnamefont {Hollerith}}, \bibinfo {author} {\bibfnamefont {P.~L.}\ \bibnamefont {Stroganov}}, \bibinfo {author} {\bibfnamefont {M.}~\bibnamefont {Kalinowski}}, \bibinfo {author} {\bibfnamefont {A.~A.}\ \bibnamefont {Geim}}, \bibinfo {author} {\bibfnamefont {S.~J.}\ \bibnamefont {Evered}}, \bibinfo {author} {\bibfnamefont {S.~H.}\ \bibnamefont {Li}}, \bibinfo {author} {\bibfnamefont {X.}~\bibnamefont {Lyu}}, \bibinfo {author} {\bibfnamefont {L.~M.}\ \bibnamefont {Peters}}, \bibinfo {author} {\bibfnamefont {D.}~\bibnamefont {Bluvstein}}, \bibinfo {author} {\bibfnamefont {T.~T.}\ \bibnamefont {Wang}}, \bibinfo {author} {\bibfnamefont
  {M.}~\bibnamefont {Greiner}}, \bibinfo {author} {\bibfnamefont {V.}~\bibnamefont {Vuleti{\'c}}},\ and\ \bibinfo {author} {\bibfnamefont {M.~D.}\ \bibnamefont {Lukin}},\ }\bibfield  {title} {\bibinfo {title} {Continuous operation of a coherent 3,000-qubit system},\ }\href {https://doi.org/10.1038/s41586-025-09596-6} {\bibfield  {journal} {\bibinfo  {journal} {Nature}\ }\textbf {\bibinfo {volume} {646}},\ \bibinfo {pages} {1075} (\bibinfo {year} {2025})}\BibitemShut {NoStop}%
\bibitem [{\citenamefont {Zhou}\ \emph {et~al.}(2025)\citenamefont {Zhou}, \citenamefont {Duckering}, \citenamefont {Zhao}, \citenamefont {Bluvstein}, \citenamefont {Cain}, \citenamefont {Kubica}, \citenamefont {Wang},\ and\ \citenamefont {Lukin}}]{zhou2025resource}%
  \BibitemOpen
  \bibfield  {author} {\bibinfo {author} {\bibfnamefont {H.}~\bibnamefont {Zhou}}, \bibinfo {author} {\bibfnamefont {C.}~\bibnamefont {Duckering}}, \bibinfo {author} {\bibfnamefont {C.}~\bibnamefont {Zhao}}, \bibinfo {author} {\bibfnamefont {D.}~\bibnamefont {Bluvstein}}, \bibinfo {author} {\bibfnamefont {M.}~\bibnamefont {Cain}}, \bibinfo {author} {\bibfnamefont {A.}~\bibnamefont {Kubica}}, \bibinfo {author} {\bibfnamefont {S.-T.}\ \bibnamefont {Wang}},\ and\ \bibinfo {author} {\bibfnamefont {M.~D.}\ \bibnamefont {Lukin}},\ }\bibfield  {title} {\bibinfo {title} {Resource {{Analysis}} of {{Low-Overhead Transversal Architectures}} for {{Reconfigurable Atom Arrays}}},\ }in\ \href {https://doi.org/10.1145/3695053.3731039} {\emph {\bibinfo {booktitle} {Proceedings of the 52nd {{Annual International Symposium}} on {{Computer Architecture}}}}},\ \bibinfo {series and number} {{{ISCA}} '25}\ (\bibinfo  {publisher} {Association for Computing Machinery},\ \bibinfo {address} {New York, NY},\ \bibinfo {year} {2025})\
  pp.\ \bibinfo {pages} {1432--1448}\BibitemShut {NoStop}%
\bibitem [{\citenamefont {Fowler}\ \emph {et~al.}(2012)\citenamefont {Fowler}, \citenamefont {Mariantoni}, \citenamefont {Martinis},\ and\ \citenamefont {Cleland}}]{fowler2012surface}%
  \BibitemOpen
  \bibfield  {author} {\bibinfo {author} {\bibfnamefont {A.~G.}\ \bibnamefont {Fowler}}, \bibinfo {author} {\bibfnamefont {M.}~\bibnamefont {Mariantoni}}, \bibinfo {author} {\bibfnamefont {J.~M.}\ \bibnamefont {Martinis}},\ and\ \bibinfo {author} {\bibfnamefont {A.~N.}\ \bibnamefont {Cleland}},\ }\bibfield  {title} {\bibinfo {title} {Surface codes: {{Towards}} practical large-scale quantum computation},\ }\href {https://doi.org/10.1103/PhysRevA.86.032324} {\bibfield  {journal} {\bibinfo  {journal} {Phys. Rev. A}\ }\textbf {\bibinfo {volume} {86}},\ \bibinfo {pages} {032324} (\bibinfo {year} {2012})}\BibitemShut {NoStop}%
\bibitem [{\citenamefont {O'Gorman}\ and\ \citenamefont {Campbell}(2017)}]{ogorman2017quantum}%
  \BibitemOpen
  \bibfield  {author} {\bibinfo {author} {\bibfnamefont {J.}~\bibnamefont {O'Gorman}}\ and\ \bibinfo {author} {\bibfnamefont {E.~T.}\ \bibnamefont {Campbell}},\ }\bibfield  {title} {\bibinfo {title} {Quantum computation with realistic magic-state factories},\ }\href {https://doi.org/10.1103/PhysRevA.95.032338} {\bibfield  {journal} {\bibinfo  {journal} {Phys. Rev. A}\ }\textbf {\bibinfo {volume} {95}},\ \bibinfo {pages} {032338} (\bibinfo {year} {2017})}\BibitemShut {NoStop}%
\bibitem [{\citenamefont {Gidney}\ and\ \citenamefont {Eker{\aa}}(2021)}]{gidney2021how}%
  \BibitemOpen
  \bibfield  {author} {\bibinfo {author} {\bibfnamefont {C.}~\bibnamefont {Gidney}}\ and\ \bibinfo {author} {\bibfnamefont {M.}~\bibnamefont {Eker{\aa}}},\ }\bibfield  {title} {\bibinfo {title} {How to factor 2048 bit {{RSA}} integers in 8 hours using 20 million noisy qubits},\ }\href {https://doi.org/10.22331/q-2021-04-15-433} {\bibfield  {journal} {\bibinfo  {journal} {Quantum}\ }\textbf {\bibinfo {volume} {5}},\ \bibinfo {pages} {433} (\bibinfo {year} {2021})}\BibitemShut {NoStop}%
\bibitem [{\citenamefont {Gidney}(2025)}]{gidney2025how}%
  \BibitemOpen
  \bibfield  {author} {\bibinfo {author} {\bibfnamefont {C.}~\bibnamefont {Gidney}},\ }\bibfield  {title} {\bibinfo {title} {How to factor 2048 bit {{RSA}} integers with less than a million noisy qubits},\ }\href {https://doi.org/10.48550/arXiv.2505.15917} {\bibfield  {journal} {\bibinfo  {journal} {arXiv:2505.15917}\ } (\bibinfo {year} {2025})}\BibitemShut {NoStop}%
\bibitem [{\citenamefont {Preskill}(1998)}]{preskill1998reliable}%
  \BibitemOpen
  \bibfield  {author} {\bibinfo {author} {\bibfnamefont {J.}~\bibnamefont {Preskill}},\ }\bibfield  {title} {\bibinfo {title} {Reliable quantum computers},\ }\href {https://doi.org/10.1098/rspa.1998.0167} {\bibfield  {journal} {\bibinfo  {journal} {Proc. R. Soc. London, Ser. A}\ }\textbf {\bibinfo {volume} {454}},\ \bibinfo {pages} {385} (\bibinfo {year} {1998})}\BibitemShut {NoStop}%
\bibitem [{\citenamefont {Terhal}(2015)}]{terhal2015quantum}%
  \BibitemOpen
  \bibfield  {author} {\bibinfo {author} {\bibfnamefont {B.~M.}\ \bibnamefont {Terhal}},\ }\bibfield  {title} {\bibinfo {title} {Quantum error correction for quantum memories},\ }\href {https://doi.org/10.1103/RevModPhys.87.307} {\bibfield  {journal} {\bibinfo  {journal} {Rev. Mod. Phys.}\ }\textbf {\bibinfo {volume} {87}},\ \bibinfo {pages} {307} (\bibinfo {year} {2015})}\BibitemShut {NoStop}%
\bibitem [{\citenamefont {Lis}\ \emph {et~al.}(2023)\citenamefont {Lis}, \citenamefont {Senoo}, \citenamefont {McGrew}, \citenamefont {R{\"o}nchen}, \citenamefont {Jenkins},\ and\ \citenamefont {Kaufman}}]{lis2023midcircuit}%
  \BibitemOpen
  \bibfield  {author} {\bibinfo {author} {\bibfnamefont {J.~W.}\ \bibnamefont {Lis}}, \bibinfo {author} {\bibfnamefont {A.}~\bibnamefont {Senoo}}, \bibinfo {author} {\bibfnamefont {W.~F.}\ \bibnamefont {McGrew}}, \bibinfo {author} {\bibfnamefont {F.}~\bibnamefont {R{\"o}nchen}}, \bibinfo {author} {\bibfnamefont {A.}~\bibnamefont {Jenkins}},\ and\ \bibinfo {author} {\bibfnamefont {A.~M.}\ \bibnamefont {Kaufman}},\ }\bibfield  {title} {\bibinfo {title} {Midcircuit {{Operations Using}} the omg {{Architecture}} in {{Neutral Atom Arrays}}},\ }\href {https://doi.org/10.1103/PhysRevX.13.041035} {\bibfield  {journal} {\bibinfo  {journal} {Phys. Rev. X}\ }\textbf {\bibinfo {volume} {13}},\ \bibinfo {pages} {041035} (\bibinfo {year} {2023})}\BibitemShut {NoStop}%
\bibitem [{\citenamefont {Norcia}\ \emph {et~al.}(2023)\citenamefont {Norcia}, \citenamefont {Cairncross}, \citenamefont {Barnes}, \citenamefont {Battaglino}, \citenamefont {Brown}, \citenamefont {Brown}, \citenamefont {Cassella}, \citenamefont {Chen}, \citenamefont {Coxe}, \citenamefont {Crow}, \citenamefont {Epstein}, \citenamefont {Griger}, \citenamefont {Jones}, \citenamefont {Kim}, \citenamefont {Kindem}, \citenamefont {King}, \citenamefont {Kondov}, \citenamefont {Kotru}, \citenamefont {Lauigan}, \citenamefont {Li}, \citenamefont {Lu}, \citenamefont {Megidish}, \citenamefont {Marjanovic}, \citenamefont {McDonald}, \citenamefont {Mittiga}, \citenamefont {Muniz}, \citenamefont {Narayanaswami}, \citenamefont {Nishiguchi}, \citenamefont {Notermans}, \citenamefont {Paule}, \citenamefont {Pawlak}, \citenamefont {Peng}, \citenamefont {Ryou}, \citenamefont {Smull}, \citenamefont {Stack}, \citenamefont {Stone}, \citenamefont {Sucich}, \citenamefont {Urbanek}, \citenamefont {Van De~Veerdonk}, \citenamefont
  {Vendeiro}, \citenamefont {Wilkason}, \citenamefont {Wu}, \citenamefont {Xie}, \citenamefont {Zhang},\ and\ \citenamefont {Bloom}}]{norcia2023midcircuit}%
  \BibitemOpen
  \bibfield  {author} {\bibinfo {author} {\bibfnamefont {M.~A.}\ \bibnamefont {Norcia}}, \bibinfo {author} {\bibfnamefont {W.~B.}\ \bibnamefont {Cairncross}}, \bibinfo {author} {\bibfnamefont {K.}~\bibnamefont {Barnes}}, \bibinfo {author} {\bibfnamefont {P.}~\bibnamefont {Battaglino}}, \bibinfo {author} {\bibfnamefont {A.}~\bibnamefont {Brown}}, \bibinfo {author} {\bibfnamefont {M.~O.}\ \bibnamefont {Brown}}, \bibinfo {author} {\bibfnamefont {K.}~\bibnamefont {Cassella}}, \bibinfo {author} {\bibfnamefont {C.-A.}\ \bibnamefont {Chen}}, \bibinfo {author} {\bibfnamefont {R.}~\bibnamefont {Coxe}}, \bibinfo {author} {\bibfnamefont {D.}~\bibnamefont {Crow}}, \bibinfo {author} {\bibfnamefont {J.}~\bibnamefont {Epstein}}, \bibinfo {author} {\bibfnamefont {C.}~\bibnamefont {Griger}}, \bibinfo {author} {\bibfnamefont {A.~M.~W.}\ \bibnamefont {Jones}}, \bibinfo {author} {\bibfnamefont {H.}~\bibnamefont {Kim}}, \bibinfo {author} {\bibfnamefont {J.~M.}\ \bibnamefont {Kindem}}, \bibinfo {author} {\bibfnamefont
  {J.}~\bibnamefont {King}}, \bibinfo {author} {\bibfnamefont {S.~S.}\ \bibnamefont {Kondov}}, \bibinfo {author} {\bibfnamefont {K.}~\bibnamefont {Kotru}}, \bibinfo {author} {\bibfnamefont {J.}~\bibnamefont {Lauigan}}, \bibinfo {author} {\bibfnamefont {M.}~\bibnamefont {Li}}, \bibinfo {author} {\bibfnamefont {M.}~\bibnamefont {Lu}}, \bibinfo {author} {\bibfnamefont {E.}~\bibnamefont {Megidish}}, \bibinfo {author} {\bibfnamefont {J.}~\bibnamefont {Marjanovic}}, \bibinfo {author} {\bibfnamefont {M.}~\bibnamefont {McDonald}}, \bibinfo {author} {\bibfnamefont {T.}~\bibnamefont {Mittiga}}, \bibinfo {author} {\bibfnamefont {J.~A.}\ \bibnamefont {Muniz}}, \bibinfo {author} {\bibfnamefont {S.}~\bibnamefont {Narayanaswami}}, \bibinfo {author} {\bibfnamefont {C.}~\bibnamefont {Nishiguchi}}, \bibinfo {author} {\bibfnamefont {R.}~\bibnamefont {Notermans}}, \bibinfo {author} {\bibfnamefont {T.}~\bibnamefont {Paule}}, \bibinfo {author} {\bibfnamefont {K.~A.}\ \bibnamefont {Pawlak}}, \bibinfo {author} {\bibfnamefont
  {L.~S.}\ \bibnamefont {Peng}}, \bibinfo {author} {\bibfnamefont {A.}~\bibnamefont {Ryou}}, \bibinfo {author} {\bibfnamefont {A.}~\bibnamefont {Smull}}, \bibinfo {author} {\bibfnamefont {D.}~\bibnamefont {Stack}}, \bibinfo {author} {\bibfnamefont {M.}~\bibnamefont {Stone}}, \bibinfo {author} {\bibfnamefont {A.}~\bibnamefont {Sucich}}, \bibinfo {author} {\bibfnamefont {M.}~\bibnamefont {Urbanek}}, \bibinfo {author} {\bibfnamefont {R.~J.~M.}\ \bibnamefont {Van De~Veerdonk}}, \bibinfo {author} {\bibfnamefont {Z.}~\bibnamefont {Vendeiro}}, \bibinfo {author} {\bibfnamefont {T.}~\bibnamefont {Wilkason}}, \bibinfo {author} {\bibfnamefont {T.-Y.}\ \bibnamefont {Wu}}, \bibinfo {author} {\bibfnamefont {X.}~\bibnamefont {Xie}}, \bibinfo {author} {\bibfnamefont {X.}~\bibnamefont {Zhang}},\ and\ \bibinfo {author} {\bibfnamefont {B.~J.}\ \bibnamefont {Bloom}},\ }\bibfield  {title} {\bibinfo {title} {Midcircuit {{Qubit Measurement}} and {{Rearrangement}} in a 171 {{Yb Atomic Array}}},\ }\href
  {https://doi.org/10.1103/PhysRevX.13.041034} {\bibfield  {journal} {\bibinfo  {journal} {Phys. Rev. X}\ }\textbf {\bibinfo {volume} {13}},\ \bibinfo {pages} {041034} (\bibinfo {year} {2023})}\BibitemShut {NoStop}%
\bibitem [{\citenamefont {Cirac}\ \emph {et~al.}(1999)\citenamefont {Cirac}, \citenamefont {Ekert}, \citenamefont {Huelga},\ and\ \citenamefont {Macchiavello}}]{cirac1999distributed}%
  \BibitemOpen
  \bibfield  {author} {\bibinfo {author} {\bibfnamefont {J.~I.}\ \bibnamefont {Cirac}}, \bibinfo {author} {\bibfnamefont {A.~K.}\ \bibnamefont {Ekert}}, \bibinfo {author} {\bibfnamefont {S.~F.}\ \bibnamefont {Huelga}},\ and\ \bibinfo {author} {\bibfnamefont {C.}~\bibnamefont {Macchiavello}},\ }\bibfield  {title} {\bibinfo {title} {Distributed quantum computation over noisy channels},\ }\href {https://doi.org/10.1103/PhysRevA.59.4249} {\bibfield  {journal} {\bibinfo  {journal} {Phys. Rev. A}\ }\textbf {\bibinfo {volume} {59}},\ \bibinfo {pages} {4249} (\bibinfo {year} {1999})}\BibitemShut {NoStop}%
\bibitem [{\citenamefont {Jiang}\ \emph {et~al.}(2007)\citenamefont {Jiang}, \citenamefont {Taylor}, \citenamefont {S{\o}rensen},\ and\ \citenamefont {Lukin}}]{jiang2007distributed}%
  \BibitemOpen
  \bibfield  {author} {\bibinfo {author} {\bibfnamefont {L.}~\bibnamefont {Jiang}}, \bibinfo {author} {\bibfnamefont {J.~M.}\ \bibnamefont {Taylor}}, \bibinfo {author} {\bibfnamefont {A.~S.}\ \bibnamefont {S{\o}rensen}},\ and\ \bibinfo {author} {\bibfnamefont {M.~D.}\ \bibnamefont {Lukin}},\ }\bibfield  {title} {\bibinfo {title} {Distributed quantum computation based on small quantum registers},\ }\href {https://doi.org/10.1103/PhysRevA.76.062323} {\bibfield  {journal} {\bibinfo  {journal} {Phys. Rev. A}\ }\textbf {\bibinfo {volume} {76}},\ \bibinfo {pages} {062323} (\bibinfo {year} {2007})}\BibitemShut {NoStop}%
\bibitem [{\citenamefont {Nickerson}\ \emph {et~al.}(2013)\citenamefont {Nickerson}, \citenamefont {Li},\ and\ \citenamefont {Benjamin}}]{nickerson2013topological}%
  \BibitemOpen
  \bibfield  {author} {\bibinfo {author} {\bibfnamefont {N.~H.}\ \bibnamefont {Nickerson}}, \bibinfo {author} {\bibfnamefont {Y.}~\bibnamefont {Li}},\ and\ \bibinfo {author} {\bibfnamefont {S.~C.}\ \bibnamefont {Benjamin}},\ }\bibfield  {title} {\bibinfo {title} {Topological quantum computing with a very noisy network and local error rates approaching one percent},\ }\href {https://doi.org/10.1038/ncomms2773} {\bibfield  {journal} {\bibinfo  {journal} {Nat. Commun.}\ }\textbf {\bibinfo {volume} {4}},\ \bibinfo {pages} {1756} (\bibinfo {year} {2013})}\BibitemShut {NoStop}%
\bibitem [{\citenamefont {Nickerson}\ \emph {et~al.}(2014)\citenamefont {Nickerson}, \citenamefont {Fitzsimons},\ and\ \citenamefont {Benjamin}}]{nickerson2014freely}%
  \BibitemOpen
  \bibfield  {author} {\bibinfo {author} {\bibfnamefont {N.~H.}\ \bibnamefont {Nickerson}}, \bibinfo {author} {\bibfnamefont {J.~F.}\ \bibnamefont {Fitzsimons}},\ and\ \bibinfo {author} {\bibfnamefont {S.~C.}\ \bibnamefont {Benjamin}},\ }\bibfield  {title} {\bibinfo {title} {Freely {{Scalable Quantum Technologies Using Cells}} of 5-to-50 {{Qubits}} with {{Very Lossy}} and {{Noisy Photonic Links}}},\ }\href {https://doi.org/10.1103/PhysRevX.4.041041} {\bibfield  {journal} {\bibinfo  {journal} {Phys. Rev. X}\ }\textbf {\bibinfo {volume} {4}},\ \bibinfo {pages} {041041} (\bibinfo {year} {2014})}\BibitemShut {NoStop}%
\bibitem [{\citenamefont {Monroe}\ \emph {et~al.}(2014)\citenamefont {Monroe}, \citenamefont {Raussendorf}, \citenamefont {Ruthven}, \citenamefont {Brown}, \citenamefont {Maunz}, \citenamefont {Duan},\ and\ \citenamefont {Kim}}]{monroe2014largescale}%
  \BibitemOpen
  \bibfield  {author} {\bibinfo {author} {\bibfnamefont {C.}~\bibnamefont {Monroe}}, \bibinfo {author} {\bibfnamefont {R.}~\bibnamefont {Raussendorf}}, \bibinfo {author} {\bibfnamefont {A.}~\bibnamefont {Ruthven}}, \bibinfo {author} {\bibfnamefont {K.~R.}\ \bibnamefont {Brown}}, \bibinfo {author} {\bibfnamefont {P.}~\bibnamefont {Maunz}}, \bibinfo {author} {\bibfnamefont {L.-M.}\ \bibnamefont {Duan}},\ and\ \bibinfo {author} {\bibfnamefont {J.}~\bibnamefont {Kim}},\ }\bibfield  {title} {\bibinfo {title} {Large-scale modular quantum-computer architecture with atomic memory and photonic interconnects},\ }\href {https://doi.org/10.1103/PhysRevA.89.022317} {\bibfield  {journal} {\bibinfo  {journal} {Phys. Rev. A}\ }\textbf {\bibinfo {volume} {89}},\ \bibinfo {pages} {022317} (\bibinfo {year} {2014})}\BibitemShut {NoStop}%
\bibitem [{\citenamefont {Covey}\ \emph {et~al.}(2023)\citenamefont {Covey}, \citenamefont {Weinfurter},\ and\ \citenamefont {Bernien}}]{covey2023quantum}%
  \BibitemOpen
  \bibfield  {author} {\bibinfo {author} {\bibfnamefont {J.~P.}\ \bibnamefont {Covey}}, \bibinfo {author} {\bibfnamefont {H.}~\bibnamefont {Weinfurter}},\ and\ \bibinfo {author} {\bibfnamefont {H.}~\bibnamefont {Bernien}},\ }\bibfield  {title} {\bibinfo {title} {Quantum networks with neutral atom processing nodes},\ }\href {https://doi.org/10.1038/s41534-023-00759-9} {\bibfield  {journal} {\bibinfo  {journal} {npj Quantum Inf}\ }\textbf {\bibinfo {volume} {9}},\ \bibinfo {pages} {90} (\bibinfo {year} {2023})}\BibitemShut {NoStop}%
\bibitem [{\citenamefont {Ramette}\ \emph {et~al.}(2024)\citenamefont {Ramette}, \citenamefont {Sinclair}, \citenamefont {Breuckmann},\ and\ \citenamefont {Vuleti{\'c}}}]{ramette2024faulttolerant}%
  \BibitemOpen
  \bibfield  {author} {\bibinfo {author} {\bibfnamefont {J.}~\bibnamefont {Ramette}}, \bibinfo {author} {\bibfnamefont {J.}~\bibnamefont {Sinclair}}, \bibinfo {author} {\bibfnamefont {N.~P.}\ \bibnamefont {Breuckmann}},\ and\ \bibinfo {author} {\bibfnamefont {V.}~\bibnamefont {Vuleti{\'c}}},\ }\bibfield  {title} {\bibinfo {title} {Fault-tolerant connection of error-corrected qubits with noisy links},\ }\href {https://doi.org/10.1038/s41534-024-00855-4} {\bibfield  {journal} {\bibinfo  {journal} {npj Quantum Inf}\ }\textbf {\bibinfo {volume} {10}},\ \bibinfo {pages} {58} (\bibinfo {year} {2024})}\BibitemShut {NoStop}%
\bibitem [{\citenamefont {Main}\ \emph {et~al.}(2025)\citenamefont {Main}, \citenamefont {Drmota}, \citenamefont {Nadlinger}, \citenamefont {Ainley}, \citenamefont {Agrawal}, \citenamefont {Nichol}, \citenamefont {Srinivas}, \citenamefont {Araneda},\ and\ \citenamefont {Lucas}}]{main2025distributed}%
  \BibitemOpen
  \bibfield  {author} {\bibinfo {author} {\bibfnamefont {D.}~\bibnamefont {Main}}, \bibinfo {author} {\bibfnamefont {P.}~\bibnamefont {Drmota}}, \bibinfo {author} {\bibfnamefont {D.~P.}\ \bibnamefont {Nadlinger}}, \bibinfo {author} {\bibfnamefont {E.~M.}\ \bibnamefont {Ainley}}, \bibinfo {author} {\bibfnamefont {A.}~\bibnamefont {Agrawal}}, \bibinfo {author} {\bibfnamefont {B.~C.}\ \bibnamefont {Nichol}}, \bibinfo {author} {\bibfnamefont {R.}~\bibnamefont {Srinivas}}, \bibinfo {author} {\bibfnamefont {G.}~\bibnamefont {Araneda}},\ and\ \bibinfo {author} {\bibfnamefont {D.~M.}\ \bibnamefont {Lucas}},\ }\bibfield  {title} {\bibinfo {title} {Distributed quantum computing across an optical network link},\ }\href {https://doi.org/10.1038/s41586-024-08404-x} {\bibfield  {journal} {\bibinfo  {journal} {Nature}\ }\textbf {\bibinfo {volume} {638}},\ \bibinfo {pages} {383} (\bibinfo {year} {2025})}\BibitemShut {NoStop}%
\bibitem [{\citenamefont {Young}\ \emph {et~al.}(2022)\citenamefont {Young}, \citenamefont {Safari}, \citenamefont {Huft}, \citenamefont {Zhang}, \citenamefont {Oh}, \citenamefont {Chinnarasu},\ and\ \citenamefont {Saffman}}]{young2022architecture}%
  \BibitemOpen
  \bibfield  {author} {\bibinfo {author} {\bibfnamefont {C.~B.}\ \bibnamefont {Young}}, \bibinfo {author} {\bibfnamefont {A.}~\bibnamefont {Safari}}, \bibinfo {author} {\bibfnamefont {P.}~\bibnamefont {Huft}}, \bibinfo {author} {\bibfnamefont {J.}~\bibnamefont {Zhang}}, \bibinfo {author} {\bibfnamefont {E.}~\bibnamefont {Oh}}, \bibinfo {author} {\bibfnamefont {R.}~\bibnamefont {Chinnarasu}},\ and\ \bibinfo {author} {\bibfnamefont {M.}~\bibnamefont {Saffman}},\ }\bibfield  {title} {\bibinfo {title} {An architecture for quantum networking of neutral atom processors},\ }\href {https://doi.org/10.1007/s00340-022-07865-0} {\bibfield  {journal} {\bibinfo  {journal} {Appl. Phys. B}\ }\textbf {\bibinfo {volume} {128}},\ \bibinfo {pages} {151} (\bibinfo {year} {2022})}\BibitemShut {NoStop}%
\bibitem [{\citenamefont {Sinclair}\ \emph {et~al.}(2025)\citenamefont {Sinclair}, \citenamefont {Ramette}, \citenamefont {Grinkemeyer}, \citenamefont {Bluvstein}, \citenamefont {Lukin},\ and\ \citenamefont {Vuleti{\'c}}}]{sinclair2025faulttolerant}%
  \BibitemOpen
  \bibfield  {author} {\bibinfo {author} {\bibfnamefont {J.}~\bibnamefont {Sinclair}}, \bibinfo {author} {\bibfnamefont {J.}~\bibnamefont {Ramette}}, \bibinfo {author} {\bibfnamefont {B.}~\bibnamefont {Grinkemeyer}}, \bibinfo {author} {\bibfnamefont {D.}~\bibnamefont {Bluvstein}}, \bibinfo {author} {\bibfnamefont {M.~D.}\ \bibnamefont {Lukin}},\ and\ \bibinfo {author} {\bibfnamefont {V.}~\bibnamefont {Vuleti{\'c}}},\ }\bibfield  {title} {\bibinfo {title} {Fault-tolerant optical interconnects for neutral-atom arrays},\ }\href {https://doi.org/10.1103/PhysRevResearch.7.013313} {\bibfield  {journal} {\bibinfo  {journal} {Phys. Rev. Research}\ }\textbf {\bibinfo {volume} {7}},\ \bibinfo {pages} {013313} (\bibinfo {year} {2025})}\BibitemShut {NoStop}%
\bibitem [{\citenamefont {Purcell}(1946)}]{purcell1946proceedings}%
  \BibitemOpen
  \bibfield  {author} {\bibinfo {author} {\bibfnamefont {E.~M.}\ \bibnamefont {Purcell}},\ }\bibfield  {title} {\bibinfo {title} {Proceedings of the {{American Physical Society}}: Spontaneous emission probabilities at radio frequencies},\ }\href {https://doi.org/10.1103/PhysRev.69.674} {\bibfield  {journal} {\bibinfo  {journal} {Phys. Rev.}\ }\textbf {\bibinfo {volume} {69}},\ \bibinfo {pages} {674} (\bibinfo {year} {1946})}\BibitemShut {NoStop}%
\bibitem [{\citenamefont {Bochmann}\ \emph {et~al.}(2010)\citenamefont {Bochmann}, \citenamefont {M{\"u}cke}, \citenamefont {Guhl}, \citenamefont {Ritter}, \citenamefont {Rempe},\ and\ \citenamefont {Moehring}}]{bochmann2010lossless}%
  \BibitemOpen
  \bibfield  {author} {\bibinfo {author} {\bibfnamefont {J.}~\bibnamefont {Bochmann}}, \bibinfo {author} {\bibfnamefont {M.}~\bibnamefont {M{\"u}cke}}, \bibinfo {author} {\bibfnamefont {C.}~\bibnamefont {Guhl}}, \bibinfo {author} {\bibfnamefont {S.}~\bibnamefont {Ritter}}, \bibinfo {author} {\bibfnamefont {G.}~\bibnamefont {Rempe}},\ and\ \bibinfo {author} {\bibfnamefont {D.~L.}\ \bibnamefont {Moehring}},\ }\bibfield  {title} {\bibinfo {title} {Lossless {{State Detection}} of {{Single Neutral Atoms}}},\ }\href {https://doi.org/10.1103/PhysRevLett.104.203601} {\bibfield  {journal} {\bibinfo  {journal} {Phys. Rev. Lett.}\ }\textbf {\bibinfo {volume} {104}},\ \bibinfo {pages} {203601} (\bibinfo {year} {2010})}\BibitemShut {NoStop}%
\bibitem [{\citenamefont {Gehr}\ \emph {et~al.}(2010)\citenamefont {Gehr}, \citenamefont {Volz}, \citenamefont {Dubois}, \citenamefont {Steinmetz}, \citenamefont {Colombe}, \citenamefont {Lev}, \citenamefont {Long}, \citenamefont {Est{\`e}ve},\ and\ \citenamefont {Reichel}}]{gehr2010cavitybased}%
  \BibitemOpen
  \bibfield  {author} {\bibinfo {author} {\bibfnamefont {R.}~\bibnamefont {Gehr}}, \bibinfo {author} {\bibfnamefont {J.}~\bibnamefont {Volz}}, \bibinfo {author} {\bibfnamefont {G.}~\bibnamefont {Dubois}}, \bibinfo {author} {\bibfnamefont {T.}~\bibnamefont {Steinmetz}}, \bibinfo {author} {\bibfnamefont {Y.}~\bibnamefont {Colombe}}, \bibinfo {author} {\bibfnamefont {B.~L.}\ \bibnamefont {Lev}}, \bibinfo {author} {\bibfnamefont {R.}~\bibnamefont {Long}}, \bibinfo {author} {\bibfnamefont {J.}~\bibnamefont {Est{\`e}ve}},\ and\ \bibinfo {author} {\bibfnamefont {J.}~\bibnamefont {Reichel}},\ }\bibfield  {title} {\bibinfo {title} {Cavity-{{Based Single Atom Preparation}} and {{High-Fidelity Hyperfine State Readout}}},\ }\href {https://doi.org/10.1103/PhysRevLett.104.203602} {\bibfield  {journal} {\bibinfo  {journal} {Phys. Rev. Lett.}\ }\textbf {\bibinfo {volume} {104}},\ \bibinfo {pages} {203602} (\bibinfo {year} {2010})}\BibitemShut {NoStop}%
\bibitem [{\citenamefont {Deist}\ \emph {et~al.}(2022)\citenamefont {Deist}, \citenamefont {Lu}, \citenamefont {Ho}, \citenamefont {Pasha}, \citenamefont {Zeiher}, \citenamefont {Yan},\ and\ \citenamefont {{Stamper-Kurn}}}]{deist2022midcircuit}%
  \BibitemOpen
  \bibfield  {author} {\bibinfo {author} {\bibfnamefont {E.}~\bibnamefont {Deist}}, \bibinfo {author} {\bibfnamefont {Y.-H.}\ \bibnamefont {Lu}}, \bibinfo {author} {\bibfnamefont {J.}~\bibnamefont {Ho}}, \bibinfo {author} {\bibfnamefont {M.~K.}\ \bibnamefont {Pasha}}, \bibinfo {author} {\bibfnamefont {J.}~\bibnamefont {Zeiher}}, \bibinfo {author} {\bibfnamefont {Z.}~\bibnamefont {Yan}},\ and\ \bibinfo {author} {\bibfnamefont {D.~M.}\ \bibnamefont {{Stamper-Kurn}}},\ }\bibfield  {title} {\bibinfo {title} {Mid-{{Circuit Cavity Measurement}} in a {{Neutral Atom Array}}},\ }\href {https://doi.org/10.1103/PhysRevLett.129.203602} {\bibfield  {journal} {\bibinfo  {journal} {Phys. Rev. Lett.}\ }\textbf {\bibinfo {volume} {129}},\ \bibinfo {pages} {203602} (\bibinfo {year} {2022})}\BibitemShut {NoStop}%
\bibitem [{\citenamefont {Hu}\ \emph {et~al.}(2025)\citenamefont {Hu}, \citenamefont {Sinclair}, \citenamefont {Bytyqi}, \citenamefont {Chong}, \citenamefont {Rudelis}, \citenamefont {Ramette}, \citenamefont {Vendeiro},\ and\ \citenamefont {Vuleti{\'c}}}]{hu2025siteselective}%
  \BibitemOpen
  \bibfield  {author} {\bibinfo {author} {\bibfnamefont {B.}~\bibnamefont {Hu}}, \bibinfo {author} {\bibfnamefont {J.}~\bibnamefont {Sinclair}}, \bibinfo {author} {\bibfnamefont {E.}~\bibnamefont {Bytyqi}}, \bibinfo {author} {\bibfnamefont {M.}~\bibnamefont {Chong}}, \bibinfo {author} {\bibfnamefont {A.}~\bibnamefont {Rudelis}}, \bibinfo {author} {\bibfnamefont {J.}~\bibnamefont {Ramette}}, \bibinfo {author} {\bibfnamefont {Z.}~\bibnamefont {Vendeiro}},\ and\ \bibinfo {author} {\bibfnamefont {V.}~\bibnamefont {Vuleti{\'c}}},\ }\bibfield  {title} {\bibinfo {title} {Site-{{Selective Cavity Readout}} and {{Classical Error Correction}} of a 5-{{Bit Atomic Register}}},\ }\href {https://doi.org/10.1103/PhysRevLett.134.120801} {\bibfield  {journal} {\bibinfo  {journal} {Phys. Rev. Lett.}\ }\textbf {\bibinfo {volume} {134}},\ \bibinfo {pages} {120801} (\bibinfo {year} {2025})}\BibitemShut {NoStop}%
\bibitem [{\citenamefont {Grinkemeyer}\ \emph {et~al.}(2025)\citenamefont {Grinkemeyer}, \citenamefont {{Guardado-Sanchez}}, \citenamefont {Dimitrova}, \citenamefont {Shchepanovich}, \citenamefont {Mandopoulou}, \citenamefont {Borregaard}, \citenamefont {Vuleti{\'c}},\ and\ \citenamefont {Lukin}}]{grinkemeyer2025errordetected}%
  \BibitemOpen
  \bibfield  {author} {\bibinfo {author} {\bibfnamefont {B.}~\bibnamefont {Grinkemeyer}}, \bibinfo {author} {\bibfnamefont {E.}~\bibnamefont {{Guardado-Sanchez}}}, \bibinfo {author} {\bibfnamefont {I.}~\bibnamefont {Dimitrova}}, \bibinfo {author} {\bibfnamefont {D.}~\bibnamefont {Shchepanovich}}, \bibinfo {author} {\bibfnamefont {G.~E.}\ \bibnamefont {Mandopoulou}}, \bibinfo {author} {\bibfnamefont {J.}~\bibnamefont {Borregaard}}, \bibinfo {author} {\bibfnamefont {V.}~\bibnamefont {Vuleti{\'c}}},\ and\ \bibinfo {author} {\bibfnamefont {M.~D.}\ \bibnamefont {Lukin}},\ }\bibfield  {title} {\bibinfo {title} {Error-detected quantum operations with neutral atoms mediated by an optical cavity},\ }\href {https://doi.org/10.1126/science.adr7075} {\bibfield  {journal} {\bibinfo  {journal} {Science}\ }\textbf {\bibinfo {volume} {387}},\ \bibinfo {pages} {1301} (\bibinfo {year} {2025})}\BibitemShut {NoStop}%
\bibitem [{\citenamefont {Wilk}\ \emph {et~al.}(2007)\citenamefont {Wilk}, \citenamefont {Webster}, \citenamefont {Kuhn},\ and\ \citenamefont {Rempe}}]{wilk2007singleatom}%
  \BibitemOpen
  \bibfield  {author} {\bibinfo {author} {\bibfnamefont {T.}~\bibnamefont {Wilk}}, \bibinfo {author} {\bibfnamefont {S.~C.}\ \bibnamefont {Webster}}, \bibinfo {author} {\bibfnamefont {A.}~\bibnamefont {Kuhn}},\ and\ \bibinfo {author} {\bibfnamefont {G.}~\bibnamefont {Rempe}},\ }\bibfield  {title} {\bibinfo {title} {Single-{{Atom Single-Photon Quantum Interface}}},\ }\href {https://doi.org/10.1126/science.1143835} {\bibfield  {journal} {\bibinfo  {journal} {Science}\ }\textbf {\bibinfo {volume} {317}},\ \bibinfo {pages} {488} (\bibinfo {year} {2007})}\BibitemShut {NoStop}%
\bibitem [{\citenamefont {Weber}\ \emph {et~al.}(2009)\citenamefont {Weber}, \citenamefont {Specht}, \citenamefont {M{\"u}ller}, \citenamefont {Bochmann}, \citenamefont {M{\"u}cke}, \citenamefont {Moehring},\ and\ \citenamefont {Rempe}}]{weber2009photonphoton}%
  \BibitemOpen
  \bibfield  {author} {\bibinfo {author} {\bibfnamefont {B.}~\bibnamefont {Weber}}, \bibinfo {author} {\bibfnamefont {H.~P.}\ \bibnamefont {Specht}}, \bibinfo {author} {\bibfnamefont {T.}~\bibnamefont {M{\"u}ller}}, \bibinfo {author} {\bibfnamefont {J.}~\bibnamefont {Bochmann}}, \bibinfo {author} {\bibfnamefont {M.}~\bibnamefont {M{\"u}cke}}, \bibinfo {author} {\bibfnamefont {D.~L.}\ \bibnamefont {Moehring}},\ and\ \bibinfo {author} {\bibfnamefont {G.}~\bibnamefont {Rempe}},\ }\bibfield  {title} {\bibinfo {title} {Photon-{{Photon Entanglement}} with a {{Single Trapped Atom}}},\ }\href {https://doi.org/10.1103/PhysRevLett.102.030501} {\bibfield  {journal} {\bibinfo  {journal} {Phys. Rev. Lett.}\ }\textbf {\bibinfo {volume} {102}},\ \bibinfo {pages} {030501} (\bibinfo {year} {2009})}\BibitemShut {NoStop}%
\bibitem [{\citenamefont {Lettner}\ \emph {et~al.}(2011)\citenamefont {Lettner}, \citenamefont {M{\"u}cke}, \citenamefont {Riedl}, \citenamefont {Vo}, \citenamefont {Hahn}, \citenamefont {Baur}, \citenamefont {Bochmann}, \citenamefont {Ritter}, \citenamefont {D{\"u}rr},\ and\ \citenamefont {Rempe}}]{lettner2011remote}%
  \BibitemOpen
  \bibfield  {author} {\bibinfo {author} {\bibfnamefont {M.}~\bibnamefont {Lettner}}, \bibinfo {author} {\bibfnamefont {M.}~\bibnamefont {M{\"u}cke}}, \bibinfo {author} {\bibfnamefont {S.}~\bibnamefont {Riedl}}, \bibinfo {author} {\bibfnamefont {C.}~\bibnamefont {Vo}}, \bibinfo {author} {\bibfnamefont {C.}~\bibnamefont {Hahn}}, \bibinfo {author} {\bibfnamefont {S.}~\bibnamefont {Baur}}, \bibinfo {author} {\bibfnamefont {J.}~\bibnamefont {Bochmann}}, \bibinfo {author} {\bibfnamefont {S.}~\bibnamefont {Ritter}}, \bibinfo {author} {\bibfnamefont {S.}~\bibnamefont {D{\"u}rr}},\ and\ \bibinfo {author} {\bibfnamefont {G.}~\bibnamefont {Rempe}},\ }\bibfield  {title} {\bibinfo {title} {Remote {{Entanglement}} between a {{Single Atom}} and a {{Bose-Einstein Condensate}}},\ }\href {https://doi.org/10.1103/PhysRevLett.106.210503} {\bibfield  {journal} {\bibinfo  {journal} {Phys. Rev. Lett.}\ }\textbf {\bibinfo {volume} {106}},\ \bibinfo {pages} {210503} (\bibinfo {year} {2011})}\BibitemShut {NoStop}%
\bibitem [{\citenamefont {Reiserer}\ \emph {et~al.}(2014)\citenamefont {Reiserer}, \citenamefont {Kalb}, \citenamefont {Rempe},\ and\ \citenamefont {Ritter}}]{reiserer2014quantum}%
  \BibitemOpen
  \bibfield  {author} {\bibinfo {author} {\bibfnamefont {A.}~\bibnamefont {Reiserer}}, \bibinfo {author} {\bibfnamefont {N.}~\bibnamefont {Kalb}}, \bibinfo {author} {\bibfnamefont {G.}~\bibnamefont {Rempe}},\ and\ \bibinfo {author} {\bibfnamefont {S.}~\bibnamefont {Ritter}},\ }\bibfield  {title} {\bibinfo {title} {A quantum gate between a flying optical photon and a single trapped atom},\ }\href {https://doi.org/10.1038/nature13177} {\bibfield  {journal} {\bibinfo  {journal} {Nature}\ }\textbf {\bibinfo {volume} {508}},\ \bibinfo {pages} {237} (\bibinfo {year} {2014})}\BibitemShut {NoStop}%
\bibitem [{\citenamefont {Reiserer}\ and\ \citenamefont {Rempe}(2015)}]{reiserer2015cavitybased}%
  \BibitemOpen
  \bibfield  {author} {\bibinfo {author} {\bibfnamefont {A.}~\bibnamefont {Reiserer}}\ and\ \bibinfo {author} {\bibfnamefont {G.}~\bibnamefont {Rempe}},\ }\bibfield  {title} {\bibinfo {title} {Cavity-based quantum networks with single atoms and optical photons},\ }\href {https://doi.org/10.1103/RevModPhys.87.1379} {\bibfield  {journal} {\bibinfo  {journal} {Rev. Mod. Phys.}\ }\textbf {\bibinfo {volume} {87}},\ \bibinfo {pages} {1379} (\bibinfo {year} {2015})}\BibitemShut {NoStop}%
\bibitem [{\citenamefont {Ritter}\ \emph {et~al.}(2012)\citenamefont {Ritter}, \citenamefont {N{\"o}lleke}, \citenamefont {Hahn}, \citenamefont {Reiserer}, \citenamefont {Neuzner}, \citenamefont {Uphoff}, \citenamefont {M{\"u}cke}, \citenamefont {Figueroa}, \citenamefont {Bochmann},\ and\ \citenamefont {Rempe}}]{ritter2012elementary}%
  \BibitemOpen
  \bibfield  {author} {\bibinfo {author} {\bibfnamefont {S.}~\bibnamefont {Ritter}}, \bibinfo {author} {\bibfnamefont {C.}~\bibnamefont {N{\"o}lleke}}, \bibinfo {author} {\bibfnamefont {C.}~\bibnamefont {Hahn}}, \bibinfo {author} {\bibfnamefont {A.}~\bibnamefont {Reiserer}}, \bibinfo {author} {\bibfnamefont {A.}~\bibnamefont {Neuzner}}, \bibinfo {author} {\bibfnamefont {M.}~\bibnamefont {Uphoff}}, \bibinfo {author} {\bibfnamefont {M.}~\bibnamefont {M{\"u}cke}}, \bibinfo {author} {\bibfnamefont {E.}~\bibnamefont {Figueroa}}, \bibinfo {author} {\bibfnamefont {J.}~\bibnamefont {Bochmann}},\ and\ \bibinfo {author} {\bibfnamefont {G.}~\bibnamefont {Rempe}},\ }\bibfield  {title} {\bibinfo {title} {An elementary quantum network of single atoms in optical cavities},\ }\href {https://doi.org/10.1038/nature11023} {\bibfield  {journal} {\bibinfo  {journal} {Nature}\ }\textbf {\bibinfo {volume} {484}},\ \bibinfo {pages} {195} (\bibinfo {year} {2012})}\BibitemShut {NoStop}%
\bibitem [{\citenamefont {N{\"o}lleke}\ \emph {et~al.}(2013)\citenamefont {N{\"o}lleke}, \citenamefont {Neuzner}, \citenamefont {Reiserer}, \citenamefont {Hahn}, \citenamefont {Rempe},\ and\ \citenamefont {Ritter}}]{nolleke2013efficient}%
  \BibitemOpen
  \bibfield  {author} {\bibinfo {author} {\bibfnamefont {C.}~\bibnamefont {N{\"o}lleke}}, \bibinfo {author} {\bibfnamefont {A.}~\bibnamefont {Neuzner}}, \bibinfo {author} {\bibfnamefont {A.}~\bibnamefont {Reiserer}}, \bibinfo {author} {\bibfnamefont {C.}~\bibnamefont {Hahn}}, \bibinfo {author} {\bibfnamefont {G.}~\bibnamefont {Rempe}},\ and\ \bibinfo {author} {\bibfnamefont {S.}~\bibnamefont {Ritter}},\ }\bibfield  {title} {\bibinfo {title} {Efficient {{Teleportation Between Remote Single-Atom Quantum Memories}}},\ }\href {https://doi.org/10.1103/PhysRevLett.110.140403} {\bibfield  {journal} {\bibinfo  {journal} {Phys. Rev. Lett.}\ }\textbf {\bibinfo {volume} {110}},\ \bibinfo {pages} {140403} (\bibinfo {year} {2013})}\BibitemShut {NoStop}%
\bibitem [{\citenamefont {Langenfeld}\ \emph {et~al.}(2021)\citenamefont {Langenfeld}, \citenamefont {Welte}, \citenamefont {Hartung}, \citenamefont {Daiss}, \citenamefont {Thomas}, \citenamefont {Morin}, \citenamefont {Distante},\ and\ \citenamefont {Rempe}}]{langenfeld2021quantum}%
  \BibitemOpen
  \bibfield  {author} {\bibinfo {author} {\bibfnamefont {S.}~\bibnamefont {Langenfeld}}, \bibinfo {author} {\bibfnamefont {S.}~\bibnamefont {Welte}}, \bibinfo {author} {\bibfnamefont {L.}~\bibnamefont {Hartung}}, \bibinfo {author} {\bibfnamefont {S.}~\bibnamefont {Daiss}}, \bibinfo {author} {\bibfnamefont {P.}~\bibnamefont {Thomas}}, \bibinfo {author} {\bibfnamefont {O.}~\bibnamefont {Morin}}, \bibinfo {author} {\bibfnamefont {E.}~\bibnamefont {Distante}},\ and\ \bibinfo {author} {\bibfnamefont {G.}~\bibnamefont {Rempe}},\ }\bibfield  {title} {\bibinfo {title} {Quantum {{Teleportation}} between {{Remote Qubit Memories}} with {{Only}} a {{Single Photon}} as a {{Resource}}},\ }\href {https://doi.org/10.1103/PhysRevLett.126.130502} {\bibfield  {journal} {\bibinfo  {journal} {Phys. Rev. Lett.}\ }\textbf {\bibinfo {volume} {126}},\ \bibinfo {pages} {130502} (\bibinfo {year} {2021})}\BibitemShut {NoStop}%
\bibitem [{\citenamefont {Daiss}\ \emph {et~al.}(2021)\citenamefont {Daiss}, \citenamefont {Langenfeld}, \citenamefont {Welte}, \citenamefont {Distante}, \citenamefont {Thomas}, \citenamefont {Hartung}, \citenamefont {Morin},\ and\ \citenamefont {Rempe}}]{daiss2021quantumlogic}%
  \BibitemOpen
  \bibfield  {author} {\bibinfo {author} {\bibfnamefont {S.}~\bibnamefont {Daiss}}, \bibinfo {author} {\bibfnamefont {S.}~\bibnamefont {Langenfeld}}, \bibinfo {author} {\bibfnamefont {S.}~\bibnamefont {Welte}}, \bibinfo {author} {\bibfnamefont {E.}~\bibnamefont {Distante}}, \bibinfo {author} {\bibfnamefont {P.}~\bibnamefont {Thomas}}, \bibinfo {author} {\bibfnamefont {L.}~\bibnamefont {Hartung}}, \bibinfo {author} {\bibfnamefont {O.}~\bibnamefont {Morin}},\ and\ \bibinfo {author} {\bibfnamefont {G.}~\bibnamefont {Rempe}},\ }\bibfield  {title} {\bibinfo {title} {A quantum-logic gate between distant quantum-network modules},\ }\href {https://doi.org/10.1126/science.abe3150} {\bibfield  {journal} {\bibinfo  {journal} {Science}\ }\textbf {\bibinfo {volume} {371}},\ \bibinfo {pages} {614} (\bibinfo {year} {2021})}\BibitemShut {NoStop}%
\bibitem [{\citenamefont {Stute}\ \emph {et~al.}(2012)\citenamefont {Stute}, \citenamefont {Casabone}, \citenamefont {Schindler}, \citenamefont {Monz}, \citenamefont {Schmidt}, \citenamefont {Brandst{\"a}tter}, \citenamefont {Northup},\ and\ \citenamefont {Blatt}}]{stute2012tunable}%
  \BibitemOpen
  \bibfield  {author} {\bibinfo {author} {\bibfnamefont {A.}~\bibnamefont {Stute}}, \bibinfo {author} {\bibfnamefont {B.}~\bibnamefont {Casabone}}, \bibinfo {author} {\bibfnamefont {P.}~\bibnamefont {Schindler}}, \bibinfo {author} {\bibfnamefont {T.}~\bibnamefont {Monz}}, \bibinfo {author} {\bibfnamefont {P.~O.}\ \bibnamefont {Schmidt}}, \bibinfo {author} {\bibfnamefont {B.}~\bibnamefont {Brandst{\"a}tter}}, \bibinfo {author} {\bibfnamefont {T.~E.}\ \bibnamefont {Northup}},\ and\ \bibinfo {author} {\bibfnamefont {R.}~\bibnamefont {Blatt}},\ }\bibfield  {title} {\bibinfo {title} {Tunable ion--photon entanglement in an optical cavity},\ }\href {https://doi.org/10.1038/nature11120} {\bibfield  {journal} {\bibinfo  {journal} {Nature}\ }\textbf {\bibinfo {volume} {485}},\ \bibinfo {pages} {482} (\bibinfo {year} {2012})}\BibitemShut {NoStop}%
\bibitem [{\citenamefont {Krutyanskiy}\ \emph {et~al.}(2019)\citenamefont {Krutyanskiy}, \citenamefont {Meraner}, \citenamefont {Schupp}, \citenamefont {Krcmarsky}, \citenamefont {Hainzer},\ and\ \citenamefont {Lanyon}}]{krutyanskiy2019lightmatter}%
  \BibitemOpen
  \bibfield  {author} {\bibinfo {author} {\bibfnamefont {V.}~\bibnamefont {Krutyanskiy}}, \bibinfo {author} {\bibfnamefont {M.}~\bibnamefont {Meraner}}, \bibinfo {author} {\bibfnamefont {J.}~\bibnamefont {Schupp}}, \bibinfo {author} {\bibfnamefont {V.}~\bibnamefont {Krcmarsky}}, \bibinfo {author} {\bibfnamefont {H.}~\bibnamefont {Hainzer}},\ and\ \bibinfo {author} {\bibfnamefont {B.~P.}\ \bibnamefont {Lanyon}},\ }\bibfield  {title} {\bibinfo {title} {Light-matter entanglement over 50 km of optical fibre},\ }\href {https://doi.org/10.1038/s41534-019-0186-3} {\bibfield  {journal} {\bibinfo  {journal} {npj Quantum Inf}\ }\textbf {\bibinfo {volume} {5}},\ \bibinfo {pages} {72} (\bibinfo {year} {2019})}\BibitemShut {NoStop}%
\bibitem [{\citenamefont {Schupp}\ \emph {et~al.}(2021)\citenamefont {Schupp}, \citenamefont {Krcmarsky}, \citenamefont {Krutyanskiy}, \citenamefont {Meraner}, \citenamefont {Northup},\ and\ \citenamefont {Lanyon}}]{schupp2021interface}%
  \BibitemOpen
  \bibfield  {author} {\bibinfo {author} {\bibfnamefont {J.}~\bibnamefont {Schupp}}, \bibinfo {author} {\bibfnamefont {V.}~\bibnamefont {Krcmarsky}}, \bibinfo {author} {\bibfnamefont {V.}~\bibnamefont {Krutyanskiy}}, \bibinfo {author} {\bibfnamefont {M.}~\bibnamefont {Meraner}}, \bibinfo {author} {\bibfnamefont {T.}~\bibnamefont {Northup}},\ and\ \bibinfo {author} {\bibfnamefont {B.}~\bibnamefont {Lanyon}},\ }\bibfield  {title} {\bibinfo {title} {Interface between {{Trapped-Ion Qubits}} and {{Traveling Photons}} with {{Close-to-Optimal Efficiency}}},\ }\href {https://doi.org/10.1103/PRXQuantum.2.020331} {\bibfield  {journal} {\bibinfo  {journal} {PRX Quantum}\ }\textbf {\bibinfo {volume} {2}},\ \bibinfo {pages} {020331} (\bibinfo {year} {2021})}\BibitemShut {NoStop}%
\bibitem [{\citenamefont {Kobel}\ \emph {et~al.}(2021)\citenamefont {Kobel}, \citenamefont {Breyer},\ and\ \citenamefont {K{\"o}hl}}]{kobel2021deterministic}%
  \BibitemOpen
  \bibfield  {author} {\bibinfo {author} {\bibfnamefont {P.}~\bibnamefont {Kobel}}, \bibinfo {author} {\bibfnamefont {M.}~\bibnamefont {Breyer}},\ and\ \bibinfo {author} {\bibfnamefont {M.}~\bibnamefont {K{\"o}hl}},\ }\bibfield  {title} {\bibinfo {title} {Deterministic spin-photon entanglement from a trapped ion in a fiber {{Fabry}}--{{Perot}} cavity},\ }\href {https://doi.org/10.1038/s41534-020-00338-2} {\bibfield  {journal} {\bibinfo  {journal} {npj Quantum Inf}\ }\textbf {\bibinfo {volume} {7}},\ \bibinfo {pages} {6} (\bibinfo {year} {2021})}\BibitemShut {NoStop}%
\bibitem [{\citenamefont {Krutyanskiy}\ \emph {et~al.}(2023{\natexlab{a}})\citenamefont {Krutyanskiy}, \citenamefont {Galli}, \citenamefont {Krcmarsky}, \citenamefont {Baier}, \citenamefont {Fioretto}, \citenamefont {Pu}, \citenamefont {Mazloom}, \citenamefont {Sekatski}, \citenamefont {Canteri}, \citenamefont {Teller}, \citenamefont {Schupp}, \citenamefont {Bate}, \citenamefont {Meraner}, \citenamefont {Sangouard}, \citenamefont {Lanyon},\ and\ \citenamefont {Northup}}]{krutyanskiy2023entanglement}%
  \BibitemOpen
  \bibfield  {author} {\bibinfo {author} {\bibfnamefont {V.}~\bibnamefont {Krutyanskiy}}, \bibinfo {author} {\bibfnamefont {M.}~\bibnamefont {Galli}}, \bibinfo {author} {\bibfnamefont {V.}~\bibnamefont {Krcmarsky}}, \bibinfo {author} {\bibfnamefont {S.}~\bibnamefont {Baier}}, \bibinfo {author} {\bibfnamefont {D.~A.}\ \bibnamefont {Fioretto}}, \bibinfo {author} {\bibfnamefont {Y.}~\bibnamefont {Pu}}, \bibinfo {author} {\bibfnamefont {A.}~\bibnamefont {Mazloom}}, \bibinfo {author} {\bibfnamefont {P.}~\bibnamefont {Sekatski}}, \bibinfo {author} {\bibfnamefont {M.}~\bibnamefont {Canteri}}, \bibinfo {author} {\bibfnamefont {M.}~\bibnamefont {Teller}}, \bibinfo {author} {\bibfnamefont {J.}~\bibnamefont {Schupp}}, \bibinfo {author} {\bibfnamefont {J.}~\bibnamefont {Bate}}, \bibinfo {author} {\bibfnamefont {M.}~\bibnamefont {Meraner}}, \bibinfo {author} {\bibfnamefont {N.}~\bibnamefont {Sangouard}}, \bibinfo {author} {\bibfnamefont {B.~P.}\ \bibnamefont {Lanyon}},\ and\ \bibinfo {author} {\bibfnamefont {T.~E.}\
  \bibnamefont {Northup}},\ }\bibfield  {title} {\bibinfo {title} {Entanglement of {{Trapped-Ion Qubits Separated}} by 230 {{Meters}}},\ }\href {https://doi.org/10.1103/PhysRevLett.130.050803} {\bibfield  {journal} {\bibinfo  {journal} {Phys. Rev. Lett.}\ }\textbf {\bibinfo {volume} {130}},\ \bibinfo {pages} {050803} (\bibinfo {year} {2023}{\natexlab{a}})}\BibitemShut {NoStop}%
\bibitem [{\citenamefont {Krutyanskiy}\ \emph {et~al.}(2023{\natexlab{b}})\citenamefont {Krutyanskiy}, \citenamefont {Canteri}, \citenamefont {Meraner}, \citenamefont {Bate}, \citenamefont {Krcmarsky}, \citenamefont {Schupp}, \citenamefont {Sangouard},\ and\ \citenamefont {Lanyon}}]{krutyanskiy2023telecomwavelength}%
  \BibitemOpen
  \bibfield  {author} {\bibinfo {author} {\bibfnamefont {V.}~\bibnamefont {Krutyanskiy}}, \bibinfo {author} {\bibfnamefont {M.}~\bibnamefont {Canteri}}, \bibinfo {author} {\bibfnamefont {M.}~\bibnamefont {Meraner}}, \bibinfo {author} {\bibfnamefont {J.}~\bibnamefont {Bate}}, \bibinfo {author} {\bibfnamefont {V.}~\bibnamefont {Krcmarsky}}, \bibinfo {author} {\bibfnamefont {J.}~\bibnamefont {Schupp}}, \bibinfo {author} {\bibfnamefont {N.}~\bibnamefont {Sangouard}},\ and\ \bibinfo {author} {\bibfnamefont {B.~P.}\ \bibnamefont {Lanyon}},\ }\bibfield  {title} {\bibinfo {title} {Telecom-{{Wavelength Quantum Repeater Node Based}} on a {{Trapped-Ion Processor}}},\ }\href {https://doi.org/10.1103/PhysRevLett.130.213601} {\bibfield  {journal} {\bibinfo  {journal} {Phys. Rev. Lett.}\ }\textbf {\bibinfo {volume} {130}},\ \bibinfo {pages} {213601} (\bibinfo {year} {2023}{\natexlab{b}})}\BibitemShut {NoStop}%
\bibitem [{\citenamefont {Krutyanskiy}\ \emph {et~al.}(2024)\citenamefont {Krutyanskiy}, \citenamefont {Canteri}, \citenamefont {Meraner}, \citenamefont {Krcmarsky},\ and\ \citenamefont {Lanyon}}]{krutyanskiy2024multimode}%
  \BibitemOpen
  \bibfield  {author} {\bibinfo {author} {\bibfnamefont {V.}~\bibnamefont {Krutyanskiy}}, \bibinfo {author} {\bibfnamefont {M.}~\bibnamefont {Canteri}}, \bibinfo {author} {\bibfnamefont {M.}~\bibnamefont {Meraner}}, \bibinfo {author} {\bibfnamefont {V.}~\bibnamefont {Krcmarsky}},\ and\ \bibinfo {author} {\bibfnamefont {B.}~\bibnamefont {Lanyon}},\ }\bibfield  {title} {\bibinfo {title} {Multimode {{Ion-Photon Entanglement}} over 101 {{Kilometers}}},\ }\href {https://doi.org/10.1103/PRXQuantum.5.020308} {\bibfield  {journal} {\bibinfo  {journal} {PRX Quantum}\ }\textbf {\bibinfo {volume} {5}},\ \bibinfo {pages} {020308} (\bibinfo {year} {2024})}\BibitemShut {NoStop}%
\bibitem [{\citenamefont {Gao}\ \emph {et~al.}(2023)\citenamefont {Gao}, \citenamefont {Blackmore}, \citenamefont {Hughes}, \citenamefont {Doherty},\ and\ \citenamefont {Goodwin}}]{gao2023optimization}%
  \BibitemOpen
  \bibfield  {author} {\bibinfo {author} {\bibfnamefont {S.}~\bibnamefont {Gao}}, \bibinfo {author} {\bibfnamefont {J.~A.}\ \bibnamefont {Blackmore}}, \bibinfo {author} {\bibfnamefont {W.~J.}\ \bibnamefont {Hughes}}, \bibinfo {author} {\bibfnamefont {T.~H.}\ \bibnamefont {Doherty}},\ and\ \bibinfo {author} {\bibfnamefont {J.~F.}\ \bibnamefont {Goodwin}},\ }\bibfield  {title} {\bibinfo {title} {Optimization of {{Scalable Ion-Cavity Interfaces}} for {{Quantum Photonic Networks}}},\ }\href {https://doi.org/10.1103/PhysRevApplied.19.014033} {\bibfield  {journal} {\bibinfo  {journal} {Phys. Rev. Applied}\ }\textbf {\bibinfo {volume} {19}},\ \bibinfo {pages} {014033} (\bibinfo {year} {2023})}\BibitemShut {NoStop}%
\bibitem [{\citenamefont {Li}\ and\ \citenamefont {Thompson}(2024)}]{li2024highrate}%
  \BibitemOpen
  \bibfield  {author} {\bibinfo {author} {\bibfnamefont {Y.}~\bibnamefont {Li}}\ and\ \bibinfo {author} {\bibfnamefont {J.~D.}\ \bibnamefont {Thompson}},\ }\bibfield  {title} {\bibinfo {title} {High-{{Rate}} and {{High-Fidelity Modular Interconnects}} between {{Neutral Atom Quantum Processors}}},\ }\href {https://doi.org/10.1103/PRXQuantum.5.020363} {\bibfield  {journal} {\bibinfo  {journal} {PRX Quantum}\ }\textbf {\bibinfo {volume} {5}},\ \bibinfo {pages} {020363} (\bibinfo {year} {2024})}\BibitemShut {NoStop}%
\bibitem [{\citenamefont {Menon}\ \emph {et~al.}(2024)\citenamefont {Menon}, \citenamefont {Glachman}, \citenamefont {Pompili}, \citenamefont {Dibos},\ and\ \citenamefont {Bernien}}]{menon2024integrated}%
  \BibitemOpen
  \bibfield  {author} {\bibinfo {author} {\bibfnamefont {S.~G.}\ \bibnamefont {Menon}}, \bibinfo {author} {\bibfnamefont {N.}~\bibnamefont {Glachman}}, \bibinfo {author} {\bibfnamefont {M.}~\bibnamefont {Pompili}}, \bibinfo {author} {\bibfnamefont {A.}~\bibnamefont {Dibos}},\ and\ \bibinfo {author} {\bibfnamefont {H.}~\bibnamefont {Bernien}},\ }\bibfield  {title} {\bibinfo {title} {An integrated atom array-nanophotonic chip platform with background-free imaging},\ }\href {https://doi.org/10.1038/s41467-024-50355-4} {\bibfield  {journal} {\bibinfo  {journal} {Nat. Commun.}\ }\textbf {\bibinfo {volume} {15}},\ \bibinfo {pages} {6156} (\bibinfo {year} {2024})}\BibitemShut {NoStop}%
\bibitem [{\citenamefont {Shaw}\ \emph {et~al.}(2025)\citenamefont {Shaw}, \citenamefont {Soper}, \citenamefont {Shadmany}, \citenamefont {Kumar}, \citenamefont {Palm}, \citenamefont {Koh}, \citenamefont {Kaxiras}, \citenamefont {Taneja}, \citenamefont {Jaffe}, \citenamefont {Schuster},\ and\ \citenamefont {Simon}}]{shaw2025cavity}%
  \BibitemOpen
  \bibfield  {author} {\bibinfo {author} {\bibfnamefont {A.~L.}\ \bibnamefont {Shaw}}, \bibinfo {author} {\bibfnamefont {A.}~\bibnamefont {Soper}}, \bibinfo {author} {\bibfnamefont {D.}~\bibnamefont {Shadmany}}, \bibinfo {author} {\bibfnamefont {A.}~\bibnamefont {Kumar}}, \bibinfo {author} {\bibfnamefont {L.}~\bibnamefont {Palm}}, \bibinfo {author} {\bibfnamefont {D.-Y.}\ \bibnamefont {Koh}}, \bibinfo {author} {\bibfnamefont {V.}~\bibnamefont {Kaxiras}}, \bibinfo {author} {\bibfnamefont {L.}~\bibnamefont {Taneja}}, \bibinfo {author} {\bibfnamefont {M.}~\bibnamefont {Jaffe}}, \bibinfo {author} {\bibfnamefont {D.~I.}\ \bibnamefont {Schuster}},\ and\ \bibinfo {author} {\bibfnamefont {J.}~\bibnamefont {Simon}},\ }\bibfield  {title} {\bibinfo {title} {A cavity array microscope for parallel single-atom interfacing},\ }\href {https://doi.org/10.48550/arXiv.2506.10919} {\bibfield  {journal} {\bibinfo  {journal} {arXiv:2506.10919}\ } (\bibinfo {year} {2025})}\BibitemShut {NoStop}%
\bibitem [{\citenamefont {K{\"u}bler}\ \emph {et~al.}(2010)\citenamefont {K{\"u}bler}, \citenamefont {Shaffer}, \citenamefont {Baluktsian}, \citenamefont {L{\"o}w},\ and\ \citenamefont {Pfau}}]{kubler2010coherent}%
  \BibitemOpen
  \bibfield  {author} {\bibinfo {author} {\bibfnamefont {H.}~\bibnamefont {K{\"u}bler}}, \bibinfo {author} {\bibfnamefont {J.~P.}\ \bibnamefont {Shaffer}}, \bibinfo {author} {\bibfnamefont {T.}~\bibnamefont {Baluktsian}}, \bibinfo {author} {\bibfnamefont {R.}~\bibnamefont {L{\"o}w}},\ and\ \bibinfo {author} {\bibfnamefont {T.}~\bibnamefont {Pfau}},\ }\bibfield  {title} {\bibinfo {title} {Coherent excitation of {{Rydberg}} atoms in micrometre-sized atomic vapour cells},\ }\href {https://doi.org/10.1038/nphoton.2009.260} {\bibfield  {journal} {\bibinfo  {journal} {Nat. Photonics}\ }\textbf {\bibinfo {volume} {4}},\ \bibinfo {pages} {112} (\bibinfo {year} {2010})}\BibitemShut {NoStop}%
\bibitem [{\citenamefont {Wang}\ \emph {et~al.}(2025)\citenamefont {Wang}, \citenamefont {Fomra}, \citenamefont {Agrawal}, \citenamefont {Lezec},\ and\ \citenamefont {Britton}}]{wang2025can}%
  \BibitemOpen
  \bibfield  {author} {\bibinfo {author} {\bibfnamefont {W.}~\bibnamefont {Wang}}, \bibinfo {author} {\bibfnamefont {D.}~\bibnamefont {Fomra}}, \bibinfo {author} {\bibfnamefont {A.}~\bibnamefont {Agrawal}}, \bibinfo {author} {\bibfnamefont {H.~J.}\ \bibnamefont {Lezec}},\ and\ \bibinfo {author} {\bibfnamefont {J.~W.}\ \bibnamefont {Britton}},\ }\bibfield  {title} {\bibinfo {title} {Can {{TCOs Transform Cavity-QED}}?},\ }\href {https://doi.org/10.48550/arXiv.2506.02501} {\bibfield  {journal} {\bibinfo  {journal} {arXiv:2506.02501}\ } (\bibinfo {year} {2025})}\BibitemShut {NoStop}%
\bibitem [{\citenamefont {Horak}\ \emph {et~al.}(2002)\citenamefont {Horak}, \citenamefont {Ritsch}, \citenamefont {Fischer}, \citenamefont {Maunz}, \citenamefont {Puppe}, \citenamefont {Pinkse},\ and\ \citenamefont {Rempe}}]{horak2002optical}%
  \BibitemOpen
  \bibfield  {author} {\bibinfo {author} {\bibfnamefont {P.}~\bibnamefont {Horak}}, \bibinfo {author} {\bibfnamefont {H.}~\bibnamefont {Ritsch}}, \bibinfo {author} {\bibfnamefont {T.}~\bibnamefont {Fischer}}, \bibinfo {author} {\bibfnamefont {P.}~\bibnamefont {Maunz}}, \bibinfo {author} {\bibfnamefont {T.}~\bibnamefont {Puppe}}, \bibinfo {author} {\bibfnamefont {P.~W.~H.}\ \bibnamefont {Pinkse}},\ and\ \bibinfo {author} {\bibfnamefont {G.}~\bibnamefont {Rempe}},\ }\bibfield  {title} {\bibinfo {title} {Optical {{Kaleidoscope Using}} a {{Single Atom}}},\ }\href {https://doi.org/10.1103/PhysRevLett.88.043601} {\bibfield  {journal} {\bibinfo  {journal} {Phys. Rev. Lett.}\ }\textbf {\bibinfo {volume} {88}},\ \bibinfo {pages} {043601} (\bibinfo {year} {2002})}\BibitemShut {NoStop}%
\bibitem [{\citenamefont {Puppe}\ \emph {et~al.}(2004)\citenamefont {Puppe}, \citenamefont {Maunz}, \citenamefont {Fischer}, \citenamefont {Pinkse},\ and\ \citenamefont {Rempe}}]{puppe2004singleatom}%
  \BibitemOpen
  \bibfield  {author} {\bibinfo {author} {\bibfnamefont {T.}~\bibnamefont {Puppe}}, \bibinfo {author} {\bibfnamefont {P.}~\bibnamefont {Maunz}}, \bibinfo {author} {\bibfnamefont {T.}~\bibnamefont {Fischer}}, \bibinfo {author} {\bibfnamefont {P.}~\bibnamefont {Pinkse}},\ and\ \bibinfo {author} {\bibfnamefont {G.}~\bibnamefont {Rempe}},\ }\bibfield  {title} {\bibinfo {title} {Single-{{Atom Trajectories}} in {{Higher-Order Transverse Modes}} of a {{High-Finesse Optical Cavity}}},\ }\href {https://doi.org/10.1238/Physica.Topical.112a00007} {\bibfield  {journal} {\bibinfo  {journal} {Phys. Scr.}\ }\textbf {\bibinfo {volume} {T112}},\ \bibinfo {pages} {7} (\bibinfo {year} {2004})}\BibitemShut {NoStop}%
\bibitem [{\citenamefont {Vaidya}\ \emph {et~al.}(2018)\citenamefont {Vaidya}, \citenamefont {Guo}, \citenamefont {Kroeze}, \citenamefont {Ballantine}, \citenamefont {Koll{\'a}r}, \citenamefont {Keeling},\ and\ \citenamefont {Lev}}]{vaidya2018tunablerange}%
  \BibitemOpen
  \bibfield  {author} {\bibinfo {author} {\bibfnamefont {V.~D.}\ \bibnamefont {Vaidya}}, \bibinfo {author} {\bibfnamefont {Y.}~\bibnamefont {Guo}}, \bibinfo {author} {\bibfnamefont {R.~M.}\ \bibnamefont {Kroeze}}, \bibinfo {author} {\bibfnamefont {K.~E.}\ \bibnamefont {Ballantine}}, \bibinfo {author} {\bibfnamefont {A.~J.}\ \bibnamefont {Koll{\'a}r}}, \bibinfo {author} {\bibfnamefont {J.}~\bibnamefont {Keeling}},\ and\ \bibinfo {author} {\bibfnamefont {B.~L.}\ \bibnamefont {Lev}},\ }\bibfield  {title} {\bibinfo {title} {Tunable-{{Range}}, {{Photon-Mediated Atomic Interactions}} in {{Multimode Cavity QED}}},\ }\href {https://doi.org/10.1103/PhysRevX.8.011002} {\bibfield  {journal} {\bibinfo  {journal} {Phys. Rev. X}\ }\textbf {\bibinfo {volume} {8}},\ \bibinfo {pages} {011002} (\bibinfo {year} {2018})}\BibitemShut {NoStop}%
\bibitem [{\citenamefont {Kroeze}\ \emph {et~al.}(2025)\citenamefont {Kroeze}, \citenamefont {Marsh}, \citenamefont {Atri~Schuller}, \citenamefont {Hunt}, \citenamefont {Bourzutschky}, \citenamefont {Winer}, \citenamefont {Gopalakrishnan}, \citenamefont {Keeling},\ and\ \citenamefont {Lev}}]{kroeze2025directly}%
  \BibitemOpen
  \bibfield  {author} {\bibinfo {author} {\bibfnamefont {R.~M.}\ \bibnamefont {Kroeze}}, \bibinfo {author} {\bibfnamefont {B.~P.}\ \bibnamefont {Marsh}}, \bibinfo {author} {\bibfnamefont {D.}~\bibnamefont {Atri~Schuller}}, \bibinfo {author} {\bibfnamefont {H.~S.}\ \bibnamefont {Hunt}}, \bibinfo {author} {\bibfnamefont {A.~N.}\ \bibnamefont {Bourzutschky}}, \bibinfo {author} {\bibfnamefont {M.}~\bibnamefont {Winer}}, \bibinfo {author} {\bibfnamefont {S.}~\bibnamefont {Gopalakrishnan}}, \bibinfo {author} {\bibfnamefont {J.}~\bibnamefont {Keeling}},\ and\ \bibinfo {author} {\bibfnamefont {B.~L.}\ \bibnamefont {Lev}},\ }\bibfield  {title} {\bibinfo {title} {Directly observing replica symmetry breaking in a vector quantum-optical spin glass},\ }\href {https://doi.org/10.1126/science.adu7710} {\bibfield  {journal} {\bibinfo  {journal} {Science}\ }\textbf {\bibinfo {volume} {389}},\ \bibinfo {pages} {1122} (\bibinfo {year} {2025})}\BibitemShut {NoStop}%
\bibitem [{\citenamefont {Clark}\ \emph {et~al.}(2020)\citenamefont {Clark}, \citenamefont {Schine}, \citenamefont {Baum}, \citenamefont {Jia},\ and\ \citenamefont {Simon}}]{clark2020observation}%
  \BibitemOpen
  \bibfield  {author} {\bibinfo {author} {\bibfnamefont {L.~W.}\ \bibnamefont {Clark}}, \bibinfo {author} {\bibfnamefont {N.}~\bibnamefont {Schine}}, \bibinfo {author} {\bibfnamefont {C.}~\bibnamefont {Baum}}, \bibinfo {author} {\bibfnamefont {N.}~\bibnamefont {Jia}},\ and\ \bibinfo {author} {\bibfnamefont {J.}~\bibnamefont {Simon}},\ }\bibfield  {title} {\bibinfo {title} {Observation of {{Laughlin}} states made of light},\ }\href {https://doi.org/10.1038/s41586-020-2318-5} {\bibfield  {journal} {\bibinfo  {journal} {Nature}\ }\textbf {\bibinfo {volume} {582}},\ \bibinfo {pages} {41} (\bibinfo {year} {2020})}\BibitemShut {NoStop}%
\bibitem [{\citenamefont {Kroeze}\ \emph {et~al.}(2023)\citenamefont {Kroeze}, \citenamefont {Marsh}, \citenamefont {Lin}, \citenamefont {Keeling},\ and\ \citenamefont {Lev}}]{kroeze2023high}%
  \BibitemOpen
  \bibfield  {author} {\bibinfo {author} {\bibfnamefont {R.~M.}\ \bibnamefont {Kroeze}}, \bibinfo {author} {\bibfnamefont {B.~P.}\ \bibnamefont {Marsh}}, \bibinfo {author} {\bibfnamefont {K.-Y.}\ \bibnamefont {Lin}}, \bibinfo {author} {\bibfnamefont {J.}~\bibnamefont {Keeling}},\ and\ \bibinfo {author} {\bibfnamefont {B.~L.}\ \bibnamefont {Lev}},\ }\bibfield  {title} {\bibinfo {title} {High {{Cooperativity Using}} a {{Confocal-Cavity}}--{{QED Microscope}}},\ }\href {https://doi.org/10.1103/PRXQuantum.4.020326} {\bibfield  {journal} {\bibinfo  {journal} {PRX Quantum}\ }\textbf {\bibinfo {volume} {4}},\ \bibinfo {pages} {020326} (\bibinfo {year} {2023})}\BibitemShut {NoStop}%
\bibitem [{\citenamefont {Siegman}(1986)}]{siegman1986lasers}%
  \BibitemOpen
  \bibfield  {author} {\bibinfo {author} {\bibfnamefont {A.~E.}\ \bibnamefont {Siegman}},\ }\href@noop {} {\emph {\bibinfo {title} {Lasers}}}\ (\bibinfo  {publisher} {University Science Books},\ \bibinfo {address} {Mill Valley, CA},\ \bibinfo {year} {1986})\BibitemShut {NoStop}%
\bibitem [{\citenamefont {Burgers}\ \emph {et~al.}(2022)\citenamefont {Burgers}, \citenamefont {Ma}, \citenamefont {Saskin}, \citenamefont {Wilson}, \citenamefont {Alarc{\'o}n}, \citenamefont {Greene},\ and\ \citenamefont {Thompson}}]{burgers2022controlling}%
  \BibitemOpen
  \bibfield  {author} {\bibinfo {author} {\bibfnamefont {A.~P.}\ \bibnamefont {Burgers}}, \bibinfo {author} {\bibfnamefont {S.}~\bibnamefont {Ma}}, \bibinfo {author} {\bibfnamefont {S.}~\bibnamefont {Saskin}}, \bibinfo {author} {\bibfnamefont {J.}~\bibnamefont {Wilson}}, \bibinfo {author} {\bibfnamefont {M.~A.}\ \bibnamefont {Alarc{\'o}n}}, \bibinfo {author} {\bibfnamefont {C.~H.}\ \bibnamefont {Greene}},\ and\ \bibinfo {author} {\bibfnamefont {J.~D.}\ \bibnamefont {Thompson}},\ }\bibfield  {title} {\bibinfo {title} {Controlling {{Rydberg Excitations Using Ion-Core Transitions}} in {{Alkaline-Earth Atom-Tweezer Arrays}}},\ }\href {https://doi.org/10.1103/PRXQuantum.3.020326} {\bibfield  {journal} {\bibinfo  {journal} {PRX Quantum}\ }\textbf {\bibinfo {volume} {3}},\ \bibinfo {pages} {020326} (\bibinfo {year} {2022})}\BibitemShut {NoStop}%
\bibitem [{\citenamefont {Baum}\ \emph {et~al.}(2023)\citenamefont {Baum}, \citenamefont {Jaffe}, \citenamefont {Palm}, \citenamefont {Kumar},\ and\ \citenamefont {Simon}}]{baum2023optical}%
  \BibitemOpen
  \bibfield  {author} {\bibinfo {author} {\bibfnamefont {C.}~\bibnamefont {Baum}}, \bibinfo {author} {\bibfnamefont {M.}~\bibnamefont {Jaffe}}, \bibinfo {author} {\bibfnamefont {L.}~\bibnamefont {Palm}}, \bibinfo {author} {\bibfnamefont {A.}~\bibnamefont {Kumar}},\ and\ \bibinfo {author} {\bibfnamefont {J.}~\bibnamefont {Simon}},\ }\bibfield  {title} {\bibinfo {title} {Optical mode conversion via spatiotemporally modulated atomic susceptibility},\ }\href {https://doi.org/10.1364/OE.476638} {\bibfield  {journal} {\bibinfo  {journal} {Opt. Express}\ }\textbf {\bibinfo {volume} {31}},\ \bibinfo {pages} {528} (\bibinfo {year} {2023})}\BibitemShut {NoStop}%
\bibitem [{\citenamefont {Bornet}\ \emph {et~al.}(2024)\citenamefont {Bornet}, \citenamefont {Emperauger}, \citenamefont {Chen}, \citenamefont {Machado}, \citenamefont {Chern}, \citenamefont {Leclerc}, \citenamefont {G{\'e}ly}, \citenamefont {Chew}, \citenamefont {Barredo}, \citenamefont {Lahaye}, \citenamefont {Yao},\ and\ \citenamefont {Browaeys}}]{bornet2024enhancing}%
  \BibitemOpen
  \bibfield  {author} {\bibinfo {author} {\bibfnamefont {G.}~\bibnamefont {Bornet}}, \bibinfo {author} {\bibfnamefont {G.}~\bibnamefont {Emperauger}}, \bibinfo {author} {\bibfnamefont {C.}~\bibnamefont {Chen}}, \bibinfo {author} {\bibfnamefont {F.}~\bibnamefont {Machado}}, \bibinfo {author} {\bibfnamefont {S.}~\bibnamefont {Chern}}, \bibinfo {author} {\bibfnamefont {L.}~\bibnamefont {Leclerc}}, \bibinfo {author} {\bibfnamefont {B.}~\bibnamefont {G{\'e}ly}}, \bibinfo {author} {\bibfnamefont {Y.~T.}\ \bibnamefont {Chew}}, \bibinfo {author} {\bibfnamefont {D.}~\bibnamefont {Barredo}}, \bibinfo {author} {\bibfnamefont {T.}~\bibnamefont {Lahaye}}, \bibinfo {author} {\bibfnamefont {N.~Y.}\ \bibnamefont {Yao}},\ and\ \bibinfo {author} {\bibfnamefont {A.}~\bibnamefont {Browaeys}},\ }\bibfield  {title} {\bibinfo {title} {Enhancing a {{Many-Body Dipolar Rydberg Tweezer Array}} with {{Arbitrary Local Controls}}},\ }\href {https://doi.org/10.1103/PhysRevLett.132.263601} {\bibfield  {journal} {\bibinfo  {journal} {Phys.
  Rev. Lett.}\ }\textbf {\bibinfo {volume} {132}},\ \bibinfo {pages} {263601} (\bibinfo {year} {2024})}\BibitemShut {NoStop}%
\bibitem [{\citenamefont {Orsi}\ \emph {et~al.}(2024)\citenamefont {Orsi}, \citenamefont {Sauerwein}, \citenamefont {Bhatt}, \citenamefont {Faltinath}, \citenamefont {Fedotova}, \citenamefont {Reiter}, \citenamefont {{Cantat-Moltrecht}},\ and\ \citenamefont {Brantut}}]{orsi2024cavity}%
  \BibitemOpen
  \bibfield  {author} {\bibinfo {author} {\bibfnamefont {F.}~\bibnamefont {Orsi}}, \bibinfo {author} {\bibfnamefont {N.}~\bibnamefont {Sauerwein}}, \bibinfo {author} {\bibfnamefont {R.~P.}\ \bibnamefont {Bhatt}}, \bibinfo {author} {\bibfnamefont {J.}~\bibnamefont {Faltinath}}, \bibinfo {author} {\bibfnamefont {E.}~\bibnamefont {Fedotova}}, \bibinfo {author} {\bibfnamefont {N.}~\bibnamefont {Reiter}}, \bibinfo {author} {\bibfnamefont {T.}~\bibnamefont {{Cantat-Moltrecht}}},\ and\ \bibinfo {author} {\bibfnamefont {J.-P.}\ \bibnamefont {Brantut}},\ }\bibfield  {title} {\bibinfo {title} {Cavity {{Microscope}} for {{Micrometer-Scale Control}} of {{Atom-Photon Interactions}}},\ }\href {https://doi.org/10.1103/PRXQuantum.5.040333} {\bibfield  {journal} {\bibinfo  {journal} {PRX Quantum}\ }\textbf {\bibinfo {volume} {5}},\ \bibinfo {pages} {040333} (\bibinfo {year} {2024})}\BibitemShut {NoStop}%
\bibitem [{\citenamefont {Aqua}\ and\ \citenamefont {Dayan}(2025)}]{aqua2025atommediated}%
  \BibitemOpen
  \bibfield  {author} {\bibinfo {author} {\bibfnamefont {Z.}~\bibnamefont {Aqua}}\ and\ \bibinfo {author} {\bibfnamefont {B.}~\bibnamefont {Dayan}},\ }\bibfield  {title} {\bibinfo {title} {Atom-{{Mediated Deterministic Generation}} and {{Stitching}} of {{Photonic Graph States}}},\ }\href {https://doi.org/10.1103/PRXQuantum.6.010340} {\bibfield  {journal} {\bibinfo  {journal} {PRX Quantum}\ }\textbf {\bibinfo {volume} {6}},\ \bibinfo {pages} {010340} (\bibinfo {year} {2025})}\BibitemShut {NoStop}%
\bibitem [{\citenamefont {Morizur}\ \emph {et~al.}(2010)\citenamefont {Morizur}, \citenamefont {Nicholls}, \citenamefont {Jian}, \citenamefont {Armstrong}, \citenamefont {Treps}, \citenamefont {Hage}, \citenamefont {Hsu}, \citenamefont {Bowen}, \citenamefont {Janousek},\ and\ \citenamefont {Bachor}}]{morizur2010programmable}%
  \BibitemOpen
  \bibfield  {author} {\bibinfo {author} {\bibfnamefont {J.-F.}\ \bibnamefont {Morizur}}, \bibinfo {author} {\bibfnamefont {L.}~\bibnamefont {Nicholls}}, \bibinfo {author} {\bibfnamefont {P.}~\bibnamefont {Jian}}, \bibinfo {author} {\bibfnamefont {S.}~\bibnamefont {Armstrong}}, \bibinfo {author} {\bibfnamefont {N.}~\bibnamefont {Treps}}, \bibinfo {author} {\bibfnamefont {B.}~\bibnamefont {Hage}}, \bibinfo {author} {\bibfnamefont {M.}~\bibnamefont {Hsu}}, \bibinfo {author} {\bibfnamefont {W.}~\bibnamefont {Bowen}}, \bibinfo {author} {\bibfnamefont {J.}~\bibnamefont {Janousek}},\ and\ \bibinfo {author} {\bibfnamefont {H.-A.}\ \bibnamefont {Bachor}},\ }\bibfield  {title} {\bibinfo {title} {Programmable unitary spatial mode manipulation},\ }\href {https://doi.org/10.1364/JOSAA.27.002524} {\bibfield  {journal} {\bibinfo  {journal} {JOSA A}\ }\textbf {\bibinfo {volume} {27}},\ \bibinfo {pages} {2524} (\bibinfo {year} {2010})}\BibitemShut {NoStop}%
\bibitem [{\citenamefont {Fontaine}\ \emph {et~al.}(2019)\citenamefont {Fontaine}, \citenamefont {Ryf}, \citenamefont {Chen}, \citenamefont {Neilson}, \citenamefont {Kim},\ and\ \citenamefont {Carpenter}}]{fontaine2019laguerregaussian}%
  \BibitemOpen
  \bibfield  {author} {\bibinfo {author} {\bibfnamefont {N.~K.}\ \bibnamefont {Fontaine}}, \bibinfo {author} {\bibfnamefont {R.}~\bibnamefont {Ryf}}, \bibinfo {author} {\bibfnamefont {H.}~\bibnamefont {Chen}}, \bibinfo {author} {\bibfnamefont {D.~T.}\ \bibnamefont {Neilson}}, \bibinfo {author} {\bibfnamefont {K.}~\bibnamefont {Kim}},\ and\ \bibinfo {author} {\bibfnamefont {J.}~\bibnamefont {Carpenter}},\ }\bibfield  {title} {\bibinfo {title} {Laguerre-{{Gaussian}} mode sorter},\ }\href {https://doi.org/10.1038/s41467-019-09840-4} {\bibfield  {journal} {\bibinfo  {journal} {Nat. Commun.}\ }\textbf {\bibinfo {volume} {10}},\ \bibinfo {pages} {1865} (\bibinfo {year} {2019})}\BibitemShut {NoStop}%
\bibitem [{\citenamefont {Fontaine}\ \emph {et~al.}(2021)\citenamefont {Fontaine}, \citenamefont {Chen}, \citenamefont {Mazur}, \citenamefont {Dallachiesa}, \citenamefont {Kim}, \citenamefont {Ryf}, \citenamefont {Neilson},\ and\ \citenamefont {Carpenter}}]{fontaine2021hermitegaussian}%
  \BibitemOpen
  \bibfield  {author} {\bibinfo {author} {\bibfnamefont {N.~K.}\ \bibnamefont {Fontaine}}, \bibinfo {author} {\bibfnamefont {H.}~\bibnamefont {Chen}}, \bibinfo {author} {\bibfnamefont {M.}~\bibnamefont {Mazur}}, \bibinfo {author} {\bibfnamefont {L.}~\bibnamefont {Dallachiesa}}, \bibinfo {author} {\bibfnamefont {K.}~\bibnamefont {Kim}}, \bibinfo {author} {\bibfnamefont {R.}~\bibnamefont {Ryf}}, \bibinfo {author} {\bibfnamefont {D.}~\bibnamefont {Neilson}},\ and\ \bibinfo {author} {\bibfnamefont {J.}~\bibnamefont {Carpenter}},\ }\bibfield  {title} {\bibinfo {title} {Hermite-{{Gaussian}} mode multiplexer supporting 1035 modes},\ }in\ \href {https://doi.org/10.1364/OFC.2021.M3D.4} {\emph {\bibinfo {booktitle} {Optical {{Fiber Communication Conference}} ({{OFC}}) 2021}}}\ (\bibinfo  {publisher} {Optica Publishing Group},\ \bibinfo {address} {Washington, DC},\ \bibinfo {year} {2021})\ p.\ \bibinfo {pages} {M3D.4}\BibitemShut {NoStop}%
\bibitem [{\citenamefont {Zhang}\ and\ \citenamefont {Fontaine}(2023)}]{zhang2023multiplane}%
  \BibitemOpen
  \bibfield  {author} {\bibinfo {author} {\bibfnamefont {Y.}~\bibnamefont {Zhang}}\ and\ \bibinfo {author} {\bibfnamefont {N.~K.}\ \bibnamefont {Fontaine}},\ }\bibfield  {title} {\bibinfo {title} {Multi-{{Plane Light Conversion}}: {{A Practical Tutorial}}},\ }\href {https://doi.org/10.48550/arXiv.2304.11323} {\bibfield  {journal} {\bibinfo  {journal} {arXiv:2304.11323}\ } (\bibinfo {year} {2023})}\BibitemShut {NoStop}%
\bibitem [{\citenamefont {Choi}\ \emph {et~al.}(2024)\citenamefont {Choi}, \citenamefont {Pluchar}, \citenamefont {He}, \citenamefont {Guha},\ and\ \citenamefont {Wilson}}]{choi2024quantum}%
  \BibitemOpen
  \bibfield  {author} {\bibinfo {author} {\bibfnamefont {M.}~\bibnamefont {Choi}}, \bibinfo {author} {\bibfnamefont {C.}~\bibnamefont {Pluchar}}, \bibinfo {author} {\bibfnamefont {W.}~\bibnamefont {He}}, \bibinfo {author} {\bibfnamefont {S.}~\bibnamefont {Guha}},\ and\ \bibinfo {author} {\bibfnamefont {D.}~\bibnamefont {Wilson}},\ }\bibfield  {title} {\bibinfo {title} {Quantum limited imaging of a nanomechanical resonator with a spatial mode sorter},\ }\href {https://doi.org/10.48550/arXiv.2411.04980} {\bibfield  {journal} {\bibinfo  {journal} {arXiv:2411.04980}\ } (\bibinfo {year} {2024})}\BibitemShut {NoStop}%
\bibitem [{\citenamefont {Shirasaki}(1996)}]{shirasaki1996large}%
  \BibitemOpen
  \bibfield  {author} {\bibinfo {author} {\bibfnamefont {M.}~\bibnamefont {Shirasaki}},\ }\bibfield  {title} {\bibinfo {title} {Large angular dispersion by a virtually imaged phased array and its application to a wavelength demultiplexer},\ }\href {https://doi.org/10.1364/OL.21.000366} {\bibfield  {journal} {\bibinfo  {journal} {Opt. Lett.}\ }\textbf {\bibinfo {volume} {21}},\ \bibinfo {pages} {366} (\bibinfo {year} {1996})}\BibitemShut {NoStop}%
\bibitem [{\citenamefont {Xiao}\ \emph {et~al.}(2004)\citenamefont {Xiao}, \citenamefont {Weiner},\ and\ \citenamefont {Lin}}]{xiao2004dispersion}%
  \BibitemOpen
  \bibfield  {author} {\bibinfo {author} {\bibfnamefont {S.}~\bibnamefont {Xiao}}, \bibinfo {author} {\bibfnamefont {A.}~\bibnamefont {Weiner}},\ and\ \bibinfo {author} {\bibfnamefont {C.}~\bibnamefont {Lin}},\ }\bibfield  {title} {\bibinfo {title} {A dispersion law for virtually imaged phased-array spectral dispersers based on paraxial wave theory},\ }\href {https://doi.org/10.1109/JQE.2004.825210} {\bibfield  {journal} {\bibinfo  {journal} {IEEE Journal of Quantum Electronics}\ }\textbf {\bibinfo {volume} {40}},\ \bibinfo {pages} {420} (\bibinfo {year} {2004})}\BibitemShut {NoStop}%
\bibitem [{\citenamefont {Limbach}(2019)}]{limbach2019fully}%
  \BibitemOpen
  \bibfield  {author} {\bibinfo {author} {\bibfnamefont {C.~M.}\ \bibnamefont {Limbach}},\ }\bibfield  {title} {\bibinfo {title} {Fully resolved lineshape measurement of a seeded and unseeded optical parametric oscillator using a virtually imaged phased array spectrometer},\ }\href {https://doi.org/10.1364/OL.44.003821} {\bibfield  {journal} {\bibinfo  {journal} {Opt. Lett.}\ }\textbf {\bibinfo {volume} {44}},\ \bibinfo {pages} {3821} (\bibinfo {year} {2019})}\BibitemShut {NoStop}%
\bibitem [{\citenamefont {Leonov}\ \emph {et~al.}(2021)\citenamefont {Leonov}, \citenamefont {Rad}, \citenamefont {Wu},\ and\ \citenamefont {Limbach}}]{leonov2021application}%
  \BibitemOpen
  \bibfield  {author} {\bibinfo {author} {\bibfnamefont {B.~S.}\ \bibnamefont {Leonov}}, \bibinfo {author} {\bibfnamefont {A.~A.}\ \bibnamefont {Rad}}, \bibinfo {author} {\bibfnamefont {Y.}~\bibnamefont {Wu}},\ and\ \bibinfo {author} {\bibfnamefont {C.~M.}\ \bibnamefont {Limbach}},\ }\bibfield  {title} {\bibinfo {title} {Application of a pulsed {{OPO}} seeded by a {{CW OPO}} to investigation of linewidth effects on {{TALIF}} excitation efficiency},\ }\href {https://doi.org/10.1016/j.optlastec.2021.107370} {\bibfield  {journal} {\bibinfo  {journal} {Opt. Laser Technol.}\ }\textbf {\bibinfo {volume} {144}},\ \bibinfo {pages} {107370} (\bibinfo {year} {2021})}\BibitemShut {NoStop}%
\bibitem [{\citenamefont {Wei}\ \emph {et~al.}(2026)\citenamefont {Wei}, \citenamefont {Li}, \citenamefont {Karve}, \citenamefont {Shaw}, \citenamefont {Schuster},\ and\ \citenamefont {Simon}}]{wei202610}%
  \BibitemOpen
  \bibfield  {author} {\bibinfo {author} {\bibfnamefont {X.}~\bibnamefont {Wei}}, \bibinfo {author} {\bibfnamefont {Z.}~\bibnamefont {Li}}, \bibinfo {author} {\bibfnamefont {A.~V.}\ \bibnamefont {Karve}}, \bibinfo {author} {\bibfnamefont {A.~L.}\ \bibnamefont {Shaw}}, \bibinfo {author} {\bibfnamefont {D.~I.}\ \bibnamefont {Schuster}},\ and\ \bibinfo {author} {\bibfnamefont {J.}~\bibnamefont {Simon}},\ }\bibfield  {title} {\bibinfo {title} {A 10 {{Megahertz Spatial Light Modulator}}},\ }\bibfield  {journal} {\bibinfo  {journal} {arXiv:2601.08906}\ }\href {https://doi.org/https://doi.org/10.48550/arXiv:2601.08906} {https://doi.org/10.48550/arXiv:2601.08906} (\bibinfo {year} {2026})\BibitemShut {NoStop}%
\bibitem [{\citenamefont {Boozer}\ \emph {et~al.}(2006)\citenamefont {Boozer}, \citenamefont {Boca}, \citenamefont {Miller}, \citenamefont {Northup},\ and\ \citenamefont {Kimble}}]{boozer2006cooling}%
  \BibitemOpen
  \bibfield  {author} {\bibinfo {author} {\bibfnamefont {A.~D.}\ \bibnamefont {Boozer}}, \bibinfo {author} {\bibfnamefont {A.}~\bibnamefont {Boca}}, \bibinfo {author} {\bibfnamefont {R.}~\bibnamefont {Miller}}, \bibinfo {author} {\bibfnamefont {T.~E.}\ \bibnamefont {Northup}},\ and\ \bibinfo {author} {\bibfnamefont {H.~J.}\ \bibnamefont {Kimble}},\ }\bibfield  {title} {\bibinfo {title} {Cooling to the {{Ground State}} of {{Axial Motion}} for {{One Atom Strongly Coupled}} to an {{Optical Cavity}}},\ }\href {https://doi.org/10.1103/PhysRevLett.97.083602} {\bibfield  {journal} {\bibinfo  {journal} {Phys. Rev. Lett.}\ }\textbf {\bibinfo {volume} {97}},\ \bibinfo {pages} {083602} (\bibinfo {year} {2006})}\BibitemShut {NoStop}%
\bibitem [{\citenamefont {Knoernschild}\ \emph {et~al.}(2010)\citenamefont {Knoernschild}, \citenamefont {Zhang}, \citenamefont {Isenhower}, \citenamefont {Gill}, \citenamefont {Lu}, \citenamefont {Saffman},\ and\ \citenamefont {Kim}}]{knoernschild2010independent}%
  \BibitemOpen
  \bibfield  {author} {\bibinfo {author} {\bibfnamefont {C.}~\bibnamefont {Knoernschild}}, \bibinfo {author} {\bibfnamefont {X.~L.}\ \bibnamefont {Zhang}}, \bibinfo {author} {\bibfnamefont {L.}~\bibnamefont {Isenhower}}, \bibinfo {author} {\bibfnamefont {A.~T.}\ \bibnamefont {Gill}}, \bibinfo {author} {\bibfnamefont {F.~P.}\ \bibnamefont {Lu}}, \bibinfo {author} {\bibfnamefont {M.}~\bibnamefont {Saffman}},\ and\ \bibinfo {author} {\bibfnamefont {J.}~\bibnamefont {Kim}},\ }\bibfield  {title} {\bibinfo {title} {Independent individual addressing of multiple neutral atom qubits with a micromirror-based beam steering system},\ }\href {https://doi.org/10.1063/1.3494526} {\bibfield  {journal} {\bibinfo  {journal} {Appl. Phys. Lett.}\ }\textbf {\bibinfo {volume} {97}},\ \bibinfo {pages} {134101} (\bibinfo {year} {2010})}\BibitemShut {NoStop}%
\bibitem [{\citenamefont {Graham}\ \emph {et~al.}(2023)\citenamefont {Graham}, \citenamefont {Oh},\ and\ \citenamefont {Saffman}}]{graham2023multiscale}%
  \BibitemOpen
  \bibfield  {author} {\bibinfo {author} {\bibfnamefont {T.~M.}\ \bibnamefont {Graham}}, \bibinfo {author} {\bibfnamefont {E.}~\bibnamefont {Oh}},\ and\ \bibinfo {author} {\bibfnamefont {M.}~\bibnamefont {Saffman}},\ }\bibfield  {title} {\bibinfo {title} {Multiscale architecture for fast optical addressing and control of large-scale qubit arrays},\ }\href {https://doi.org/10.1364/AO.484367} {\bibfield  {journal} {\bibinfo  {journal} {Appl. Opt.}\ }\textbf {\bibinfo {volume} {62}},\ \bibinfo {pages} {3242} (\bibinfo {year} {2023})}\BibitemShut {NoStop}%
\bibitem [{\citenamefont {Menssen}\ \emph {et~al.}(2023)\citenamefont {Menssen}, \citenamefont {Hermans}, \citenamefont {Christen}, \citenamefont {Propson}, \citenamefont {Li}, \citenamefont {Leenheer}, \citenamefont {Zimmermann}, \citenamefont {Dong}, \citenamefont {Larocque}, \citenamefont {Raniwala}, \citenamefont {Gilbert}, \citenamefont {Eichenfield},\ and\ \citenamefont {Englund}}]{menssen2023scalable}%
  \BibitemOpen
  \bibfield  {author} {\bibinfo {author} {\bibfnamefont {A.~J.}\ \bibnamefont {Menssen}}, \bibinfo {author} {\bibfnamefont {A.}~\bibnamefont {Hermans}}, \bibinfo {author} {\bibfnamefont {I.}~\bibnamefont {Christen}}, \bibinfo {author} {\bibfnamefont {T.}~\bibnamefont {Propson}}, \bibinfo {author} {\bibfnamefont {C.}~\bibnamefont {Li}}, \bibinfo {author} {\bibfnamefont {A.~J.}\ \bibnamefont {Leenheer}}, \bibinfo {author} {\bibfnamefont {M.}~\bibnamefont {Zimmermann}}, \bibinfo {author} {\bibfnamefont {M.}~\bibnamefont {Dong}}, \bibinfo {author} {\bibfnamefont {H.}~\bibnamefont {Larocque}}, \bibinfo {author} {\bibfnamefont {H.}~\bibnamefont {Raniwala}}, \bibinfo {author} {\bibfnamefont {G.}~\bibnamefont {Gilbert}}, \bibinfo {author} {\bibfnamefont {M.}~\bibnamefont {Eichenfield}},\ and\ \bibinfo {author} {\bibfnamefont {D.~R.}\ \bibnamefont {Englund}},\ }\bibfield  {title} {\bibinfo {title} {Scalable photonic integrated circuits for high-fidelity light control},\ }\href {https://doi.org/10.1364/OPTICA.489504}
  {\bibfield  {journal} {\bibinfo  {journal} {Optica}\ }\textbf {\bibinfo {volume} {10}},\ \bibinfo {pages} {1366} (\bibinfo {year} {2023})}\BibitemShut {NoStop}%
\bibitem [{\citenamefont {Zhang}\ \emph {et~al.}(2024)\citenamefont {Zhang}, \citenamefont {Peng}, \citenamefont {Paul},\ and\ \citenamefont {Thompson}}]{zhang2024scaled}%
  \BibitemOpen
  \bibfield  {author} {\bibinfo {author} {\bibfnamefont {B.}~\bibnamefont {Zhang}}, \bibinfo {author} {\bibfnamefont {P.}~\bibnamefont {Peng}}, \bibinfo {author} {\bibfnamefont {A.}~\bibnamefont {Paul}},\ and\ \bibinfo {author} {\bibfnamefont {J.~D.}\ \bibnamefont {Thompson}},\ }\bibfield  {title} {\bibinfo {title} {Scaled local gate controller for optically addressed qubits},\ }\href {https://doi.org/10.1364/OPTICA.512155} {\bibfield  {journal} {\bibinfo  {journal} {Optica}\ }\textbf {\bibinfo {volume} {11}},\ \bibinfo {pages} {227} (\bibinfo {year} {2024})}\BibitemShut {NoStop}%
\bibitem [{\citenamefont {Christen}\ \emph {et~al.}(2025)\citenamefont {Christen}, \citenamefont {Propson}, \citenamefont {Sutula}, \citenamefont {Sattari}, \citenamefont {Choong}, \citenamefont {Panuski}, \citenamefont {Melville}, \citenamefont {Mallek}, \citenamefont {Brabec}, \citenamefont {Hamilton}, \citenamefont {Dixon}, \citenamefont {Menssen}, \citenamefont {Braje}, \citenamefont {Ghadimi},\ and\ \citenamefont {Englund}}]{christen2025integrated}%
  \BibitemOpen
  \bibfield  {author} {\bibinfo {author} {\bibfnamefont {I.}~\bibnamefont {Christen}}, \bibinfo {author} {\bibfnamefont {T.}~\bibnamefont {Propson}}, \bibinfo {author} {\bibfnamefont {M.}~\bibnamefont {Sutula}}, \bibinfo {author} {\bibfnamefont {H.}~\bibnamefont {Sattari}}, \bibinfo {author} {\bibfnamefont {G.}~\bibnamefont {Choong}}, \bibinfo {author} {\bibfnamefont {C.}~\bibnamefont {Panuski}}, \bibinfo {author} {\bibfnamefont {A.}~\bibnamefont {Melville}}, \bibinfo {author} {\bibfnamefont {J.}~\bibnamefont {Mallek}}, \bibinfo {author} {\bibfnamefont {C.}~\bibnamefont {Brabec}}, \bibinfo {author} {\bibfnamefont {S.}~\bibnamefont {Hamilton}}, \bibinfo {author} {\bibfnamefont {P.~B.}\ \bibnamefont {Dixon}}, \bibinfo {author} {\bibfnamefont {A.~J.}\ \bibnamefont {Menssen}}, \bibinfo {author} {\bibfnamefont {D.}~\bibnamefont {Braje}}, \bibinfo {author} {\bibfnamefont {A.~H.}\ \bibnamefont {Ghadimi}},\ and\ \bibinfo {author} {\bibfnamefont {D.}~\bibnamefont {Englund}},\ }\bibfield  {title} {\bibinfo {title} {An
  integrated photonic engine for programmable atomic control},\ }\href {https://doi.org/10.1038/s41467-024-55423-3} {\bibfield  {journal} {\bibinfo  {journal} {Nat. Commun.}\ }\textbf {\bibinfo {volume} {16}},\ \bibinfo {pages} {82} (\bibinfo {year} {2025})}\BibitemShut {NoStop}%
\bibitem [{\citenamefont {Lin}\ \emph {et~al.}(2025)\citenamefont {Lin}, \citenamefont {Fang}, \citenamefont {Yu}, \citenamefont {Xi}, \citenamefont {Shao}, \citenamefont {Li},\ and\ \citenamefont {Li}}]{lin2025optical}%
  \BibitemOpen
  \bibfield  {author} {\bibinfo {author} {\bibfnamefont {Q.}~\bibnamefont {Lin}}, \bibinfo {author} {\bibfnamefont {S.}~\bibnamefont {Fang}}, \bibinfo {author} {\bibfnamefont {Y.}~\bibnamefont {Yu}}, \bibinfo {author} {\bibfnamefont {Z.}~\bibnamefont {Xi}}, \bibinfo {author} {\bibfnamefont {L.}~\bibnamefont {Shao}}, \bibinfo {author} {\bibfnamefont {B.}~\bibnamefont {Li}},\ and\ \bibinfo {author} {\bibfnamefont {M.}~\bibnamefont {Li}},\ }\bibfield  {title} {\bibinfo {title} {Optical multi-beam steering and communication using integrated acousto-optics arrays},\ }\href {https://doi.org/10.1038/s41467-025-59831-x} {\bibfield  {journal} {\bibinfo  {journal} {Nat. Commun.}\ }\textbf {\bibinfo {volume} {16}},\ \bibinfo {pages} {4501} (\bibinfo {year} {2025})}\BibitemShut {NoStop}%
\bibitem [{\citenamefont {Zhao}\ \emph {et~al.}(2025)\citenamefont {Zhao}, \citenamefont {Singh}, \citenamefont {Singh}, \citenamefont {Thoreen}, \citenamefont {DeAngelo}, \citenamefont {Dominguez}, \citenamefont {Leenheer}, \citenamefont {Peyskens}, \citenamefont {Lukin}, \citenamefont {Englund}, \citenamefont {Eichenfield}, \citenamefont {Gemelke},\ and\ \citenamefont {Wan}}]{zhao2025integrated}%
  \BibitemOpen
  \bibfield  {author} {\bibinfo {author} {\bibfnamefont {M.}~\bibnamefont {Zhao}}, \bibinfo {author} {\bibfnamefont {M.}~\bibnamefont {Singh}}, \bibinfo {author} {\bibfnamefont {A.}~\bibnamefont {Singh}}, \bibinfo {author} {\bibfnamefont {H.}~\bibnamefont {Thoreen}}, \bibinfo {author} {\bibfnamefont {R.~J.}\ \bibnamefont {DeAngelo}}, \bibinfo {author} {\bibfnamefont {D.}~\bibnamefont {Dominguez}}, \bibinfo {author} {\bibfnamefont {A.}~\bibnamefont {Leenheer}}, \bibinfo {author} {\bibfnamefont {F.}~\bibnamefont {Peyskens}}, \bibinfo {author} {\bibfnamefont {A.}~\bibnamefont {Lukin}}, \bibinfo {author} {\bibfnamefont {D.}~\bibnamefont {Englund}}, \bibinfo {author} {\bibfnamefont {M.}~\bibnamefont {Eichenfield}}, \bibinfo {author} {\bibfnamefont {N.}~\bibnamefont {Gemelke}},\ and\ \bibinfo {author} {\bibfnamefont {N.~H.}\ \bibnamefont {Wan}},\ }\bibfield  {title} {\bibinfo {title} {An integrated photonics platform for high-speed, ultrahigh-extinction, many-channel quantum control},\ }\href
  {https://doi.org/10.48550/arXiv.2508.09920} {\bibfield  {journal} {\bibinfo  {journal} {arXiv:2508.09920}\ } (\bibinfo {year} {2025})}\BibitemShut {NoStop}%
\bibitem [{\citenamefont {Li}\ \emph {et~al.}(2025)\citenamefont {Li}, \citenamefont {Hou}, \citenamefont {Wang}, \citenamefont {Wang}, \citenamefont {He}, \citenamefont {Zhou}, \citenamefont {Wang}, \citenamefont {Liu}, \citenamefont {Wang}, \citenamefont {Xu},\ and\ \citenamefont {Zhan}}]{li2025fiber}%
  \BibitemOpen
  \bibfield  {author} {\bibinfo {author} {\bibfnamefont {X.}~\bibnamefont {Li}}, \bibinfo {author} {\bibfnamefont {J.-Y.}\ \bibnamefont {Hou}}, \bibinfo {author} {\bibfnamefont {J.-C.}\ \bibnamefont {Wang}}, \bibinfo {author} {\bibfnamefont {G.-W.}\ \bibnamefont {Wang}}, \bibinfo {author} {\bibfnamefont {X.-D.}\ \bibnamefont {He}}, \bibinfo {author} {\bibfnamefont {F.}~\bibnamefont {Zhou}}, \bibinfo {author} {\bibfnamefont {Y.-B.}\ \bibnamefont {Wang}}, \bibinfo {author} {\bibfnamefont {M.}~\bibnamefont {Liu}}, \bibinfo {author} {\bibfnamefont {J.}~\bibnamefont {Wang}}, \bibinfo {author} {\bibfnamefont {P.}~\bibnamefont {Xu}},\ and\ \bibinfo {author} {\bibfnamefont {M.-S.}\ \bibnamefont {Zhan}},\ }\bibfield  {title} {\bibinfo {title} {A fiber array architecture for atom quantum computing},\ }\href {https://doi.org/10.1038/s41467-025-64738-8} {\bibfield  {journal} {\bibinfo  {journal} {Nat. Commun.}\ }\textbf {\bibinfo {volume} {16}},\ \bibinfo {pages} {9728} (\bibinfo {year} {2025})}\BibitemShut {NoStop}%
\bibitem [{\citenamefont {Maunz}\ \emph {et~al.}(2009)\citenamefont {Maunz}, \citenamefont {Olmschenk}, \citenamefont {Hayes}, \citenamefont {Matsukevich}, \citenamefont {Duan},\ and\ \citenamefont {Monroe}}]{maunz2009heralded}%
  \BibitemOpen
  \bibfield  {author} {\bibinfo {author} {\bibfnamefont {P.}~\bibnamefont {Maunz}}, \bibinfo {author} {\bibfnamefont {S.}~\bibnamefont {Olmschenk}}, \bibinfo {author} {\bibfnamefont {D.}~\bibnamefont {Hayes}}, \bibinfo {author} {\bibfnamefont {D.~N.}\ \bibnamefont {Matsukevich}}, \bibinfo {author} {\bibfnamefont {L.-M.}\ \bibnamefont {Duan}},\ and\ \bibinfo {author} {\bibfnamefont {C.}~\bibnamefont {Monroe}},\ }\bibfield  {title} {\bibinfo {title} {Heralded {{Quantum Gate}} between {{Remote Quantum Memories}}},\ }\href {https://doi.org/10.1103/PhysRevLett.102.250502} {\bibfield  {journal} {\bibinfo  {journal} {Phys. Rev. Lett.}\ }\textbf {\bibinfo {volume} {102}},\ \bibinfo {pages} {250502} (\bibinfo {year} {2009})}\BibitemShut {NoStop}%
\bibitem [{\citenamefont {Gottesman}\ and\ \citenamefont {Chuang}(1999)}]{gottesman1999demonstrating}%
  \BibitemOpen
  \bibfield  {author} {\bibinfo {author} {\bibfnamefont {D.}~\bibnamefont {Gottesman}}\ and\ \bibinfo {author} {\bibfnamefont {I.~L.}\ \bibnamefont {Chuang}},\ }\bibfield  {title} {\bibinfo {title} {Demonstrating the viability of universal quantum computation using teleportation and single-qubit operations},\ }\href {https://doi.org/10.1038/46503} {\bibfield  {journal} {\bibinfo  {journal} {Nature}\ }\textbf {\bibinfo {volume} {402}},\ \bibinfo {pages} {390} (\bibinfo {year} {1999})}\BibitemShut {NoStop}%
\bibitem [{\citenamefont {{van Leent}}\ \emph {et~al.}(2022)\citenamefont {{van Leent}}, \citenamefont {Bock}, \citenamefont {Fertig}, \citenamefont {Garthoff}, \citenamefont {Eppelt}, \citenamefont {Zhou}, \citenamefont {Malik}, \citenamefont {Seubert}, \citenamefont {Bauer}, \citenamefont {Rosenfeld}, \citenamefont {Zhang}, \citenamefont {Becher},\ and\ \citenamefont {Weinfurter}}]{vanleent2022entangling}%
  \BibitemOpen
  \bibfield  {author} {\bibinfo {author} {\bibfnamefont {T.}~\bibnamefont {{van Leent}}}, \bibinfo {author} {\bibfnamefont {M.}~\bibnamefont {Bock}}, \bibinfo {author} {\bibfnamefont {F.}~\bibnamefont {Fertig}}, \bibinfo {author} {\bibfnamefont {R.}~\bibnamefont {Garthoff}}, \bibinfo {author} {\bibfnamefont {S.}~\bibnamefont {Eppelt}}, \bibinfo {author} {\bibfnamefont {Y.}~\bibnamefont {Zhou}}, \bibinfo {author} {\bibfnamefont {P.}~\bibnamefont {Malik}}, \bibinfo {author} {\bibfnamefont {M.}~\bibnamefont {Seubert}}, \bibinfo {author} {\bibfnamefont {T.}~\bibnamefont {Bauer}}, \bibinfo {author} {\bibfnamefont {W.}~\bibnamefont {Rosenfeld}}, \bibinfo {author} {\bibfnamefont {W.}~\bibnamefont {Zhang}}, \bibinfo {author} {\bibfnamefont {C.}~\bibnamefont {Becher}},\ and\ \bibinfo {author} {\bibfnamefont {H.}~\bibnamefont {Weinfurter}},\ }\bibfield  {title} {\bibinfo {title} {Entangling single atoms over 33 km telecom fibre},\ }\href {https://doi.org/10.1038/s41586-022-04764-4} {\bibfield  {journal} {\bibinfo
  {journal} {Nature}\ }\textbf {\bibinfo {volume} {607}},\ \bibinfo {pages} {69} (\bibinfo {year} {2022})}\BibitemShut {NoStop}%
\bibitem [{\citenamefont {Singh}\ \emph {et~al.}(2024)\citenamefont {Singh}, \citenamefont {Gu}, \citenamefont {de~Bone}, \citenamefont {Villase{\~n}or}, \citenamefont {Elkouss},\ and\ \citenamefont {Borregaard}}]{singh2024modular}%
  \BibitemOpen
  \bibfield  {author} {\bibinfo {author} {\bibfnamefont {S.}~\bibnamefont {Singh}}, \bibinfo {author} {\bibfnamefont {F.}~\bibnamefont {Gu}}, \bibinfo {author} {\bibfnamefont {S.}~\bibnamefont {de~Bone}}, \bibinfo {author} {\bibfnamefont {E.}~\bibnamefont {Villase{\~n}or}}, \bibinfo {author} {\bibfnamefont {D.}~\bibnamefont {Elkouss}},\ and\ \bibinfo {author} {\bibfnamefont {J.}~\bibnamefont {Borregaard}},\ }\bibfield  {title} {\bibinfo {title} {Modular {{Architectures}} and {{Entanglement Schemes}} for {{Error-Corrected Distributed Quantum Computation}}},\ }\href {https://doi.org/10.48550/arXiv.2408.02837} {\bibfield  {journal} {\bibinfo  {journal} {arXiv:2408.02837}\ } (\bibinfo {year} {2024})}\BibitemShut {NoStop}%
\bibitem [{\citenamefont {Hofmann}\ \emph {et~al.}(2012)\citenamefont {Hofmann}, \citenamefont {Krug}, \citenamefont {Ortegel}, \citenamefont {G{\'e}rard}, \citenamefont {Weber}, \citenamefont {Rosenfeld},\ and\ \citenamefont {Weinfurter}}]{hofmann2012heralded}%
  \BibitemOpen
  \bibfield  {author} {\bibinfo {author} {\bibfnamefont {J.}~\bibnamefont {Hofmann}}, \bibinfo {author} {\bibfnamefont {M.}~\bibnamefont {Krug}}, \bibinfo {author} {\bibfnamefont {N.}~\bibnamefont {Ortegel}}, \bibinfo {author} {\bibfnamefont {L.}~\bibnamefont {G{\'e}rard}}, \bibinfo {author} {\bibfnamefont {M.}~\bibnamefont {Weber}}, \bibinfo {author} {\bibfnamefont {W.}~\bibnamefont {Rosenfeld}},\ and\ \bibinfo {author} {\bibfnamefont {H.}~\bibnamefont {Weinfurter}},\ }\bibfield  {title} {\bibinfo {title} {Heralded {{Entanglement Between Widely Separated Atoms}}},\ }\href {https://doi.org/10.1126/science.1221856} {\bibfield  {journal} {\bibinfo  {journal} {Science}\ }\textbf {\bibinfo {volume} {337}},\ \bibinfo {pages} {72} (\bibinfo {year} {2012})}\BibitemShut {NoStop}%
\bibitem [{\citenamefont {Rosenfeld}\ \emph {et~al.}(2017)\citenamefont {Rosenfeld}, \citenamefont {Burchardt}, \citenamefont {Garthoff}, \citenamefont {Redeker}, \citenamefont {Ortegel}, \citenamefont {Rau},\ and\ \citenamefont {Weinfurter}}]{rosenfeld2017eventready}%
  \BibitemOpen
  \bibfield  {author} {\bibinfo {author} {\bibfnamefont {W.}~\bibnamefont {Rosenfeld}}, \bibinfo {author} {\bibfnamefont {D.}~\bibnamefont {Burchardt}}, \bibinfo {author} {\bibfnamefont {R.}~\bibnamefont {Garthoff}}, \bibinfo {author} {\bibfnamefont {K.}~\bibnamefont {Redeker}}, \bibinfo {author} {\bibfnamefont {N.}~\bibnamefont {Ortegel}}, \bibinfo {author} {\bibfnamefont {M.}~\bibnamefont {Rau}},\ and\ \bibinfo {author} {\bibfnamefont {H.}~\bibnamefont {Weinfurter}},\ }\bibfield  {title} {\bibinfo {title} {Event-{{Ready Bell Test Using Entangled Atoms Simultaneously Closing Detection}} and {{Locality Loopholes}}},\ }\href {https://doi.org/10.1103/PhysRevLett.119.010402} {\bibfield  {journal} {\bibinfo  {journal} {Phys. Rev. Lett.}\ }\textbf {\bibinfo {volume} {119}},\ \bibinfo {pages} {010402} (\bibinfo {year} {2017})}\BibitemShut {NoStop}%
\bibitem [{\citenamefont {Stephenson}\ \emph {et~al.}(2020)\citenamefont {Stephenson}, \citenamefont {Nadlinger}, \citenamefont {Nichol}, \citenamefont {An}, \citenamefont {Drmota}, \citenamefont {Ballance}, \citenamefont {Thirumalai}, \citenamefont {Goodwin}, \citenamefont {Lucas},\ and\ \citenamefont {Ballance}}]{stephenson2020highrate}%
  \BibitemOpen
  \bibfield  {author} {\bibinfo {author} {\bibfnamefont {L.~J.}\ \bibnamefont {Stephenson}}, \bibinfo {author} {\bibfnamefont {D.~P.}\ \bibnamefont {Nadlinger}}, \bibinfo {author} {\bibfnamefont {B.~C.}\ \bibnamefont {Nichol}}, \bibinfo {author} {\bibfnamefont {S.}~\bibnamefont {An}}, \bibinfo {author} {\bibfnamefont {P.}~\bibnamefont {Drmota}}, \bibinfo {author} {\bibfnamefont {T.~G.}\ \bibnamefont {Ballance}}, \bibinfo {author} {\bibfnamefont {K.}~\bibnamefont {Thirumalai}}, \bibinfo {author} {\bibfnamefont {J.~F.}\ \bibnamefont {Goodwin}}, \bibinfo {author} {\bibfnamefont {D.~M.}\ \bibnamefont {Lucas}},\ and\ \bibinfo {author} {\bibfnamefont {C.~J.}\ \bibnamefont {Ballance}},\ }\bibfield  {title} {\bibinfo {title} {High-{{Rate}}, {{High-Fidelity Entanglement}} of {{Qubits Across}} an {{Elementary Quantum Network}}},\ }\href {https://doi.org/10.1103/PhysRevLett.124.110501} {\bibfield  {journal} {\bibinfo  {journal} {Phys. Rev. Lett.}\ }\textbf {\bibinfo {volume} {124}},\ \bibinfo {pages} {110501} (\bibinfo
  {year} {2020})}\BibitemShut {NoStop}%
\bibitem [{\citenamefont {O'Reilly}\ \emph {et~al.}(2024)\citenamefont {O'Reilly}, \citenamefont {Toh}, \citenamefont {Goetting}, \citenamefont {Saha}, \citenamefont {Shalaev}, \citenamefont {Carter}, \citenamefont {Risinger}, \citenamefont {Kalakuntla}, \citenamefont {Li}, \citenamefont {Verma},\ and\ \citenamefont {Monroe}}]{oreilly2024fast}%
  \BibitemOpen
  \bibfield  {author} {\bibinfo {author} {\bibfnamefont {J.}~\bibnamefont {O'Reilly}}, \bibinfo {author} {\bibfnamefont {G.}~\bibnamefont {Toh}}, \bibinfo {author} {\bibfnamefont {I.}~\bibnamefont {Goetting}}, \bibinfo {author} {\bibfnamefont {S.}~\bibnamefont {Saha}}, \bibinfo {author} {\bibfnamefont {M.}~\bibnamefont {Shalaev}}, \bibinfo {author} {\bibfnamefont {A.~L.}\ \bibnamefont {Carter}}, \bibinfo {author} {\bibfnamefont {A.}~\bibnamefont {Risinger}}, \bibinfo {author} {\bibfnamefont {A.}~\bibnamefont {Kalakuntla}}, \bibinfo {author} {\bibfnamefont {T.}~\bibnamefont {Li}}, \bibinfo {author} {\bibfnamefont {A.}~\bibnamefont {Verma}},\ and\ \bibinfo {author} {\bibfnamefont {C.}~\bibnamefont {Monroe}},\ }\bibfield  {title} {\bibinfo {title} {Fast {{Photon-Mediated Entanglement}} of {{Continuously Cooled Trapped Ions}} for {{Quantum Networking}}},\ }\href {https://doi.org/10.1103/PhysRevLett.133.090802} {\bibfield  {journal} {\bibinfo  {journal} {Phys. Rev. Lett.}\ }\textbf {\bibinfo {volume} {133}},\
  \bibinfo {pages} {090802} (\bibinfo {year} {2024})}\BibitemShut {NoStop}%
\bibitem [{\citenamefont {Thomas}\ \emph {et~al.}(2022)\citenamefont {Thomas}, \citenamefont {Ruscio}, \citenamefont {Morin},\ and\ \citenamefont {Rempe}}]{thomas2022efficient}%
  \BibitemOpen
  \bibfield  {author} {\bibinfo {author} {\bibfnamefont {P.}~\bibnamefont {Thomas}}, \bibinfo {author} {\bibfnamefont {L.}~\bibnamefont {Ruscio}}, \bibinfo {author} {\bibfnamefont {O.}~\bibnamefont {Morin}},\ and\ \bibinfo {author} {\bibfnamefont {G.}~\bibnamefont {Rempe}},\ }\bibfield  {title} {\bibinfo {title} {Efficient generation of entangled multiphoton graph states from a single atom},\ }\href {https://doi.org/10.1038/s41586-022-04987-5} {\bibfield  {journal} {\bibinfo  {journal} {Nature}\ }\textbf {\bibinfo {volume} {608}},\ \bibinfo {pages} {677} (\bibinfo {year} {2022})}\BibitemShut {NoStop}%
\bibitem [{\citenamefont {Morin}\ \emph {et~al.}(2019)\citenamefont {Morin}, \citenamefont {K{\"o}rber}, \citenamefont {Langenfeld},\ and\ \citenamefont {Rempe}}]{morin2019deterministic}%
  \BibitemOpen
  \bibfield  {author} {\bibinfo {author} {\bibfnamefont {O.}~\bibnamefont {Morin}}, \bibinfo {author} {\bibfnamefont {M.}~\bibnamefont {K{\"o}rber}}, \bibinfo {author} {\bibfnamefont {S.}~\bibnamefont {Langenfeld}},\ and\ \bibinfo {author} {\bibfnamefont {G.}~\bibnamefont {Rempe}},\ }\bibfield  {title} {\bibinfo {title} {Deterministic {{Shaping}} and {{Reshaping}} of {{Single-Photon Temporal Wave Functions}}},\ }\href {https://doi.org/10.1103/PhysRevLett.123.133602} {\bibfield  {journal} {\bibinfo  {journal} {Phys. Rev. Lett.}\ }\textbf {\bibinfo {volume} {123}},\ \bibinfo {pages} {133602} (\bibinfo {year} {2019})}\BibitemShut {NoStop}%
\bibitem [{\citenamefont {Berthusen}\ \emph {et~al.}(2025)\citenamefont {Berthusen}, \citenamefont {Tan}, \citenamefont {Huang},\ and\ \citenamefont {Gottesman}}]{berthusen2025adaptive}%
  \BibitemOpen
  \bibfield  {author} {\bibinfo {author} {\bibfnamefont {N.}~\bibnamefont {Berthusen}}, \bibinfo {author} {\bibfnamefont {S.~J.~S.}\ \bibnamefont {Tan}}, \bibinfo {author} {\bibfnamefont {E.}~\bibnamefont {Huang}},\ and\ \bibinfo {author} {\bibfnamefont {D.}~\bibnamefont {Gottesman}},\ }\bibfield  {title} {\bibinfo {title} {Adaptive {{Syndrome Extraction}}},\ }\href {https://doi.org/10.1103/ps3r-wf84} {\bibfield  {journal} {\bibinfo  {journal} {PRX Quantum}\ }\textbf {\bibinfo {volume} {6}},\ \bibinfo {pages} {030307} (\bibinfo {year} {2025})}\BibitemShut {NoStop}%
\bibitem [{\citenamefont {Utama}\ \emph {et~al.}(2021)\citenamefont {Utama}, \citenamefont {Chow}, \citenamefont {Nguyen},\ and\ \citenamefont {Kurtsiefer}}]{utama2021coupling}%
  \BibitemOpen
  \bibfield  {author} {\bibinfo {author} {\bibfnamefont {A.~N.}\ \bibnamefont {Utama}}, \bibinfo {author} {\bibfnamefont {C.~H.}\ \bibnamefont {Chow}}, \bibinfo {author} {\bibfnamefont {C.~H.}\ \bibnamefont {Nguyen}},\ and\ \bibinfo {author} {\bibfnamefont {C.}~\bibnamefont {Kurtsiefer}},\ }\bibfield  {title} {\bibinfo {title} {Coupling light to higher order transverse modes of a near-concentric optical cavity},\ }\href {https://doi.org/10.1364/OE.413737} {\bibfield  {journal} {\bibinfo  {journal} {Opt. Express}\ }\textbf {\bibinfo {volume} {29}},\ \bibinfo {pages} {8130} (\bibinfo {year} {2021})}\BibitemShut {NoStop}%
\bibitem [{\citenamefont {Dalibard}\ and\ \citenamefont {Cohen-Tannoudji}(1985)}]{Dalibard}%
  \BibitemOpen
  \bibfield  {author} {\bibinfo {author} {\bibfnamefont {J.}~\bibnamefont {Dalibard}}\ and\ \bibinfo {author} {\bibfnamefont {C.}~\bibnamefont {Cohen-Tannoudji}},\ }\bibfield  {title} {\bibinfo {title} {Dressed-atom approach to atomic motion in laser light: the dipole force revisited},\ }\href {https://doi.org/10.1364/JOSAB.2.001707} {\bibfield  {journal} {\bibinfo  {journal} {J. Opt. Soc. Am. B}\ }\textbf {\bibinfo {volume} {2}},\ \bibinfo {pages} {1707} (\bibinfo {year} {1985})}\BibitemShut {NoStop}%
\bibitem [{\citenamefont {Schmiegelow}\ \emph {et~al.}(2016)\citenamefont {Schmiegelow}, \citenamefont {Kaufmann}, \citenamefont {Ruster}, \citenamefont {Schulz}, \citenamefont {Kaushal}, \citenamefont {Hettrich}, \citenamefont {Schmidt-Kaler},\ and\ \citenamefont {Poschinger}}]{poschinger2016ion6nm}%
  \BibitemOpen
  \bibfield  {author} {\bibinfo {author} {\bibfnamefont {C.~T.}\ \bibnamefont {Schmiegelow}}, \bibinfo {author} {\bibfnamefont {H.}~\bibnamefont {Kaufmann}}, \bibinfo {author} {\bibfnamefont {T.}~\bibnamefont {Ruster}}, \bibinfo {author} {\bibfnamefont {J.}~\bibnamefont {Schulz}}, \bibinfo {author} {\bibfnamefont {V.}~\bibnamefont {Kaushal}}, \bibinfo {author} {\bibfnamefont {M.}~\bibnamefont {Hettrich}}, \bibinfo {author} {\bibfnamefont {F.}~\bibnamefont {Schmidt-Kaler}},\ and\ \bibinfo {author} {\bibfnamefont {U.~G.}\ \bibnamefont {Poschinger}},\ }\bibfield  {title} {\bibinfo {title} {Phase-stable free-space optical lattices for trapped ions},\ }\href {https://doi.org/10.1103/PhysRevLett.116.033002} {\bibfield  {journal} {\bibinfo  {journal} {Phys. Rev. Lett.}\ }\textbf {\bibinfo {volume} {116}},\ \bibinfo {pages} {033002} (\bibinfo {year} {2016})}\BibitemShut {NoStop}%
\bibitem [{\citenamefont {Clements}\ \emph {et~al.}(2026)\citenamefont {Clements}, \citenamefont {Knollmann}, \citenamefont {Corsetti}, \citenamefont {Li}, \citenamefont {Hattori}, \citenamefont {Notaros}, \citenamefont {Swint}, \citenamefont {Sneh}, \citenamefont {Kim}, \citenamefont {Leu}, \citenamefont {Callahan}, \citenamefont {Mahony}, \citenamefont {West}, \citenamefont {Sorace-Agaskar}, \citenamefont {Kharas}, \citenamefont {McConnell}, \citenamefont {Bruzewicz}, \citenamefont {Chuang}, \citenamefont {Notaros},\ and\ \citenamefont {Chiaverini}}]{felix2026ion10nm}%
  \BibitemOpen
  \bibfield  {author} {\bibinfo {author} {\bibfnamefont {E.}~\bibnamefont {Clements}}, \bibinfo {author} {\bibfnamefont {F.~W.}\ \bibnamefont {Knollmann}}, \bibinfo {author} {\bibfnamefont {S.}~\bibnamefont {Corsetti}}, \bibinfo {author} {\bibfnamefont {Z.}~\bibnamefont {Li}}, \bibinfo {author} {\bibfnamefont {A.}~\bibnamefont {Hattori}}, \bibinfo {author} {\bibfnamefont {M.}~\bibnamefont {Notaros}}, \bibinfo {author} {\bibfnamefont {R.}~\bibnamefont {Swint}}, \bibinfo {author} {\bibfnamefont {T.}~\bibnamefont {Sneh}}, \bibinfo {author} {\bibfnamefont {M.~E.}\ \bibnamefont {Kim}}, \bibinfo {author} {\bibfnamefont {A.~D.}\ \bibnamefont {Leu}}, \bibinfo {author} {\bibfnamefont {P.}~\bibnamefont {Callahan}}, \bibinfo {author} {\bibfnamefont {T.}~\bibnamefont {Mahony}}, \bibinfo {author} {\bibfnamefont {G.~N.}\ \bibnamefont {West}}, \bibinfo {author} {\bibfnamefont {C.}~\bibnamefont {Sorace-Agaskar}}, \bibinfo {author} {\bibfnamefont {D.}~\bibnamefont {Kharas}}, \bibinfo {author} {\bibfnamefont {R.}~\bibnamefont
  {McConnell}}, \bibinfo {author} {\bibfnamefont {C.~D.}\ \bibnamefont {Bruzewicz}}, \bibinfo {author} {\bibfnamefont {I.~L.}\ \bibnamefont {Chuang}}, \bibinfo {author} {\bibfnamefont {J.}~\bibnamefont {Notaros}},\ and\ \bibinfo {author} {\bibfnamefont {J.}~\bibnamefont {Chiaverini}},\ }\bibfield  {title} {\bibinfo {title} {Sub-doppler cooling of a trapped ion in a phase-stable polarization gradient},\ }\href {https://doi.org/10.1103/fy3t-f1hz} {\bibfield  {journal} {\bibinfo  {journal} {Phys. Rev. Lett.}\ }\textbf {\bibinfo {volume} {136}},\ \bibinfo {pages} {023201} (\bibinfo {year} {2026})}\BibitemShut {NoStop}%
\bibitem [{\citenamefont {Karpa}\ \emph {et~al.}(2013)\citenamefont {Karpa}, \citenamefont {Bylinskii}, \citenamefont {Gangloff}, \citenamefont {Cetina},\ and\ \citenamefont {Vuleti\ifmmode~\acute{c}\else \'{c}\fi{}}}]{vladan2013ion10nm}%
  \BibitemOpen
  \bibfield  {author} {\bibinfo {author} {\bibfnamefont {L.}~\bibnamefont {Karpa}}, \bibinfo {author} {\bibfnamefont {A.}~\bibnamefont {Bylinskii}}, \bibinfo {author} {\bibfnamefont {D.}~\bibnamefont {Gangloff}}, \bibinfo {author} {\bibfnamefont {M.}~\bibnamefont {Cetina}},\ and\ \bibinfo {author} {\bibfnamefont {V.}~\bibnamefont {Vuleti\ifmmode~\acute{c}\else \'{c}\fi{}}},\ }\bibfield  {title} {\bibinfo {title} {Suppression of ion transport due to long-lived subwavelength localization by an optical lattice},\ }\href {https://doi.org/10.1103/PhysRevLett.111.163002} {\bibfield  {journal} {\bibinfo  {journal} {Phys. Rev. Lett.}\ }\textbf {\bibinfo {volume} {111}},\ \bibinfo {pages} {163002} (\bibinfo {year} {2013})}\BibitemShut {NoStop}%
\bibitem [{\citenamefont {Shaw}\ \emph {et~al.}(2024)\citenamefont {Shaw}, \citenamefont {Finkelstein}, \citenamefont {Tsai}, \citenamefont {Scholl}, \citenamefont {Yoon}, \citenamefont {Choi},\ and\ \citenamefont {Endres}}]{Shaw2024MultiEnsemble}%
  \BibitemOpen
  \bibfield  {author} {\bibinfo {author} {\bibfnamefont {A.~L.}\ \bibnamefont {Shaw}}, \bibinfo {author} {\bibfnamefont {R.}~\bibnamefont {Finkelstein}}, \bibinfo {author} {\bibfnamefont {R.~B.-S.}\ \bibnamefont {Tsai}}, \bibinfo {author} {\bibfnamefont {P.}~\bibnamefont {Scholl}}, \bibinfo {author} {\bibfnamefont {T.~H.}\ \bibnamefont {Yoon}}, \bibinfo {author} {\bibfnamefont {J.}~\bibnamefont {Choi}},\ and\ \bibinfo {author} {\bibfnamefont {M.}~\bibnamefont {Endres}},\ }\bibfield  {title} {\bibinfo {title} {Multi-ensemble metrology by programming local rotations with atom movements},\ }\href {https://doi.org/10.1038/s41567-023-02323-w} {\bibfield  {journal} {\bibinfo  {journal} {Nat. Phys.}\ }\textbf {\bibinfo {volume} {20}},\ \bibinfo {pages} {195} (\bibinfo {year} {2024})}\BibitemShut {NoStop}%
\bibitem [{\citenamefont {Gillen-Christandl}\ \emph {et~al.}(2016)\citenamefont {Gillen-Christandl}, \citenamefont {Gillen}, \citenamefont {Piotrowicz},\ and\ \citenamefont {Saffman}}]{gillen2016supergaussian}%
  \BibitemOpen
  \bibfield  {author} {\bibinfo {author} {\bibfnamefont {K.}~\bibnamefont {Gillen-Christandl}}, \bibinfo {author} {\bibfnamefont {G.~D.}\ \bibnamefont {Gillen}}, \bibinfo {author} {\bibfnamefont {M.~J.}\ \bibnamefont {Piotrowicz}},\ and\ \bibinfo {author} {\bibfnamefont {M.}~\bibnamefont {Saffman}},\ }\bibfield  {title} {\bibinfo {title} {Comparison of gaussian and super gaussian laser beams for addressing atomic qubits},\ }\href {https://doi.org/10.1007/s00340-016-6407-y} {\bibfield  {journal} {\bibinfo  {journal} {Appl. Phys. B}\ }\textbf {\bibinfo {volume} {122}},\ \bibinfo {pages} {131} (\bibinfo {year} {2016})}\BibitemShut {NoStop}%
\end{thebibliography}%

\end{document}